\documentclass{report}

\newcommand{\version}{arxiv}

\usepackage{etoolbox}                              
\usepackage{calc}                                  
\usepackage{geometry,fancyhdr,setspace,emptypage}  
\usepackage{lmodern}                               
\usepackage{amsmath}                               
\usepackage{mathtools}                             
\usepackage{tensor}                                
\usepackage{tocloft}                               
\usepackage{enumitem}                              
\usepackage{booktabs,caption}                      
\usepackage{natbib}                                
\usepackage{textcomp}                              
\usepackage{hyperref}                              

\makeatletter
\patchcmd{\@makechapterhead}{\vspace*{50\p@}}{\vspace*{10pt}}{}{}
\patchcmd{\@makeschapterhead}{\vspace*{50\p@}}{\vspace*{10pt}}{}{}
\makeatother
\expandafter\ifstrequal\expandafter{\version}{arxiv}{%
  \geometry{margin=1in}}{%
\expandafter\ifstrequal\expandafter{\version}{bound}{%
  \geometry{paperwidth=8in,paperheight=10.25in,twoside,bindingoffset=0.5in,hmargin=0.75in,vmargin=1in,headsep=0.15in,footskip=0.4in}
  \newlength{\rulewidth}
  \setlength{\rulewidth}{0.4pt}
  \fancyhf{}
  \fancyhead[OR,EL]{\thepage}
  \fancyhead[OC]{\nouppercase{\rightmark}}
  \fancyhead[EC]{\nouppercase{\leftmark}}

  \fancypagestyle{plain}{%
    \fancyhf{}
    \fancyfoot[C]{\thepage}
    
    }
  \pagestyle{fancy}
  
  \renewcommand{\sectionmark}[1]{}}{%
\expandafter\ifstrequal\expandafter{\version}{library-electronic}{%
  \geometry{margin=1in,footskip=0.25in}
  \AtBeginDocument{\fontsize{12}{14}\selectfont}
  \doublespacing
  \setlength{\parindent}{2\parindent}}{%
\expandafter\ifstrequal\expandafter{\version}{library-hard}{%
  \geometry{bindingoffset=0.5in,margin=1in,footskip=0.25in}
  \AtBeginDocument{\fontsize{12}{14}\selectfont}
  \setlength{\parindent}{2\parindent}}{}}}}

\expandafter\ifstrequal\expandafter{\version}{library-hard}{%
  \setlength{\cftbeforechapskip}{0.7\baselineskip}}{}

\allowdisplaybreaks[1]

\expandafter\ifstrequal\expandafter{\version}{library-electronic}{%
  \setlist[itemize]{leftmargin=\parindent}
  \setlist[enumerate]{leftmargin=\parindent}}{%
\expandafter\ifstrequal\expandafter{\version}{library-hard}{%
  \setlist[itemize]{leftmargin=\parindent}
  \setlist[enumerate]{leftmargin=\parindent}}{}}

\hypersetup{hidelinks}

\makeatletter
\newcommand{\extref}[1]{\textup{\tagform@{#1}}}
\makeatother

\newcommand{\ee}{\mathrm{e}}
\newcommand{\dd}{\mathrm{d}}
\newcommand{\dt}{\dd t}
\newcommand{\dx}{\dd x}
\newcommand{\dr}{\dd r}
\newcommand{\dth}{\dd\theta}
\newcommand{\dph}{\dd\phi}
\newcommand{\pp}{\partial}
\newcommand{\del}{\nabla}
\DeclareMathOperator{\tr}{tr}
\DeclareMathOperator{\sgn}{sgn}

\newcommand{\altvec}[1]{\mathbf{#1}}
\DeclarePairedDelimiter{\paren}{\lparen}{\rparen}
\DeclarePairedDelimiter{\parenopen}{\lparen}{.}
\DeclarePairedDelimiter{\parenclose}{.}{\rparen}
\DeclarePairedDelimiter{\abs}{\lvert}{\rvert}
\DeclarePairedDelimiter{\ave}{\langle}{\rangle}
\DeclarePairedDelimiter{\set}{\lbrace}{\rbrace}

\newcommand{\cs}{c_\mathrm{s}}
\newcommand{\pgas}{p_\mathrm{gas}}
\newcommand{\pmag}{p_\mathrm{mag}}
\newcommand{\wgas}{w_\mathrm{gas}}
\newcommand{\wtot}{w_\mathrm{tot}}
\newcommand{\rcrit}{r_\mathrm{c}}
\newcommand{\ucrit}{u_\mathrm{c}}
\newcommand{\tcrit}{T_\mathrm{c}}
\newcommand{\rhor}{r_\mathrm{hor}}
\newcommand{\risco}{r_\mathrm{isco}}
\newcommand{\redge}{r_\mathrm{edge}}
\newcommand{\rpeak}{r_\mathrm{peak}}
\newcommand{\gammamax}{\gamma_\mathrm{max}}
\newcommand{\mprot}{m_\mathrm{p}}
\newcommand{\melec}{m_\mathrm{e}}
\newcommand{\msun}{M_\odot}
\newcommand{\cm}{\mathrm{cm}}
\newcommand{\g}{\mathrm{g}}

\newcommand{\phn}{\phantom{0}}
\newcommand{\phd}{\phantom{.}}

\begin{document}


\newlength{\titletopsep}
\newlength{\titlebottomsep}

\expandafter\ifstrequal\expandafter{\version}{arxiv}%
{
  \setlength{\titletopsep}{0.5in}
  \setlength{\titlebottomsep}{0.75in}
  \begin{titlepage}
    \begingroup \centering \Large
      \vspace*{\titletopsep}
      \textsc{Development and Application of Numerical Techniques \\ for General-Relativistic Magnetohydrodynamics \\ Simulations of Black Hole Accretion} \par
      \vspace{\stretch{1}}
      Christopher Joseph White \par
      \vspace{\stretch{1}}
      \textsc{A Dissertation Presented to \\ the Faculty of Princeton University \\ in Candidacy for the Degree of Doctor of Philosophy} \par
      \vspace{\stretch{1}}
      \textsc{Recommended for Acceptance \\ by the Department of Astrophysical Sciences} \\ Advisor: James M.\ Stone \par
      \vspace{\stretch{1}}
      September 2016 \par
      \vspace{\titlebottomsep}
    \endgroup
  \end{titlepage}
}%
{}

\expandafter\ifstrequal\expandafter{\version}{bound}%
{
  \setlength{\titletopsep}{0.5in}
  \setlength{\titlebottomsep}{0.75in}
  \begin{titlepage}
    \begingroup \centering \Large
      \vspace*{\titletopsep}
      \textsc{Development and Application \\ of Numerical Techniques for \\ General-Relativistic Magnetohydrodynamics \\ Simulations of Black Hole Accretion} \par
      \vspace{\stretch{1}}
      Christopher Joseph White \par
      \vspace{\stretch{1}}
      \textsc{A Dissertation Presented to \\ the Faculty of Princeton University \\ in Candidacy for the Degree of Doctor of Philosophy} \par
      \vspace{\stretch{1}}
      \textsc{Recommended for Acceptance \\ by the Department of Astrophysical Sciences} \\ Advisor: James M.\ Stone \par
      \vspace{\stretch{1}}
      September 2016 \par
      \vspace{\titlebottomsep}
    \endgroup
  \end{titlepage}
  \newpage
  \thispagestyle{empty}
  \hbox{}
}%
{}

\expandafter\ifstrequal\expandafter{\version}{library-electronic}%
{
  \setlength{\titletopsep}{0in}
  \setlength{\titlebottomsep}{1in}
  \begin{titlepage}
    \begingroup \centering \Large
      \vspace*{\titletopsep}
      \textsc{Development and Application of Numerical Techniques \\ for General-Relativistic Magnetohydrodynamics \\ Simulations of Black Hole Accretion} \par
      \vspace{\stretch{1}}
      Christopher Joseph White \par
      \vspace{\stretch{1}}
      \textsc{A Dissertation \\ Presented to the Faculty \\ of Princeton University \\ in Candidacy for the Degree \\ of Doctor of Philosophy} \par
      \vspace{\stretch{1}}
      \textsc{Recommended for Acceptance \\ by the Department of \\ Astrophysical Sciences} \\ Advisor: James M.\ Stone \par
      \vspace{\stretch{1}}
      September 2016 \par
      \vspace{\titlebottomsep}
    \endgroup
  \end{titlepage}
}%
{}

\expandafter\ifstrequal\expandafter{\version}{library-hard}%
{
  \setlength{\titletopsep}{0.5in}
  \setlength{\titlebottomsep}{0.75in}
  \begin{titlepage}
    \begingroup \centering \Large
      \vspace*{\titletopsep}
      \textsc{Development and Application of Numerical Techniques \\ for General-Relativistic Magnetohydrodynamics \\ Simulations of Black Hole Accretion} \par
      \vspace{\stretch{1}}
      Christopher Joseph White \par
      \vspace{\stretch{1}}
      \textsc{A Dissertation \\ Presented to the Faculty \\ of Princeton University \\ in Candidacy for the Degree \\ of Doctor of Philosophy} \par
      \vspace{\stretch{1}}
      \textsc{Recommended for Acceptance \\ by the Department of \\ Astrophysical Sciences} \\ Advisor: James M.\ Stone \par
      \vspace{\stretch{1}}
      September 2016 \par
      \vspace{\titlebottomsep}
    \endgroup
  \end{titlepage}
}%
{}


\newlength{\copyrighttopsep}
\setlength{\copyrighttopsep}{0.3\textheight}

\expandafter\ifstrequal\expandafter{\version}{arxiv}%
{}%
{}

\expandafter\ifstrequal\expandafter{\version}{bound}%
{}%
{}

\expandafter\ifstrequal\expandafter{\version}{library-electronic}%
{
  \thispagestyle{empty}
  \begingroup \centering
    \vspace*{\copyrighttopsep}
    \newcommand{\intersentencespace}{\spacefactor\sfcode`. \space}
    \textcopyright\intersentencespace Copyright by Christopher Joseph White, 2016. All rights reserved. \par
  \endgroup
}%
{}

\expandafter\ifstrequal\expandafter{\version}{library-hard}%
{
  \thispagestyle{empty}
  \begingroup \centering
    \vspace*{\copyrighttopsep}
    \newcommand{\intersentencespace}{\spacefactor\sfcode`. \space}
    \textcopyright\intersentencespace Copyright by Christopher Joseph White, 2016. All rights reserved. \par
  \endgroup
}%
{}

\pagenumbering{roman}

\chapter*{Abstract}
\addcontentsline{toc}{chapter}{\protect\numberline{}Abstract}

\expandafter\ifstrequal\expandafter{\version}{library-electronic}{\setcounter{page}{3}}{}

\expandafter\ifstrequal\expandafter{\version}{library-hard}{\setcounter{page}{3}}{}

We describe the implementation of sophisticated numerical techniques for general-relativistic magnetohydrodynamics simulations in the Athena++ code framework. Improvements over many existing codes include the use of advanced Riemann solvers and of staggered-mesh constrained transport. Combined with considerations for computational performance and parallel scalability, these allow us to investigate black hole accretion flows with unprecedented accuracy. The capability of the code is demonstrated by exploring magnetically arrested disks.


\chapter*{Acknowledgments}
\addcontentsline{toc}{chapter}{\protect\numberline{}Acknowledgments}

The success of a thesis project depends in no small way on the support and guidance received along the journey. From my parents I have received many years of encouragement, dating all the way back to the first telescope had as a child and the mindset of scientific inquiry it inspired. A number of friends have also helped me pursue my academic aspirations, at some times by listening to my ideas and at others by reminding me that taking breaks can be a good thing.

Many fruitful conversations have contributed to my knowledge and understanding in this particular field, especially at Princeton and amongst the members of the Horizon Collaboration theoretical and computational astrophysics network. In particular, Charles Gammie has provided invaluable insight into the development of general-relativistic numerical codes.

Finally, I would like to thank my advisor, Jim Stone. Three years ago he suggested that the time was right for numerical advances in relativity, fluid dynamics, and other fields of astrophysics to be applied to simulations, that black hole accretion in particular stood to gain from better computation and theory. My experience since then has shown me that this assessment was correct. This proved to be no simple undertaking -- and indeed I would have it no other way -- but Jim has ever been available to assist and counsel me.


\chapter*{Preface}
\addcontentsline{toc}{chapter}{\protect\numberline{}Preface}

There are many factors that motivate the writing of a doctoral dissertation, not the least of which is the requirement of doing so in order to obtain an advanced degree.

That said, one of the benefits of the dissertation format is the ability to expand upon ideas and to give details often omitted in other venues. This work was written with the intention of providing a clear and thorough reference for the research at hand, including not just the results but the motivation behind the project as well as a log of useful insights gained while executing it. The hope is that in the future a bright young graduate student can use this document to quickly learn enough to replicate the results. Furthermore, this serves to document the relativistic portions of the Athena++ code, whose development was a core component of this author's studies.

Especially in numerics, all too often there are critical tidbits left unspoken -- ordering of computations, magic numbers found by trial and error, conditions under which the code is not expected to work properly. Almost assuredly there are some points that have been forgotten and useful facts that have been omitted from this work. Still, perhaps enough information is presented here to eliminate much of the frustration our hypothetical student might have in learning the subject. Patience is asked of established researchers reading this document; the verbosity is intentional.

Versions of some of this writing can be found elsewhere, often in a more condensed format. In particular, part of Chapter~\ref{chap:intro} and much of Chapters~\ref{chap:method} and~\ref{chap:test} and Appendix~\ref{chap:transformation} can be found in \citet{White2016}. An extension of Chapter~\ref{chap:application} is also forthcoming.

As the work here is very much done in a relativistic setting, factors of $G$ and $c$ have been omitted from most equations, retained only in cases where they lend clarity to discussion involving conventional units. The $({-},{+},{+},{+})$ signature is used throughout. Greek indices are understood to run over all spacetime components of tensors, while Latin indices run over only spatial components.

\tableofcontents

\cleardoublepage
\pagenumbering{arabic}

\chapter[Introduction]{Introduction to the Study of Black Hole Accretion}
\label{chap:intro}

When considering black holes, an overabundance of possibilities for research avenues presents itself. Without focus there would be too much material for a single work to do the subject justice. We choose then to consider black holes in the broader context of astrophysical phenomena. In particular, we will study accretion of matter onto them, as this is a line of investigation rich with both theoretical considerations and observational constraints.

\section{Black Holes and Their Importance to Astrophysics}
\label{sec:intro:importance}

Black hole solutions to the equations of general relativity (GR) were first recognized shortly after the formulation of the theory a century ago \citep{Schwarzschild1916}. However, it was not until 1958 that these solutions were appreciated as describing actual physical phenomena \citetext{\citeauthor{Finkelstein1958}}. Moreover, the crucial ingredient of spin was only introduced in 1963 \citetext{\citeauthor{Kerr1963}}.

Since that time, however, black holes have been observed throughout the cosmos. Ranging from several to billions of times the mass of the Sun, they can be found scattered throughout our own galaxy and indeed in galaxies out to the far side of the observable universe.

\subsection{Stellar-Mass Black Holes}
\label{sec:intro:importance:stellar}

The lighter, stellar-mass black holes are now recognized as one of the common endpoints of stellar evolution \citep{Woosley1995}. The collapse of the iron cores of massive stars in some cases cannot be halted at nuclear densities, the pressure forces of the matter proving insufficient to overcome gravity.

Even were these black holes to do nothing else, the processes around their formation alone would justify considerable study. They are born not in the vacuum of space but rather inside a star. The large amount of surrounding matter is susceptible to falling into the black hole, as indeed some of it will. The small size of the hole, however, means the surrounding material will generally have too much angular momentum to fall inward, necessitating transport of angular momentum outward -- a common theme that will recur in this work. Indeed some material will accrete while the rest will be ejected, all taking place within the violent death throes of a star that is both collapsing and exploding.

The accretion onto the black hole itself is thought to power the jets we observe as long-duration gamma-ray bursts (GRBs) \citep{Woosley1993}. In this case, modeling GRBs requires understanding both supernovae and general-relativistic accretion flows.

Once the progenitor star no longer exists, its matter either being consumed or thrown off to infinity, it is possible that a stellar-mass black hole will enter a quiescent state. With no accreting matter to emit radiation, such a black hole would be the very definition of invisible. In this case, the only way to detect such objects would be through microlensing of passing starlight \citep[e.g.\ the MACHO Project,][]{Bennett2002}.

In fact, though, many stellar-mass black holes do not find themselves in such complete isolation. Stars often come in pairs and so there is often a stellar companion nearby. Though a black hole's gravity is no stronger than a star of equivalent mass, it can still find itself drawing material off the companion star. This infalling matter can form an accretion disk, and indeed such disks have been observed in what are known as X-ray binaries. This can happen either as it captures particles blown off the star in winds (high-mass X-ray binaries), or as the star evolves, swells, and overflows its Roche lobe (low-mass X-ray binaries). These systems provide ideal laboratories to study black hole accretion unobscured by surrounding matter as is found in GRBs and often with supermassive black holes.

\subsection{Supermassive Black Holes}
\label{sec:intro:importance:supermassive}

Dynamical observations of stars at the center of our galaxy \citep{Schodel2002,Gillessen2009} have shown that a very large mass, $4\times10^6\ \msun$, must be confined to a very small region, $2\times10^{15}\ \cm$. While such constraints could technically be satisfied by diffuse matter that has not collapsed to a black hole, or by a large collection of stars and other condensed objects, these alternate arrangements could only exist for short times in contrived scenarios before self-gravity collapses them. Moreover, there is a distinct lack of the visible signatures one would expect from large amounts of ordinary matter. Thus the community has concluded there must be a supermassive black hole of lying at the center of the galaxy.

Our galaxy is not unique in this regard. It is now believed that nearly all major galaxies host supermassive black holes ranging in mass from approximately $10^5\ \msun$ to $10^9\ \msun$ \citep{Ho2008}. The most energetic systems are known as active galactic nuclei (AGNs).

The very features that allow us to observe such distant black holes -- emission from hot accretion disks and relativistic jets -- hint at an important role these objects play in the evolution of galaxies. Accreting black holes emit large amounts of energy and momentum, affecting the galaxy as a whole. Ultimately, black holes provide a mechanism for extracting work from the enormous reservoir of gravitational potential energy in a relatively loosely bound system such as a galaxy (compared to the extreme of tight binding as in a black hole). That is, having a small amount of mass fall all the way to the horizon can release enough energy and momentum to unbind a significantly larger mass from the galactic potential.

This black hole feedback is now recognized as an important part of the dynamics and evolution of many galaxies \citep{Fabian2012}.

\subsection{Other Black Holes}
\label{sec:intro:importance:other}

Black holes of other mass ranges are discussed in the literature. For example, there is speculation that intermediate-mass black holes of hundreds to thousands of solar masses could exist \citep{Miller2004}. If they do occur, there is no reason to expect their accretion processes could not be modeled with the same techniques developed for lesser and greater masses. Formed from either direct collapse, black hole mergers, or accretion in dense star clusters or near supermassive black holes, such black holes even have claimed observations. However, the community has yet to reach a consensus on the interpretation of these observations or even on whether such black holes are prevalent in nature.

Another mass range of interest are very small black holes with masses less than that of the Moon. Any such black hole would presently be losing mass to Hawking radiation faster than it gains mass from cosmic microwave background photons. In fact, black holes formed in the very early universe with masses of around $10^{15}\ \g$ would be undergoing the final stages of evaporation in the present epoch \citep{MacGibbon2008}, leading to potentially observable outcomes at the ends of their lives. To date, no such phenomena have been unambiguously observed.

In the interest of connecting to observations, and recognizing that there is no shortage of science to be done with the commonly observed black holes, we restrict our attention to stellar-mass and supermassive black holes in the present work.

\section{Astrophysical Accretion}
\label{sec:intro:accretion}

Consider matter falling onto or orbiting around a black hole. Momentum transfer and/or initial conditions often work to make the angular momentum directions aligned between different parts of the flow. This results in a disk, whether it be thick or thin, at least whenever the infalling matter's density is large enough to have a large number of collisions between particles over the course of an orbit. We will be working in this regime, where the continuum fluid approximation holds.

As already mentioned, a generic feature of systems in which matter falls into black holes is that the infalling matter initially has too much angular momentum. Despite the effectiveness of angular momentum transfer in creating a disk, once such a disk is formed transfer via intermolecular viscous forces is far too slow to account for observed accretion rates. This is a general attribute of astrophysical accretion.

Despite the lack of a clear mechanism to transport angular momentum outward, models were developed based on the assumption that such a mechanism would be found. The most famous of these is the $\alpha$-disk of \citet{Shakura1973}. In that work they argue that in both turbulent and magnetic phenomena the $r\phi$-component of the stress tensor (i.e.\ the radial flux of angular momentum) is proportional to $\rho \cs^2$, where $\cs$ is the sound speed. The proportionality is effectively the turbulent Mach number in the former case and the square of the Alfv\'en Mach number in the latter. Calling this proportionality constant $\alpha$ and assuming it to be constant in a disk allows one to investigate models that capture the important features of the flow and allow for a connection to observations. Still, an ansatz is not a mechanism, and the explanation for the emergence of an effective viscosity took a number of years.

An extensive effort comprising both analytical and numerical work was dedicated to resolving this issue. A thorough review can be found in \citet{Balbus1998}. We present only the briefest summary here. The ability of turbulence to transport angular momentum (though not guaranteed in all situations) was recognized early, and so many investigations focused on demonstrating the inevitability of turbulence in astrophysical disks. Extremely high Reynolds numbers combined with the shear inherent in Keplerian motion would seem to promote turbulence. However as discussed in \citet{Balbus1996} the opposite signs of the angular velocity and angular momentum gradients actually lead to linear and nonlinear stability against turbulence. The idea of convection driving turbulence was also considered, but as shown in \citet{Ryu1992} and verified in simulations this has the tendency to transport angular inward rather than outward.

Attention was also given to global features of accretion disks, those that rely on the curvature and periodicity of the domain rather than just local shear, centrifugal, and Coriolis effects. The general results, though, were that such mechanisms do not solve the transport issue. The nonaxisymmetric instability of thick disks \citep{Papaloizou1984} was found to saturate without turbulence, and a tidally driven instability \citep{Goodman2003} also did not show clear signs of the desired angular momentum transport. Another tidal effect, that of spiral shocks, was however shown to lead to angular momentum transport, though this mechanism only applies when there is a massive companion as in a binary system.

The solution to the transport problem was discovered by \citet{Balbus1991}. It is not purely hydrodynamical in nature but rather relies on magnetic fields. Weak fields generically lead to a local instability, the magnetorotational instability (MRI). This results in turbulent flow, allowing effective angular momentum transport in disks.

The MRI can be pictured by imagining a vertical magnetic field line threading a disk. If the fluid containing the line at the disk midplane is perturbed outward, it will find itself rotating with a decreased angular velocity (by assumption; this instability depends on such an angular velocity profile). In ideal magnetohydrodynamics (MHD), flux freezing means we can consider the field line to be dragged with the fluid. The result is that the outwardly displaced part of the field line falls behind in rotation, leading to a magnetic tension that pulls the outer element forward in azimuth and the inner elements backward. This radial flux of azimuthal momentum is exactly the transport desired.

As discussed by \citeauthor{Balbus1991}, the MRI is linearly unstable:\ small perturbations grow exponentially. Moreover, it relies on a weak rather than strong field, and the above argument generalizes to other orientations of the field. As a result one generically expects it to manifest in nearly any ionized plasma in a roughly Keplerian disk.

Since its discovery, the MRI has been modeled in a large number of simulations, both local shearing boxes and global disks. It is found to lead to a turbulent state in which angular momentum is transported outward, driving accretion. The MRI eventually saturates in disks, leading to an effective viscosity.

Though initially investigated in the Newtonian regime, the general principles behind the MRI hold in a general-relativistic setting \citep{Gammie2004}. As a result, it becomes critical that electromagnetic fields be incorporated into our models. With a proper, relativistic treatment of hydrodynamics and electromagnetism, a diverse set of complex accretion phenomena can be understood.

Accretion in the regime we are considering is fundamentally a fluid dynamics problem (albeit one on a curved manifold). Moreover, we expect the systems we are considering to become turbulent. As such, purely analytic modeling can only take us so far, and numeric simulations must be utilized.

We note that for the purposes of studying black hole accretion, it is convenient that ideal systems are scale invariant. Everything scales with the mass of the central object, so for instance a given set of predictions or simulations for a black hole of mass $M$ holds with the understanding that the units of length are $GM/c^2$ and those of time are $GM/c^3$, regardless of whether $M = 1~\msun$ or $M = 10^6~\msun$. As a result, the theoretical and numerical study of supermassive black hole accretion is by no means disjoint from that of accretion onto smaller black holes. This freedom to scale results only breaks down if other assumptions are violated in the rescaling, such as if the accreting material becomes nonnegligibly self-gravitating, or if the density becomes low enough that the continuum fluid approximation no longer applies.

\section{Numerical Considerations}
\label{sec:intro:numerics}

The complex, nonlinear dynamics of most astrophysical fluids make numerical simulations a key tool in understanding them, even in flat spacetime. A number of interesting phenomena, however, occur in regions of strong enough gravity that general relativity cannot be neglected, requiring numerical algorithms capable of evolving fluids in curved spacetimes. Examples include accretion onto black holes \citep{Abramowicz2013}, collapsar models of long-duration gamma-ray bursts \citep{Woosley1993}, and merging neutron star binaries \citep{Faber2012}, among others.

Some of the most successful numerical algorithms for solving the equations of compressible fluid dynamics are finite-volume methods, primarily due to their superior accuracy and stability for shock capturing \citep{VanLeer1979}. In such methods the domain is partitioned into discrete cells and the quantities stored and evolved are cell volume averages of the conserved quantities. In Godunov schemes, fluxes of conserved quantities are determined by solving Riemann problems at each interface. In particular, one considers the interface to be a plane separating two spatially constant fluids filling all space, and the solutions of the Riemann problem determine the time-averaged fluxes across the plane in this simplified problem. The overall scheme is described in \S\ref{sec:method:algorithm}.

Two key algorithmic components of a finite-volume Godunov scheme for MHD are the Riemann solver and the method by which the divergence-free constraint on the magnetic field is enforced. We now examine each of these in turn, reviewing the options that have been developed for implementing them.

\subsection{Riemann Solvers}
\label{sec:intro:numerics:riemann}

The accuracy of a simulation depends critically on the accuracy of the Riemann solver adopted. Methods that have been developed to solve the Riemann problem include:
\begin{enumerate}
  \item Central. As described in \citet{Kurganov2000}, one can neglect the internal wave structure in the Riemann fan (equivalently the eigenstructure of the conservation equations) entirely, using only the conserved states, their corresponding fluxes, and possibly maximal wavespeeds. Conceptually central schemes proceed from the state at one time level to the state at the next without necessarily finding fluxes through interfaces; in this sense they are not true Riemann solvers. Gradients of both conserved quantities and fluxes are used to compute time evolution, but these gradients are taken componentwise, allowing the scheme to be independent of the (usually highly nonlinear) relationship between components.
  \item Roe. Developed in \citet{Roe1981} and improved in \citet{Harten1983b}, this is an exact solver in that it finds exact solutions to a particular linearized form of the equations. Accuracy comes at the cost of computational complexity, as the fluxes are determined by solving an eigenvalue problem for a $7$-dimensional system (in MHD).
  \item HLL (Harten, Lax, and van~Leer). These approximate methods are built upon the foundation laid by \citet{Harten1983a}. They assume a certain number of waves propagate from discontinuities, neglecting some of the waves. They then base the fluxes on which waves are bounding the region of interest, i.e.\ which are the slowest leftgoing and rightgoing waves. A number of different HLL Riemann solvers have been developed for relativistic fluids, including the following:
  \begin{enumerate}
    \item LLF (local Lax--Friedrichs). The largest-in-magnitude eigenvalue of the Roe matrix, i.e.\ the fastest linear wavespeed, is taken to be the signal speed on both sides of the interface. This solver is computationally inexpensive even in relativity, as it has only a single intermediate state and neglects the physical nature of the left and right states. This comes at the cost of being diffusive for flows slower than the extremal wavespeed (see \S\ref{sec:test:linear}).
    \item HLLE (Harten--Lax--van~Leer+Einfeldt). This solver is nearly identical to LLF, having just a single intermediate state, with the only difference being that the two signal speeds are allowed to be different. One uses the speed of the fastest leftgoing wave, considering both sides of the interface, as well as the fastest rightgoing wave, as suggested by \citet{Einfeldt1988}. In practice HLLE and LLF give very similar results and have only small performance differences.
    \item HLLC (HLL+contact). By resolving not only the extremal waves but also the contact discontinuity in the Riemann fan, one can get better results. Such an HLLC Riemann solver is developed for special relativity in \citet{Mignone2005}, and we consider that algorithm for hydrodynamics. \Citet{Mignone2006} also develop an HLLC solver for relativistic MHD, though we forgo implementing this in favor of HLLD.
    \item HLLD (HLL+discontinuities). This solver is developed for relativistic MHD by \citet{Mignone2009}. Given the extremal (fast magnetosonic) wavespeeds, it resolves not only the contact but also the Alfv\'en waves. It improves the accuracy of the fluxes by lessening the numerical diffusion, though at the expense of requiring nonlinear root finds. In particular, one uses the secant method to find the total pressure across the contact. In the case of a strong longitudinal magnetic field, the conserved HLLE state is constructed and inverted using a Newton--Raphson solver in order to find a total pressure to initialize the secant method.
  \end{enumerate}
\end{enumerate}

Exact Riemann solvers are computationally expensive, especially in general-relativistic MHD (GRMHD). Most of the codes written to date employ simple solvers such as LLF or HLLE. However, the simplest approximate solvers tend to be much more diffusive than their more exact counterparts for subsonic flows. This has led to our decision to work to implement HLLC and HLLD in general relativity, as described in \S\ref{sec:method:algorithm}, following the suggestion by \citet{Pons1998} and \citet{Anton2006}.

\subsection{The Divergence-Free Constraint}
\label{sec:intro:numerics:constraint}

In the case of MHD, enforcing the divergence-free constraint is important to prevent spurious production of magnetic monopoles. A number of algorithms have been developed to this end, including the following:
\begin{enumerate}
  \item Divergence cleaning. Auxiliary equations are introduced in order to damp and/or advect to the boundaries any monopoles that are developed. These equations can be elliptic \citep{Brackbill1980,Ramshaw1983}, parabolic \citep{Marder1987}, or hyperbolic \citep{Dedner2002}, with only the last naturally able to respect causality in relativistic settings.
  \item Vector potential evolution. The vector potential components $A^i$ are evolved instead of the magnetic field components $B^i$. This naturally obeys the divergence-free constraint if the vector potential is appropriately staggered, though one must be careful to choose an appropriate gauge, especially when including mesh refinement \citep{Etienne2012}.
  \item Constrained transport (CT). The magnetic field is staggered in such a way as to maintain a specific discretized version of the constraint. \Citet{Evans1988} introduce staggered-mesh CT in a general-relativistic context.
  \item Flux-CT. \Citet{Toth2000} discusses how to alter the fluxes so as to preserve the divergence-free constraint on a particular stencil of cell-centered magnetic fields.
\end{enumerate}
We choose to employ the \citeauthor{Evans1988}\ method. The details of the algorithm are developed for Cartesian grids by \citet{Gardiner2005}, where only magnetic fields (and not velocities as in the original formulation) are located at interfaces between cells. Here we extend the \citeauthor{Gardiner2005}\ algorithm, used with great success in special relativity \citep{Stone2008,Beckwith2011}, to arbitrary stationary coordinate systems.

\section{Existing Codes}
\label{sec:intro:codes}

A number of Godunov scheme codes have been developed with GRMHD capabilities on a stationary background spacetime, including Harm \citep{Gammie2003}, \citet{Komissarov2004}, the Valencia group's code \citep{Anton2006}, and ECHO \citep{DelZanna2007}. In addition, some codes combine the GRMHD and Einstein equations, evolving a self-gravitating magnetized fluid. These numerical relativity (NR) codes include the Tokyo/Kyoto group's code \citep{Shibata2005}, \citet{Anderson2006}, WhiskyMHD \citep{Giacomazzo2007}, GRHydro \citep{Mosta2014}, SpEC \citep{Muhlberger2014}, and IllinoisGRMHD \citep{Etienne2015}.

Of these ten codes, only \citet{Komissarov2004} and \citet{Anton2006} describe the use of Roe-type Riemann solvers (with the latter making the point that they can in principle use any Riemann solver), and only \citet{Shibata2005} and \citet{Anton2006} describe the use of central, non-Godunov schemes. All others restrict themselves to HLLE and LLF.

Of the aforementioned ten codes, \citet{Anderson2006} uses hyperbolic divergence cleaning, IllinoisGRMHD uses a staggered vector potential, Harm and GRHydro use flux-CT, and the remaining six use a staggered CT scheme.

The properties of these codes are summarized in Table~\ref{tab:codes}. While we are not concerned with NR here, we highlight its presence to indicate the focus of a code.

\begin{table}
  \centering
  \caption{Properties of existing codes. \label{tab:codes}}
  \begin{tabular}{cccc}
    \toprule
    Code          & NR? & Flux Scheme        & Field Scheme                   \\
    \midrule
    Harm          & No                    & LLF, HLLE          & Flux-CT                        \\
    Komissarov    & No                    & Roe                & Staggered CT                   \\
    Tokyo/Kyoto   & Yes                   & Central            & Staggered CT                   \\
    Valencia      & No                    & Central, Roe, HLLE & Staggered CT                   \\
    Anderson      & Yes                   & LLF, HLLE          & Hyperbolic divergence cleaning \\
    WhiskyMHD     & Yes                   & HLLE               & Staggered CT                   \\
    ECHO          & No                    & HLLE               & Staggered CT                   \\
    GRHydro       & Yes                   & HLLE               & Flux-CT                        \\
    SpEC          & Yes                   & HLLE               & Staggered CT                   \\
    IllinoisGRMHD & Yes                   & HLLE               & Staggered potential            \\
    \bottomrule
  \end{tabular}
\end{table}

\section{Developing a New Code}
\label{sec:intro:new_code}

The choice to develop a new code -- even one that builds upon an established code -- should not be made lightly. Often the most efficient way forward is to use an existing tool. However, we feel there are sufficient ways in which existing GRMHD codes are not ideally suited for our purposes that it becomes worthwhile to invest the effort in writing a new code. Advanced features are missing from some codes, and it should be noted that a number are not even public.

Three key considerations stand out as important for guiding the development of our code. These are \emph{accuracy}, \emph{performance}, and \emph{modularity}. While any numerical procedure will have its approximations, and even the underlying physical models will be idealized in various ways, we nonetheless strive for as accurate a representation of nature as we can given the computational resources we have available. One simple way most simulations can gain more accuracy, at least to a certain extent, is to increase the resolution at which they model a scenario. This however leads to a greater computational cost, and so we are led to performance as another guiding principle. Moreover, performance can help codes simulate systems for longer times, allowing more complex behavior to emerge. Finally, recognizing that new algorithms and models are continuously being developed, we want our code to be modular enough to be able to adopt newly developed methods. These can be both numerical insights and more detailed physics.

The new Athena++ code framework is the result of these considerations. It employs finite-volume Godunov methods to evolve the GRMHD equations on any stationary spacetime. With Athena++ in general and GRMHD in particular, we have applied our guiding principles to achieve what we consider to be valuable features. We will highlight several of these in the following sections, after which we will summarize what rewards we anticipate will follow.

\subsection{Riemann Solvers and Constrained Transport}
\label{sec:intro:new_code:riemann_ct}

We have already detailed the choices of Riemann solver (\S\ref{sec:intro:numerics:riemann}) and method for updating the magnetic field (\S\ref{sec:intro:numerics:constraint}). Both considerations directly relate to the \emph{accuracy} of the solution, minimizing numerical diffusion and maintaining the antisymmetric nature of the electromagnetic field tensor respectively. The adoption of approximate Riemann solvers also helps \emph{performance}.

For constrained transport, there is little we can vary once the general method is chosen. For example, we do not support flux-CT or divergence cleaning schemes, as these require a fundamentally different (non-staggered) grid for the magnetic field. For Riemann solvers, however, a degree of \emph{modularity} can be achieved. In a Godunov scheme, the solver has the well-defined role of taking a left state and a right state and returning the associated fluxes. The rest of the code does not care about the internal working of the solver, and so we are free to implement a variety of solvers. LLF (both frame-transforming and not), HLLE (also both frame-transforming and not), HLLC (for hydrodynamics), and HLLD (for MHD) are all implemented in Athena++, allowing one to choose on a per-simulation basis the method best suited for the task at hand.

\subsection{Mesh Refinement}
\label{sec:intro:new_code:refinement}

In many simulations the resolution requirements are not constant in space, and sometimes they even vary in time. For grids at a single, fixed resolution this leads to a choice of underresolving features of interest (harming \emph{accuracy}) or overcomputing unnecessary values (harming \emph{performance}). For example, when studying an accretion disk -- especially a thin disk as occurs with radiative cooling -- one often does not care about the near-vacuum near the poles, but one would like to see details near the midplane of the disk.

One solution is to allow for mesh refinement. Athena++ has such a scheme available, to be described more fully in a forthcoming general code method paper. Here we summarize the salient features:
\begin{enumerate}
  \item The grid is already divided into blocks of equal logical size and shape (i.e.\ the numbers of cells in each direction are fixed constants) as part of domain decomposition for parallel computing.
  \item Any block can be subdivided into $8$ blocks of the same logical size ($4$ in two dimensions) recursively.
  \item Neighboring blocks are restricted to differ by no more than one refinement level. Any blocks that share a face, an edge, or a corner are considered neighbors.
  \item Once a block is refined, it is no longer simulated at the coarser level.
  \item Prolongation and restriction operators translate information between levels. They are applied to populate ghost zones whose corresponding active zones are at a different level.
\end{enumerate}
This scheme works with \emph{modularity} in that all choices of algorithm and physics relevant to this study are compatible with such refinement. At present all refinement levels proceed in lockstep; in the future we intend to allow for hierarchical timestepping. For all simulations in this work static as opposed to adaptive mesh refinement has proved sufficient.

One particularly useful feature of mesh refinement presents itself in polar grids where the emphasis is on the equatorial regions. With polar grids the timestep is often limited by the azimuthal thickness of the cells closest to both the pole and the inner boundary. This limit in fact scales as the square of the linear resolution:\ doubling the number of azimuthal cells makes each cell half as wide, while doubling the number of polar cells moves the innermost cell centers to approximately half their distance from the pole (under the small angle approximation). (The radial coordinate of the cell centers also decreases, but because we only go as far as the horizon and not all the way to the coordinate origin, this effect is negligible.) Compare this limit to that of Cartesian grids, where the timestep decreases linearly with the linear resolution.

Without refinement, doubling the linear resolution of a polar simulation results in $32$ times the computational cost, with at most a factor of $8$ able to be weak-scaled away by using more processors in parallel (see \S\ref{sec:intro:new_code:speed}). Thus one generally expects a factor of $4$ increase in wall time for every factor of $2$ in resolution when running polar simulations. With refinement, however, one can refine only areas away from the pole. In practice even multiple levels of refinement of such regions will leave the cells wider than the original cells near the pole, meaning the timestep does not decrease at all. That is, the computational cost only scales by a factor of about $8$, all of which can be handled with more processors with no decrease in wall time.

\subsection{Polar Boundary Conditions}
\label{sec:intro:new_code:poles}

As much of our interest is in systems well represented by polar coordinates, the coordinate singularity at the poles is an issue that must be confronted. The simple way of handling the singularity, found in a number of existing codes, is to excise a small cone around the pole from the domain, placing reflecting walls at the conical boundaries. Even in the limit that the cone opening angle vanishes, this leads to an unphysical boundary that impedes the flow of material and electromagnetic fields across the pole, harming \emph{accuracy}.

Our solution is to connect cells across the pole much as we connect cells in periodic domains. That is, we avoid having a domain boundary at the pole. There are four considerations that allow us to implement this feature:
\begin{enumerate}
  \item For cells at the edge of a block along a pole, their ghost zones must match cells across the pole. In particular, a cell with azimuthal extent $\phi_- < \phi < \phi_+$ should have as its neighbor in the polar direction ($\theta$ decreasing through $0$ or increasing through $\pi$) the cell with the same radial and polar extent but covering $\phi_-+\pi < \phi < \phi_++\pi$ (module $2\pi$ as necessary). These ghost zones are needed for determining left and right states in radial Riemann problems. Importantly, there are Riemann problems affected by these ghost zones that are not lying on the pole itself, and so their fluxes cannot be neglected.
  \item Vector quantities must be appropriately reflected when copied from an active zone to a ghost zone belonging to a block on the other side of a pole. This is straightforward and little different from implementing a hard reflecting wall boundary condition.
  \item Vanishing areas force fluxes to not directly contribute to updating cells. In particular, between two opposite cells straddling a pole, the constant-$r$ interface has no area. Whatever hydrodynamical fluxes are computed from the Riemann problem, we know these will not contribute toward the update of either cell's conserved quantities. In practice, naive implementation of the update might lead to $0/0$ errors, and so one must guard against this in the code. The Riemann problem cannot be ignored, however, as the fluxes of magnetic field correspond to electric field along the pole, and this is used to update other interfaces, in particular constant-$\phi$ interfaces bordering the pole.
  \item Physically coincident electric fields must be kept the same in their various representations. As mentioned in the previous point, the electric fields along the pole are important for self-consistent evolution; they cannot be neglected or held constant. However, there will be as many representations of such values as there are cells in a ring around the pole. Roundoff error can cause these representations to differ, leading to inconsistent evolution of the magnetic fields. This leads to the production of magnetic monopoles and even numerical instability. The solution is to have all such representations of the same electric field averaged, with the average overwriting all the values, after every partial time step. Note this changes the topology of the connectivity graph for block communication.
\end{enumerate}

The complexity of polar boundary bookkeeping is simplified by forcing all blocks bordering a particular pole to be at the same level of mesh refinement.

A physical effect of properly treating the poles is that the magnetic field does not get artificially wound up as it would around an excised cone. In simulations where jet launching is being studied such artifacts are a cause for concern, as field winding can contribute to creating a jet.

\subsection{Speed and Scalability}
\label{sec:intro:new_code:speed}

Independent of the particular algorithms or problems considered, the \emph{performance} of any scientific code is grounded by the serial speed and parallel scaling of computation.

Modern processors no longer improve performance by increasing the clock frequency. Instead, pipelining, vectorization, and more efficient hardware instructions are used to reduce the wall time of computations. This means performance is no longer tied to a single number beyond all control of the programmer. While ideally compilers will optimize code to the hardware, there are often performance benefits to be had by tuning a code.

In writing Athena++ we paid close attention to such issues. The following techniques proved useful in improving performance:
\begin{itemize}
  \item Minimizing expensive evaluations and memory access. We precompute trigonometric and other such expensive functions related to the stationary metric as much as possible. At the same time, we minimize memory usage by storing only 1D and 2D arrays of precomputed values, as discussed in \S\ref{sec:method:algorithm:storage}.
  \item Sweeping through arrays as they are laid out in memory. For example, the hydrodynamical primitive variables are stored in a $5 \times N_3 \times N_2 \times N_1$ array. All nested loops range over $N_1$ in the innermost loop, avoiding strides greater than unity and thus minimizing cache misses. This includes calculating the fluxes, where even the $x^2$- and $x^3$-fluxes are evaluated in the $x^1$-direction order.
  \item Vectorizing all innermost loops. While modern compilers can often automatically vectorize loops, even when that entails inlining functions defined in other compilation units, there are several failure modes we sought to address. For one, complicated functions sometimes require manual inlining (e.g.\ in the HLLD solver, the function to evaluate the residual given a guess for the contact pressure). Additionally, vectorization cannot happen if the loop we seek to vectorize has indeterminate loops within it, as happens with algorithms that iterate until convergence. In such cases we examine how many iterations are typically used in cases were convergence happens at all, enabling us to fix a constant number of iterations.
\end{itemize}

We measure hydrodynamics and MHD performance as a function of both the metric and the Riemann solver. For special relativity, we run a shock tube in Minkowski coordinates. We then run the same shock tube using the full general relativity framework but still in Minkowski coordinates. Finally, we run a Fishbone--Moncrief torus (\S\ref{sec:test:torus}) in Kerr--Schild coordinates with nonzero spin. All tests are 3D. For the Riemann solvers we use a non-frame-transforming LLF solver (LLF-NT, as used in many GRMHD codes), a frame-transforming LLF solver (LLF-T), and either HLLC or HLLD (both frame-transforming) as appropriate. The distinction between using or not using a frame transformation (\S\ref{sec:method:algorithm:transformation}) only applies to general relativity.

Tests were performed on a single core of an Intel Xeon E5-2670 ($2.6\ \mathrm{GHz}$ Sandy Bridge) processor. We report the number of cells updated per wall time second in Table~\ref{tab:speeds}.

\begin{table}
  \centering
  \caption{Athena++ single-core performance ($10^5$ cell updates per second). \label{tab:speeds}}
  \begin{tabular}{cccc}
    \toprule
    Riemann Solver & SR & GR:\ Minkowski & GR:\ Kerr--Schild \\
    \midrule
    \multicolumn{1}{l}{Hydrodynamics} \\
    \cmidrule(r){1-1}
      LLF-NT & $12.8\phn$ & $10.4\phn$ & $\phn3.82$ \\
      LLF-T  & ---        & $10.1\phn$ & $\phn4.28$ \\
      HLLC   & $\phn8.57$ & $\phn7.36$ & $\phn3.28$ \\
    \cmidrule(r){1-1}
    \multicolumn{1}{l}{MHD} \\
    \cmidrule(r){1-1}
      LLF-NT & $\phn3.24$ & $\phn4.70$ & $\phn2.94$ \\
      LLF-T  & ---        & $\phn2.60$ & $\phn1.97$ \\
      HLLD   & $\phn1.23$ & $\phn1.13$ & $\phn1.22$ \\
    \bottomrule
  \end{tabular}
\end{table}

The decrease in performance in going from special to general relativity with the Minkowski metric reflects the cost of using the general-relativistic variable inversion and wavespeed formulas. The full cost of general relativity in realistic but reasonable metrics is represented by the final column of numbers, where nontrivial Kerr--Schild geometric factors enter into most calculations.

For a fixed geometry and Riemann solver, MHD problems run at $1/4$ to $3/4$ the speed of pure hydrodynamics problems in both special and general relativity. We also performed tests using HLLE solvers. These ran at $93\pm6\%$ the speed of the corresponding LLF tests, with similar accuracies.

Single-core optimization is however not sufficient to make use of cutting-edge computational systems, where parallel computing is the central paradigm. Athena++ is written to efficiently balance computation across over one million message-passing interface (MPI) ranks, and it can also utilize OpenMP for threading across shared memory.

\begin{figure}
  \centering
  \includegraphics[width=4in]{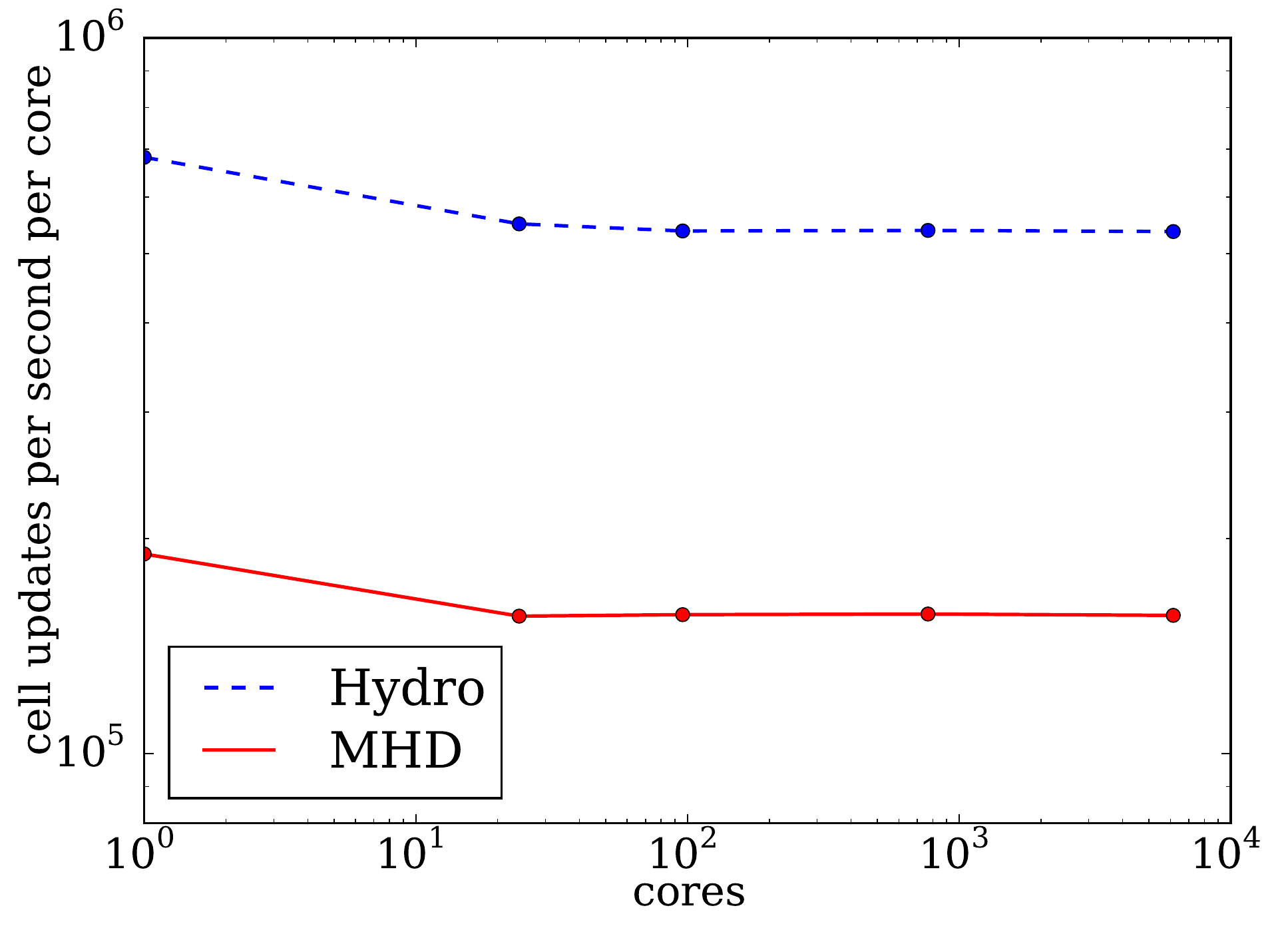}
  \caption{Performance per core running a 3D GRMHD simulation on a cluster. \label{fig:scaling}}
\end{figure}

The scaling of Athena++ was tested with a full 3D GRMHD computation using Kerr--Schild coordinates. The computation was performed on the NAOJ Cray~XC30, which has $24$ cores per node. We divide the domain into blocks of $64^3$ cells, assigning one block to each core. The scaling results out to $6144$ cores are shown in Figure~\ref{fig:scaling}. For both hydrodynamics and MHD, there is a $20\%$ per-core performance penalty to go from one core to one full node, almost certainly associated with saturating the memory bus when all cores on a node are used. However, the cost of going to many nodes from one is negligible. The performance of $6144$ cores is over $97\%$ relative to $24$ cores for hydrodynamics, and it is indistinguishable from $100\%$ for MHD.

\chapter[Numerical Solution of GRMHD]{Numerical Solution of the Equations of General-Relativistic Magnetohydrodynamics}
\label{chap:method}

With our motivations and guiding considerations established, we now turn to the mathematical and numerical details necessary for our endeavor.

\section{Differential Forms of Equations}
\label{sec:method:differential}

The relevant equations for general-relativistic magnetohydrodynamics (GRMHD) are derived in numerous sources in the literature \citep[e.g.][]{Gammie2003}. Here we omit the derivation of the governing differential equations, but we include all the necessary definitions in order to avoid notational ambiguity.

\subsection{General Equations}
\label{sec:method:differential:general}

Denote the covariant components of the spacetime metric by $g_{\mu\nu}$. Consider a perfect fluid with comoving mass density $\rho$, comoving gas pressure $\pgas$, and comoving enthalpy per unit mass $h$. For an adiabatic gas of index $\Gamma$, we have $h = 1 + \Gamma/(\Gamma-1) \times \pgas/\rho$. The fluid will have coordinate-frame $4$-velocity components $u^\mu$. There will also be magnetic field components $B^i$ in the coordinate frame.

The stress-energy tensor for the fluid has components
\begin{equation} \label{eq:stress_energy_upper}
  T^{\mu\nu} = (\rho h + b_\lambda b^\lambda) u^\mu u^\nu + \paren[\Big]{\pgas + \frac{1}{2} b_\lambda b^\lambda} g^{\mu\nu} - b^\mu b^\nu.
\end{equation}
Here
\begin{subequations} \label{eq:contravariant_b} \begin{align}
  b^0 & = g_{i\mu} B^i u^\mu, \label{eq:contravariant_b:time} \\
  b^i & = \frac{1}{u^0} (B^i + b^0 u^i) \label{eq:contravariant_b:space}
\end{align} \end{subequations}
are the components $b^\mu = u_\nu (\ast F)^{\nu\mu}$ of the contravariant comoving magnetic field in terms of the metric, $3$-magnetic field, and $4$-velocity. We note in passing that $b^\mu u_\mu = 0$. In terms of the electric and magnetic fields, the contravariant components of the dual electromagnetic field tensor are
\begin{equation} \label{eq:electromagnetic_tensor_upper}
  (\ast F)^{\mu\nu} =
  \begin{pmatrix}
    0   & -B^1           & -B^2           & -B^3           \\
    B^1 & 0              & \phantom{-}E^3 & -E^2           \\
    B^2 & -E^3           & 0              & \phantom{-}E^1 \\
    B^3 & \phantom{-}E^2 & -E^1           & 0
  \end{pmatrix},
\end{equation}
where $\mu$ indexes rows and $\nu$ columns. Note that there are different sign conventions for the field tensor.

One can verify that
\begin{equation} \label{eq:electromagnetic_tensor}
  (\ast F)^{\mu\nu} = b^\mu u^\nu - b^\nu u^\mu.
\end{equation}
As we consider only ideal MHD here, the electric fields can always be inferred from the magnetic fields and fluid velocities via \eqref{eq:contravariant_b} and \eqref{eq:electromagnetic_tensor}.

The equations of ideal MHD are
\begin{subequations} \label{eq:divergence} \begin{align}
  \del_\mu (\rho u^\mu) & = 0, \label{eq:divergence:mass} \\
  \del_\mu T^{\mu\nu} & = 0, \label{eq:divergence:stress_energy} \\
  \del_\mu (\ast F)^{\mu\nu} & = 0 \label{eq:divergence:electromagnetic}
\end{align} \end{subequations}
\citep[cf.][\extref{1}, \extref{3}, \extref{14}]{Gammie2003}. These conservation laws for mass, momentum, and magnetic field govern the time evolution of a system consisting of a magnetized fluid in a stationary spacetime.

\subsection{Rewriting Covariant Derivatives}
\label{sec:method:differential:rewriting}

In order to time-evolve our system, we need to have explicit temporal partial derivatives in our equations. Additionally, we prefer to have spatial partial derivatives. These considerations motivate the following lemma useful for rewriting covariant derivatives.

Suppose we have an arbitrary tensor with components $\tensor{A}{^{\mu\alpha_1\cdots\alpha_i}_{\beta_1\cdots\beta_j}}$. Then
\begin{equation} \begin{split} \label{eq:covariant_derivative}
  \del_\nu \tensor{A}{^{\mu\alpha_1\cdots\alpha_i}_{\beta_1\cdots\beta_j}} & = \pp_\nu \tensor{A}{^{\mu\alpha_1\cdots\alpha_i}_{\beta_1\cdots\beta_j}} + \tensor{A}{^{\sigma\alpha_1\cdots\alpha_i}_{\beta_1\cdots\beta_j}} \Gamma^\mu_{\sigma\nu} \\
  & \quad \qquad + \sum_{k=1}^i \tensor{A}{^{\mu\alpha_1\cdots\alpha_{k-1}\sigma\alpha_{k+1}\cdots\alpha_i}_{\beta_1\cdots\beta_j}} \Gamma^{\alpha_k}_{\sigma\nu} \\
  & \quad \qquad - \sum_{k=1}^j \tensor{A}{^{\mu\alpha_1\cdots\alpha_i}_{\beta_1\cdots\beta_{k-1}\sigma\beta_{k+1}\cdots\beta_j}} \Gamma^{\sigma}_{\beta_k\nu}.
\end{split} \end{equation}
The explicit summations directly result in the source terms in the equations we will be integrating.

In the case of calculating a divergence, it is possible to simplify the first two terms on the right-hand side of \eqref{eq:covariant_derivative}. For this we will suppress the spectator indices $\alpha_1, \ldots, \alpha_i$ and $\beta_1, \ldots, \beta_j$. Manipulating partial derivatives we can write
\begin{equation} \begin{split}
  \del_\mu A^\mu & = \pp_\mu A^\mu + A^\nu \Gamma^\mu_{\nu\mu} \\
  & = \frac{1}{\sqrt{-g}} \paren[\Big]{\pp_\mu (\sqrt{-g} A^\mu) - A^\mu \pp_\mu \sqrt{-g}} + A^\nu \Gamma^\mu_{\nu\mu} \\
  & = \frac{1}{\sqrt{-g}} \pp_\mu (\sqrt{-g} A^\mu) + A^\mu \paren[\bigg]{\Gamma^\nu_{\mu\nu} - \frac{1}{\sqrt{-g}} \pp_\mu \sqrt{-g}}
\end{split} \end{equation}
for any suitably differentiable strictly negative function $g$. Note that
\begin{equation}
  \frac{1}{\sqrt{-g}} \pp_\mu \sqrt{-g} = -\frac{1}{2g} \pp_\mu (-g) = \frac{1}{2g} \pp_\mu g.
\end{equation}

Now take $g$ to be the determinant of the metric. Jacobi's formula tells us that in the case of an invertible matrix $M_t$ parameterized by $t$,
\begin{equation}
  \pp_t \det(M_t) = \det(M_t) \tr(M_t^{-1} \pp_t M_t).
\end{equation}
Therefore
\begin{equation}
  \frac{1}{\sqrt{-g}} \pp_\mu \sqrt{-g} = \frac{1}{2} g^{\alpha\beta} \pp_\mu g_{\alpha\beta}.
\end{equation}
We can then write
\begin{equation}
  \del_\mu A^\mu = \frac{1}{\sqrt{-g}} \pp_\mu (\sqrt{-g} A^\mu) + A^\mu \paren[\bigg]{\frac{1}{2} g^{\nu\sigma} (\pp_\mu g_{\nu\sigma} + \pp_\nu g_{\mu\sigma} - \pp_\sigma g_{\mu\nu}) - \frac{1}{2} g^{\alpha\beta} \pp_\mu g_{\alpha\beta}}.
\end{equation}
Changing summation indices and relying on the symmetry of the metric, one can see that the first and fourth terms in parentheses cancel, as do the second and third. Thus we are left with
\begin{equation} \label{eq:vector_divergence}
  \del_\mu A^\mu = \frac{1}{\sqrt{-g}} \pp_\mu (\sqrt{-g} A^\mu),
\end{equation}
which is \extref{8.51c} of \citet{MTW}.

Restoring the suppressed indices yields the covariant-to-partial divergence rule
\begin{equation} \begin{split} \label{eq:tensor_divergence}
  \del_\mu \tensor{A}{^{\mu\alpha_1\cdots\alpha_i}_{\beta_1\cdots\beta_j}} & = \frac{1}{\sqrt{-g}} \pp_\mu \paren[\big]{\sqrt{-g} \tensor{A}{^{\mu\alpha_1\cdots\alpha_i}_{\beta_1\cdots\beta_j}}} \\
  & \quad \qquad + \sum_{k=1}^i \tensor{A}{^{\mu\alpha_1\cdots\alpha_{k-1}\sigma\alpha_{k+1}\cdots\alpha_i}_{\beta_1\cdots\beta_j}} \Gamma^{\alpha_k}_{\sigma\mu} \\
  & \quad \qquad - \sum_{k=1}^j \tensor{A}{^{\mu\alpha_1\cdots\alpha_i}_{\beta_1\cdots\beta_{k-1}\sigma\beta_{k+1}\cdots\beta_j}} \Gamma^{\sigma}_{\beta_k\mu}.
\end{split} \end{equation}

\subsection{The Equations in Terms of Partial Derivatives}
\label{sec:method:differential:partial}

We next express our equations in terms of partial derivatives, forgoing manifest covariance in favor of expediency in computational implementation.

Applying \eqref{eq:vector_divergence} to \eqref{eq:divergence:mass} yields
\begin{equation}
  \pp_0 (\sqrt{-g} \rho u^0) + \pp_j (\sqrt{-g} \rho u^j) = 0.
\end{equation}
For \eqref{eq:divergence:stress_energy}, we choose to apply \eqref{eq:tensor_divergence} to the lowered form of the equation:
\begin{equation}
  \pp_0 (\sqrt{-g}\, \tensor{T}{^0_\mu}) + \pp_i (\sqrt{-g}\, \tensor{T}{^i_\mu}) = \sqrt{-g}\, \tensor{T}{^\nu_\sigma} \Gamma^\sigma_{\mu\nu}.
\end{equation}

The last conservation law, \eqref{eq:divergence:electromagnetic}, can also be expanded via \eqref{eq:tensor_divergence}:
\begin{equation} \label{eq:em_divergence}
  \pp_0\paren[\big]{\sqrt{-g}\, (\ast F)^{0\mu}} + \pp_j\paren[\big]{\sqrt{-g}\, (\ast F)^{j\mu}} = -\sqrt{-g}\, (\ast F)^{\nu\sigma} \Gamma^\mu_{\sigma\nu}.
\end{equation}
Now the right-hand side vanishes, as it is the contraction of symmetric indices with antisymmetric ones. This is in agreement with \citet[8.51c]{MTW}. Also \eqref{eq:electromagnetic_tensor} and \eqref{eq:contravariant_b:space} tell us
\begin{equation}
  (\ast F)^{0i} = b^0 u^i - (B^i + b^0 u^i) = -B^i
\end{equation}
and
\begin{equation}
  (\ast F)^{ji} = b^j u^i - b^i u^j.
\end{equation}
Thus the spatial components of \eqref{eq:em_divergence} are equivalent to
\begin{equation}
  \pp_0 (\sqrt{-g} B^i) + \pp_j \paren[\big]{\sqrt{-g}\, (b^i u^j - b^j u^i)} = 0,
\end{equation}
while the temporal component is the constraint
\begin{equation}
  \pp_j (\sqrt{-g} B^j) = 0.
\end{equation}

Summarizing, the evolution equations for GRMHD can be written
\begin{subequations} \label{eq:differential} \begin{align}
  \pp_0 (\sqrt{-g} \rho u^0) + \pp_j (\sqrt{-g} \rho u^j) & = 0, \label{eq:differential:mass} \\
  \pp_0 (\sqrt{-g}\, \tensor{T}{^0_\mu}) + \pp_j (\sqrt{-g}\, \tensor{T}{^j_\mu}) & = \sqrt{-g}\, \tensor{T}{^\nu_\sigma} \Gamma^\sigma_{\mu\nu}, \label{eq:differential:stress_energy} \\
  \pp_0 (\sqrt{-g} B^i) + \pp_j \paren[\big]{\sqrt{-g}\, (b^i u^j - b^j u^i)} & = 0. \label{eq:differential:electromagnetic}
\end{align} \end{subequations}
These equations agree with \extref{2}, \extref{4}, and \extref{18} of \citet{Gammie2003}.

\section{Integral Forms of Equations}
\label{sec:method:integral}

The differential forms of the equations, written in terms of partial rather than covariant derivatives, would be sufficient for a finite difference scheme. However, we endeavor to develop a finite volume method, and so we must integrate our equations over regions of spacetime. This will naturally lead to equations for the evolution of volume-averaged quantities in terms of surface-averaged fluxes.

\subsection{General Equations}
\label{sec:method:integral:general}

Consider the generic form
\begin{equation} \label{eq:generic_differential}
  \pp_0 (\sqrt{-g}\, \altvec{C}) + \pp_1 (\sqrt{-g}\, \altvec{F}) + \pp_2 (\sqrt{-g}\, \altvec{G}) + \pp_3 (\sqrt{-g}\, \altvec{H}) = \sqrt{-g}\, \altvec{S}
\end{equation}
of \eqref{eq:differential}, where we group the conserved quantities, fluxes, and sources as
\begin{subequations} \label{eq:quantities} \begin{align}
  \altvec{C} & = (\rho u^0, \tensor{T}{^0_\mu}, B^i), \label{eq:quantities:conserved} \\
  \altvec{F} & = (\rho u^1, \tensor{T}{^1_\mu}, b^i u^1 - b^1 u^i), \label{eq:quantities:x} \\
  \altvec{G} & = (\rho u^2, \tensor{T}{^2_\mu}, b^i u^2 - b^2 u^i), \label{eq:quantities:y} \\
  \altvec{H} & = (\rho u^3, \tensor{T}{^3_\mu}, b^i u^3 - b^3 u^i), \label{eq:quantities:z} \\
  \altvec{S} & = (0, \tensor{T}{^\rho_\sigma} \Gamma^\sigma_{\rho\mu}, 0^i) \label{eq:quantities:sources}
\end{align} \end{subequations}
\citep[cf.][\extref{41}--\extref{43}]{Anton2006}.

\begin{figure}
  \centering
  \includegraphics[width=4in]{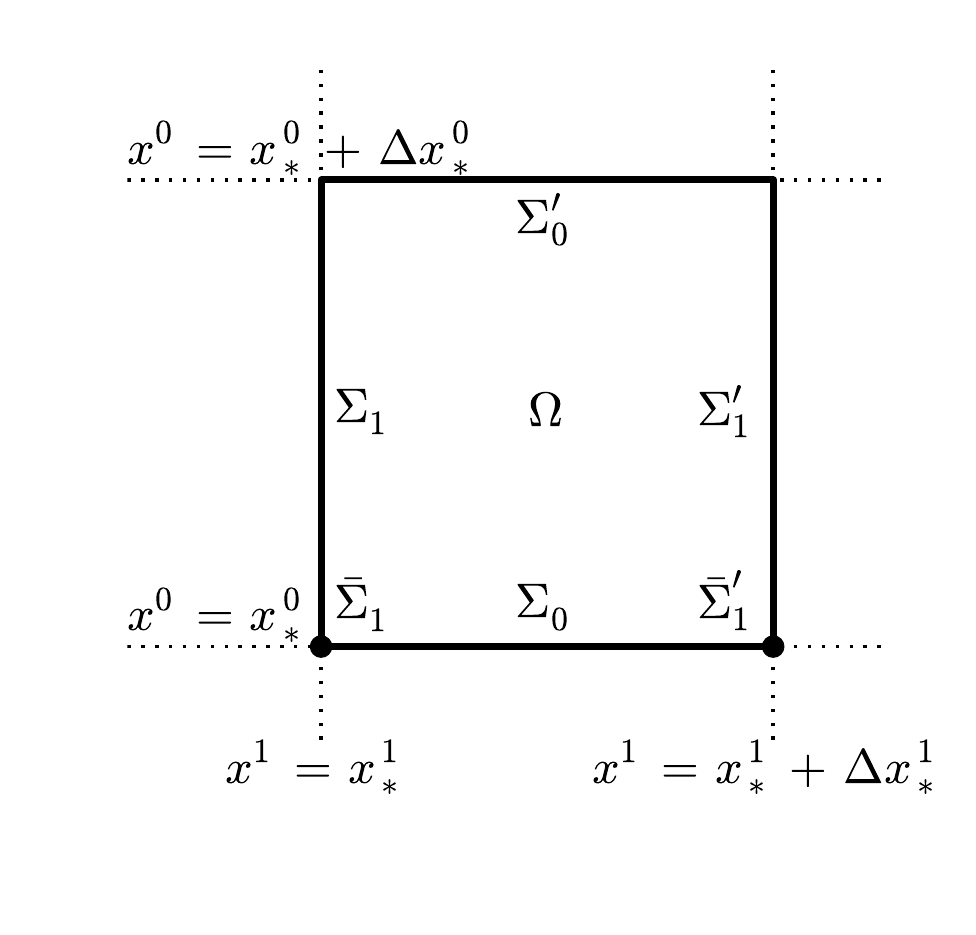}
  \caption{$1{+}1$ spacetime diagram of a volume element $\Omega$ and its bounding surfaces $\Sigma_\mu^{(\prime)}$ of constant coordinate values. \label{fig:surfaces}}
\end{figure}

Fix particular coordinates $x_*^\mu$ and positive increments $\Delta x_*^\mu$. Let $\Sigma_\mu$ be the surface $x^\mu = x_*^\mu$ for each $\mu$, and similarly let $\Sigma_\mu'$ be the surface $x^\mu = x_*^\mu + \Delta x_*^\mu$. We refer to the parallelepiped in spacetime bounded by the $3$-surfaces $\Sigma_\mu$ and $\Sigma_\mu'$ as $\Omega$, and we understand the surfaces to not extend past $\Omega$. This is illustrated in Figure~\ref{fig:surfaces}

Following \citet{Banyuls1997}, integrate \eqref{eq:generic_differential} over $\Omega$ with the volume measure $\dx^0\,\dx^1\,\dx^2\,\dx^3$ (which is not coordinate independent). Noting that the integrands on the left-hand side will be partial derivatives with respect to the coordinates, we can write
\begin{equation} \label{eq:integral} \begin{split}
  \int\limits_{\Sigma_0'} \sqrt{-g}\, \altvec{C} \, \dx^1\,\dx^2\,\dx^3 & = \int\limits_{\Sigma_0} \sqrt{-g}\, \altvec{C} \, \dx^1\,\dx^2\,\dx^3 + \int\limits_{\Omega} \sqrt{-g}\, \altvec{S} \, \dx^0\,\dx^1\,\dx^2\,\dx^3 \\
  & \quad \qquad + \int\limits_{\Sigma_1} \sqrt{-g}\, \altvec{F} \, \dx^0\,\dx^2\,\dx^3 - \int\limits_{\Sigma_1'} \sqrt{-g}\, \altvec{F} \, \dx^0\,\dx^2\,\dx^3 \\
  & \quad \qquad + \int\limits_{\Sigma_2} \sqrt{-g}\, \altvec{G} \, \dx^0\,\dx^1\,\dx^3 - \int\limits_{\Sigma_2'} \sqrt{-g}\, \altvec{G} \, \dx^0\,\dx^1\,\dx^3 \\
  & \quad \qquad + \int\limits_{\Sigma_3} \sqrt{-g}\, \altvec{H} \, \dx^0\,\dx^1\,\dx^2 - \int\limits_{\Sigma_3'} \sqrt{-g}\, \altvec{H} \, \dx^0\,\dx^1\,\dx^2.
\end{split} \end{equation}
To see that this is equivalent to \extref{28} and \extref{29} of \citeauthor{Banyuls1997}, we note two changes of notation. First, their $\dd\Omega$ is the proper volume element $\sqrt{-g}\, \dx^0\,\dx^1\,\dx^2\,\dx^3$. Second, their $\sqrt{\gamma}\, \altvec{F}^{\bar{0}} = \sqrt{-g}\, \altvec{F}^{\bar{0}}\!/\alpha$ (here we place bars over indices in their basis) is equal to our $\altvec{C}$, as their basis has as its $\bar{0}$-component the unit timelike normal
\begin{equation}
  \hat{n} = \frac{1}{\alpha} \paren[\Big]{\vec{\pp}_0 - \beta^i \vec{\pp}_i}.
\end{equation}
That is, we can write a vector
\begin{equation}
  \vec{A} = A^0 \vec{\pp}_0 + A^i \vec{\pp}_i
\end{equation}
as
\begin{equation}
  \vec{A} = A^{\bar{0}} \hat{n} + A^{\bar{\imath}} \vec{\pp}_{\bar{\imath}} = \frac{1}{\alpha} A^{\bar{0}} \vec{\pp}_0 + \paren[\bigg]{A^{\bar{\imath}} - \frac{\beta^{\bar{\imath}}}{\alpha} A^{\bar{0}}} \vec{\pp}_{\bar{\imath}},
\end{equation}
from which we conclude that $A^{\bar{0}}\!/\alpha = A^0$.

\subsection{Discretized Forms of Equations}
\label{sec:method:integral:discretized}

Let $i$ index regions bounded by surfaces of constant $x^1$, and similarly let $j$ and $k$ relate to $x^2$ and $x^3$. Let $n$ index surfaces of constant $x^0$. Fundamentally we choose to evolve the $3$-volume-averaged quantities
\begin{equation}
  \altvec{C}_{i,j,k}^n = \frac{1}{\Delta V_{i,j,k}} \int\limits_{\Sigma_0} \sqrt{-g}\, \altvec{C} \, \dx^1\,\dx^2\,\dx^3,
\end{equation}
where
\begin{equation} \label{eq:volume}
  \Delta V_{i,j,k} = \int\limits_{\Sigma_0} \sqrt{-g} \, \dx^1\,\dx^2\,\dx^3.
\end{equation}
As a result of having a stationary metric, $\Delta V_{i,j,k}$ is constant in ``time'' $x^0$.

We note that $\Delta V_{i,j,k}$ is not coordinate invariant; it would be so only if we replaced the $\sqrt{-g}$ with $\sqrt{\gamma}$, where $\gamma$ is the determinant of the $3$-metric induced on surfaces of constant $x^0$ ($\gamma_{ij} = g_{ij}$). It is however constant in ``time'' $x^0$, as a result of having a stationary metric. Also note that this formula leads to a definition of cell-centered positions. The $x^1$-coordinate of the center of cell $i,j,k$ is the value $\bar{x}^1$ such that the integral \eqref{eq:volume} limited by $x^1$ going from its minimum value to $\bar{x}^1$ is $\Delta V_{i,j,k}/2$. Similarly we can define $\bar{x}^2$ and $\bar{x}^3$.

The fluxes found along the interfaces are taken to be constant along those interfaces, denoted $\altvec{F}_{i\pm1/2,j,k}^{n+1/2}$, $\altvec{G}_{i,j\pm1/2,k}^{n+1/2}$, and $\altvec{H}_{i,j,k\pm1/2}^{n+1/2}$. The constant (coordinate-dependent) $2$-areas of the interfaces will be denoted
\begin{subequations} \label{eq:area} \begin{align}
  \Delta A_{i\pm1/2,j,k} & = \int\limits_{\bar{\Sigma}_1^{(\prime)}} \sqrt{-g} \, \dx^2\,\dx^3, \label{eq:area:1} \\
  \Delta A_{i,j\pm1/2,k} & = \int\limits_{\bar{\Sigma}_2^{(\prime)}} \sqrt{-g} \, \dx^1\,\dx^3, \label{eq:area:2} \\
  \Delta A_{i,j,k\pm1/2} & = \int\limits_{\bar{\Sigma}_3^{(\prime)}} \sqrt{-g} \, \dx^1\,\dx^2, \label{eq:area:3}
\end{align} \end{subequations}
where primes are associated with plus signs and $\bar{\Sigma}$ is the projection of the $3$-surface $\Sigma$ along coordinate $x^0$ to a $2$-surface of constant $x^0$.

We pause to also define the formulas for ``lengths'' of edges in a similar fashion:
\begin{subequations} \label{eq:length} \begin{align}
  \Delta L_{i,j\pm_a1/2,k\pm_b1/2} & = \int\limits_{\mathclap{\bar{\Sigma}_2^{(\prime_a)}\cap\bar{\Sigma}_3^{(\prime_b)}}} \sqrt{-g} \, \dx^1, \label{eq:length:1} \\
  \Delta L_{i\pm_a1/2,j,k\pm_b1/2} & = \int\limits_{\mathclap{\bar{\Sigma}_1^{(\prime_a)}\cap\bar{\Sigma}_3^{(\prime_b)}}} \sqrt{-g} \, \dx^2, \label{eq:length:2} \\
  \Delta L_{i\pm_a1/2,j\pm_b1/2,k} & = \int\limits_{\mathclap{\bar{\Sigma}_1^{(\prime_a)}\cap\bar{\Sigma}_2^{(\prime_b)}}} \sqrt{-g} \, \dx^3. \label{eq:length:3}
\end{align} \end{subequations}

\begin{figure}
  \centering
  \includegraphics[width=\textwidth]{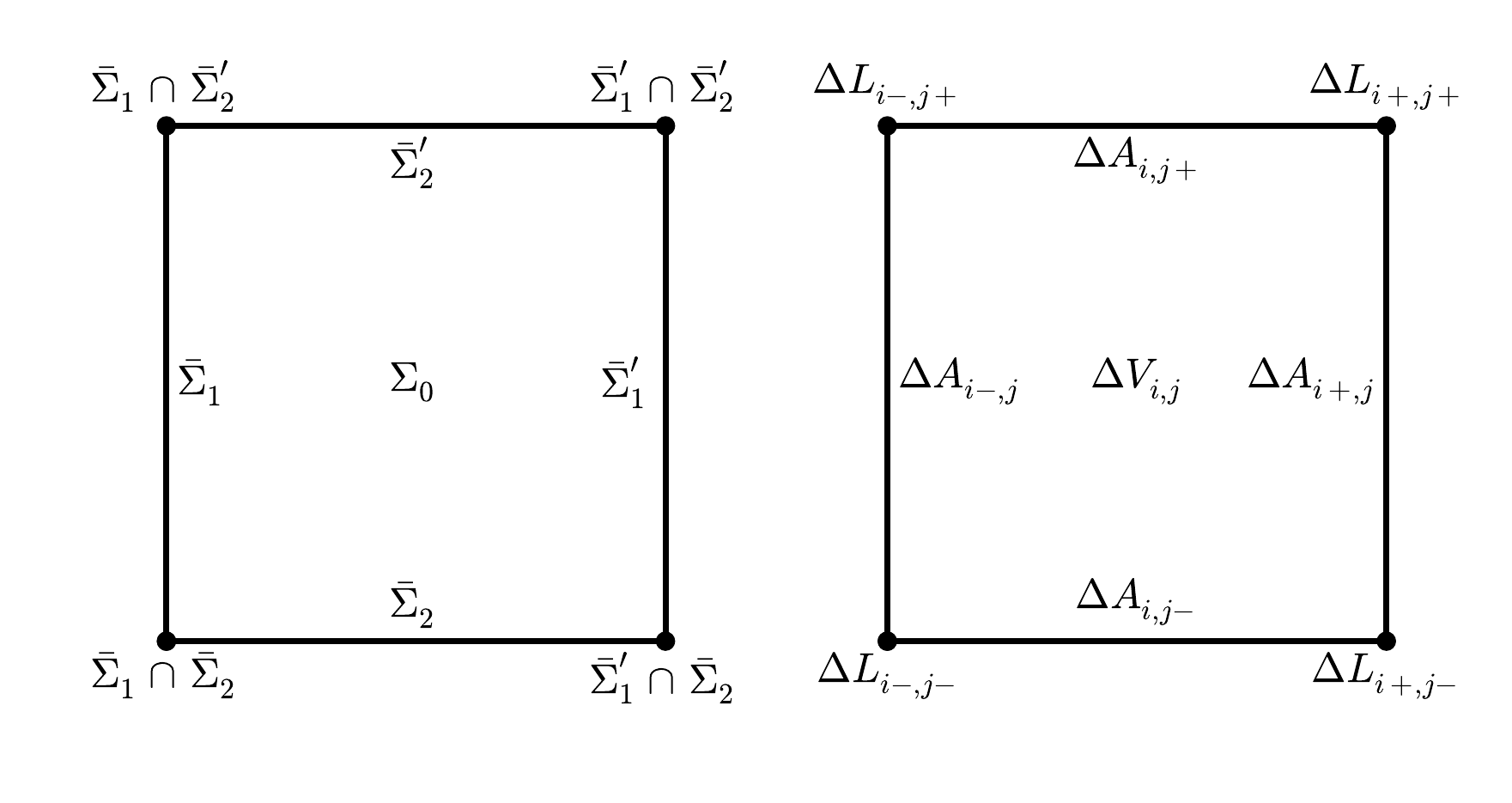}
  \caption{$2{+}0$ spacetime diagram of a single cell and its bounding surfaces and edges. The spacetime regions are labeled on the left, while on the right their corresponding sizes (integrals over them of $\sqrt{-g}$) are shown. \label{fig:volumes}}
\end{figure}

The arrangement of volumes, areas, and lengths is illustrated in Figure~\ref{fig:volumes}. The left panel labels regions of space according to the notation used in Figure~\ref{fig:surfaces}, while the right panel labels the corresponding sizes of those regions as given by \eqref{eq:volume}, \eqref{eq:area}, and \eqref{eq:length}.

The lengths are distinct from the cell widths needed to set the maximum allowable time step for stability. These latter quantities are obtained by integrating the induced $3$-metric line element on the constant-$x^0$ surface along curves of constant coordinate $x^i$. That is, let $C^1$ be the curve in $\Sigma_0$ that varies only in $x^1$, passes through the center of cell $i,j,k$, and is bounded by $\bar{\Sigma}_1$ and $\bar{\Sigma}_1'$, and similarly define $C^2$ and $C^3$. Since we have the line element $\tensor[^{(3)}]{ds}{^2} = \gamma_{ij} \, \dx^i\,\dx^j$ on $\Sigma_0$, where $\gamma_{ij} = g_{ij}$ are the $3$-metric components, we can take these cell widths to be
\begin{subequations} \label{eq:width} \begin{align}
  \Delta W_{x^1;i,j,k} & = \int\limits_{C^1} \sqrt{g_{11}} \, \dx^1, \label{eq:width:1} \\
  \Delta W_{x^2;i,j,k} & = \int\limits_{C^2} \sqrt{g_{22}} \, \dx^2, \label{eq:width:2} \\
  \Delta W_{x^3;i,j,k} & = \int\limits_{C^3} \sqrt{g_{33}} \, \dx^3. \label{eq:width:3}
\end{align} \end{subequations}

Dividing through by the volume and making the approximation that states are constant on timescales shorter than those resolved, the source term becomes
\begin{equation}
  \frac{1}{\Delta V_{i,j,k}} \int\limits_\Omega \sqrt{-g}\, \altvec{S} \, \dx^0\,\dx^1\,\dx^2\,\dx^3 = \int_{x_*^0}^{x_*^0+\Delta x_*^0} \altvec{S}_{i,j,k}^{n+1/2} \, \dx^0,
\end{equation}
where
\begin{equation}
  \altvec{S}_{i,j,k}^{n+1/2} = \frac{1}{\Delta V_{i,k,k}} \int\limits_{\Sigma_0^*} \sqrt{-g}\, \altvec{S} \, \dx^1\,\dx^2\,\dx^3.
\end{equation}
Here $\Sigma_0^*$ is the surface between $\Sigma_0$ (index $n$) and $\Sigma_0'$ (index $n+1$) corresponding to index $n+1/2$.

Now the nature of finite volume methods means we must at some point accept volume averages as the only quantities to which we have access. According to \eqref{eq:quantities:sources}, we can write the volume averages of the nonzero components of $\altvec{S}$ -- call them $S_\mu$ -- as
\begin{equation} \label{eq:source_decomposition}
  (S_\mu)_{i,j,k}^{n+1/2} \approx \tensor{T}{^\rho_\sigma}\!(\altvec{P}_{i,j,k}^{n+1/2})\, \Gamma^\sigma_{\rho\mu},
\end{equation}
where the parentheses on the right-hand side indicate the use of the pointwise function $\altvec{P} \mapsto \tensor{T}{^\mu_\nu}$ defined by lowering \eqref{eq:stress_energy_upper}, $\altvec{P}$ is the vector of primitive variables
\begin{equation} \label{eq:primitives}
  \altvec{P} = (\rho, \pgas, v^i, B^i),
\end{equation}
and we invoke the standard volume averaging notation to write
\begin{equation} \label{eq:prim_volume}
  \altvec{P}_{i,j,k}^{n+1/2} = \frac{1}{\Delta V_{i,j,k}} \int\limits_{\Sigma_0^*} \sqrt{-g}\, \altvec{P} \, \dx^1\,\dx^2\,\dx^3
\end{equation}
(and similarly for $n$ and $\Sigma_0$ and for $n+1$ and $\Sigma_0'$). Also in the same vein we will assume the pointwise correspondence $\altvec{P} \leftrightarrow \altvec{C}$ holds for $\altvec{P}_{i,j,k}^n \leftrightarrow \altvec{C}_{i,j,k}^n$ using the same cell-centered geometric quantities.

The connection coefficients appearing in \eqref{eq:source_decomposition} can be evaluated at the cell centers. The approximation of moving the coefficients outside the integral by assuming they have constant values equal to their cell-centered values is second-order accurate in space. One could imagine taking a volume average of the connection coefficients
\begin{equation} \label{eq:averaged_connection}
  \bar{\Gamma}^\sigma_{\rho\mu} = \frac{1}{\Delta V_{i,k,k}} \int\limits_{V_{i,k,k}} \sqrt{-g}\, \Gamma^\sigma_{\rho\mu} \, \dx^1\,\dx^2\,\dx^3,
\end{equation}
though this would not improve spatial accuracy. (Because a stationary metric implies a stationary connection, these values can be precomputed in either method.) Fundamentally, calculating the source term integrals to better than second order requires more information per cell. While it would be simple to evaluate the connection coefficients at more points, a higher-order quadrature would require higher-order reconstruction of either the primitives or the corresponding stress-energy tensor, utilizing information from neighboring cells.

All integrands consist of parts that are either exactly constant in $x^0$ (because the metric is assumed to be stationary) or are taken to be constant over the interval in question, and so we can perform the $x^0$-integrations separately. Piecing everything together, \eqref{eq:integral} can then be written
\begin{equation} \label{eq:finite_volume} \begin{split}
  \altvec{C}_{i,j,k}^{n+1} & = \altvec{C}_{i,j,k}^n + \Delta t \altvec{S}_{i,j,k}^{n+1/2} \\
  & \quad \qquad + \frac{\Delta t}{\Delta V_{i,j,k}} \paren[\Big]{\altvec{F}_{i-1/2,j,k}^{n+1/2} \Delta A_{i-1/2,j,k} - \altvec{F}_{i+1/2,j,k}^{n+1/2} \Delta A_{i+1/2,j,k}} \\
  & \quad \qquad + \frac{\Delta t}{\Delta V_{i,j,k}} \paren[\Big]{\altvec{G}_{i,j-1/2,k}^{n+1/2} \Delta A_{i,j-1/2,k} - \altvec{G}_{i,j+1/2,k}^{n+1/2} \Delta A_{i,j+1/2,k}} \\
  & \quad \qquad + \frac{\Delta t}{\Delta V_{i,j,k}} \paren[\Big]{\altvec{H}_{i,j,k-1/2}^{n+1/2} \Delta A_{i,j,k-1/2} - \altvec{H}_{i,j,k+1/2}^{n+1/2} \Delta A_{i,j,k+1/2}},
\end{split} \end{equation}
where $\Delta t$ is a more familiar symbol for what we earlier denoted $\Delta x_*^0$, the change in coordinate time from surface $n$ to $n+1$. Note that suitable reinterpretations of $n$, $n+1/2$, and $n+1$ yield formulas for different substeps in various integration schemes.

\subsection{Magnetic Evolution}
\label{sec:method:integral:magnetic}

Thus far we have treated the three components of the magnetic field in the same manner as the density and four components of momentum. Equation \eqref{eq:finite_volume} could be used to update volume-averaged magnetic fields in the same way as the other quantities. However, this method is known to violate the divergence-free constraint of the magnetic field $\del_\mu (\ast F)^{0\mu} = 0$. That is, even initial conditions free of magnetic monopoles develop them numerically over time.

A number of schemes have been developed to correct for spurious magnetic monopole generation as discussed in \S\ref{sec:intro:numerics:constraint}. Rather than apply post hoc fixes, however, we choose a different discretization, one that naturally preserves the constraint.

In the case of magnetic fields, we take the fundamental variables to be area averages. This is a crucial part of the constrained transport (CT) scheme of \citet{Evans1988} (see \S\ref{sec:method:algorithm:ct}). For concreteness, consider the evolution of
\begin{equation}
  (B^1)_{i-1/2,j,k}^n = \frac{1}{\Delta A_{i-1/2,j,k}} \int\limits_{A_1} B^1 \sqrt{-g} \, \dx^2\,\dx^3,
\end{equation}
where the positioning of quantities and orientations of surfaces and edges is the same as Figure~1 of \citet{Stone2008}. Given the electric fields $E^i$ defined along the edges (as discussed further in \S\ref{sec:method:algorithm:ct}), we can update the magnetic field according to
\begin{equation} \begin{split} \label{eq:b_update}
  \paren[\big]{B^1}_{i-1/2,j,k}^{n+1} & = \paren[\big]{B^1}_{i-1/2,j,k}^n \\
  & \quad \qquad + \frac{\Delta t}{\Delta A_{i-1/2,j,k}} \paren[\Big]{\Delta L_{i-1/2,j-1/2,k} \paren[\big]{E^3}_{i-1/2,j-1/2,k}^{n+1/2} \\
  & \quad \qquad \qquad - \Delta L_{i-1/2,j+1/2,k} \paren[\big]{E^3}_{i-1/2,j+1/2,k}^{n+1/2}} \\
  & \quad \qquad + \frac{\Delta t}{\Delta A_{i-1/2,j,k}} \paren[\Big]{\Delta L_{i-1/2,j,k-1/2} \paren[\big]{-E^2}_{i-1/2,j,k-1/2}^{n+1/2} \\
  & \quad \qquad \qquad - \Delta L_{i-1/2,j,k+1/2} \paren[\big]{-E^2}_{i-1/2,j,k+1/2}^{n+1/2}},
\end{split} \end{equation}
which is the discretized form of the third evolution equation \eqref{eq:differential:electromagnetic}. This replaces \eqref{eq:finite_volume} in updating the magnetic field.

\section{Numerical Algorithm}
\label{sec:method:algorithm}

With a discretization scheme in hand, we proceed to detail how quantities are evolved on the grid. The general procedure we will follow is:
\begin{enumerate}
  \item Invert the conserved variables to primitive variables. \label{enum:procedure:inversion}
  \item Reconstruct the primitive variables to interfaces. \label{enum:procedure:reconstruction}
  \item Transform the primitive variables into locally Minkowski frames. \label{enum:procedure:transformation}
  \item Solve the Riemann problem at each interface, finding fluxes. \label{enum:procedure:riemann}
  \item Transform the fluxes back into the global coordinate frame. \label{enum:procedure:inverse_transformation}
  \item Apply the CT method to calculate electric fields. \label{enum:procedure:ct}
  \item Calculate the curvature source terms from the primitive variables. \label{enum:procedure:source}
  \item Integrate forward in time. \label{enum:procedure:integrate}
\end{enumerate}

\subsection{Variable Inversion}
\label{sec:method:algorithm:inversion}

One of the most challenging aspects of numerical methods for relativistic fluid dynamics is variable inversion, the process of extracting the primitive variables from the conserved variables. The difficulty lies in the conserved variables being given by nonlinear functions of the primitive variables with no simple inverses. It is compounded by the fact that only certain primitive states are physically admissible (for example there can be no negative densities or superluminal velocities), and so a given set of conserved variables might not exactly correspond to any allowable set of primitives. Our methods described here are largely based on the $1D_W$ inversion scheme of \citet{Noble2006}.

If magnetic fields are present, they are interpolated to the volume-averaged centers as follows, using the $B^1$ case for concreteness. Given $B^1_{i\pm1/2}$ at interface coordinates $x^1_{i\pm1/2}$ and volume-averaged coordinate $x^1_i$, define $\lambda = (x^1_i - x^1_{i-1/2}) / (x^1_{i+1/2} - x^1_{i-1/2})$. Then the volume-averaged value of $B^1$ is taken to be
\begin{equation} \label{eq:b_interpolation}
  B^1_i = (1 - \lambda) B^1_{i-1/2} + \lambda B^1_{i+1/2}.
\end{equation}

In the case of general relativity, we take the primitive variables to be $\{\rho, \pgas, \tilde{u}^i\}$ and the conserved variables to be $\{\rho u^0, \tensor{T}{^0_\mu}\}$. Here we use the same velocities as in \citet{Noble2006}:\ $\tilde{u}^\mu = (\delta^\mu_\nu + n^\mu n_\nu) u^\nu$, where $n_\mu = -\alpha \delta^0_\mu$ are the components of the future-pointing unit vector normal to constant-$x^0$ hypersurfaces, and $\alpha = (-g^{00})^{-1/2}$ is the lapse. These projections have the desirable property that they describe subluminal motion no matter what values they have. Additionally, the fourth component is $\tilde{u}^0 = 0$. The Lorentz factor of the fluid as seen by the normal observer is $\gamma = (1 + g_{ij} \tilde{u}^i \tilde{u}^j)^{1/2}$. Following the $1D_W$ scheme, the Newton--Raphson method is used to solve a nonlinear equation for the relativistic gas enthalpy $W = \gamma^2 \rho h$. In the hydrodynamics case we use the same procedure as with MHD, simply setting all magnetic fields to $0$.

In special relativity, we take the primitive variables to be $\{\rho, \pgas, v^i\}$ and the conserved variables to be $\{D, E, M^i\} = \{\gamma \rho, T^{0\mu}\}$, where $\gamma = u^0$ is the fluid Lorentz factor in the coordinate frame. For MHD, we also use the Newton--Raphson method to find $W$, where we simply have $W = \gamma^2 \rho h$. The method is detailed in \citet{Mignone2007}, though we iterate on the residual of $E$ as a function of $W$ rather than $E - D$ as a function of $W - D$ as they do.

In the particular case of special-relativistic hydrodynamics, we can avoid using an iterative method and instead directly solve a quartic equation \citep{Schneider1993}. In particular, the $3$-velocity magnitude $\abs{v}$ must satisfy
\begin{equation} \label{eq:quartic_inversion}
  a_4 \abs{v}^4 + a_3 \abs{v}^3 + a_2 \abs{v}^2 + a_1 \abs{v} + a_0 = 0,
\end{equation}
where in terms of $M^2 = \eta_{ij} M^i M^j$ and $\abs{M} = \sqrt{M^2}$ we define
\begin{subequations} \begin{align}
  a_4 & = (\Gamma-1)^2 (D^2 - M^2), \\
  a_3 & = -2 \Gamma (\Gamma-1) \abs{M} E, \\
  a_2 & = \Gamma^2 E^2 + 2 (\Gamma-1) M^2 - (\Gamma-1)^2 D^2, \\
  a_1 & = -2 \Gamma \abs{M} E, \\
  a_0 & = M^2.
\end{align} \end{subequations}
Given the root $\abs{v}$, we recover the primitives
\begin{subequations} \begin{align}
  \rho & = D \sqrt{1-\abs{v}^2}, \\
  v^i & = \frac{\abs{v}}{\abs{M}} M^i, \\
  \pgas & = (\Gamma-1) (E - \eta_{ij} M^i v^j - \rho).
\end{align} \end{subequations}

As part of variable inversion we impose limits on the recovered variables. Since negative densities and pressures are unphysical, and since vanishing values cause problems in finite-volume fluid codes, we ensure $\rho > \rho_\mathrm{min}$ and $\pgas > \pgas^\mathrm{min}$, where $\rho_\mathrm{min}$ and $\pgas^\mathrm{min}$ are adjusted based on the problem. We also must obey the physical constraint $\gamma \geq 1$, and for numerical purposes we limit $\gamma < \gamma_\mathrm{max}$, where $\gamma_\mathrm{max}$ is often chosen to be $100$.

If the recovered primitives fall outside these ranges, they are modified to the nearest admissible values. The primitives are also set to the floor values in cases where the iteration does not converge. Any such modifications are then followed by reevaluating the conserved quantities in the given cell.

\subsection{Reconstruction}
\label{sec:method:algorithm:reconstruction}

As input to the Riemann problem, one needs fluid states on either side of the interface. While one can simply take the states from the two adjacent cells, this method is only first-order accurate in the size of the cells. For smooth flows, information from a larger stencil can be used to infer left and right states in a higher-order way. Reconstruction is the process of interpolating these states while preserving any actual discontinuities and not introducing spurious extrema.

For a primitive quantity $q$, we define the values on the left and right sides of the $x^1_{i-1/2}$ interface using piecewise linear reconstruction. Specifically, we use the modified van~Leer slope limiter algorithm of \citet{Mignone2014}, which reconstructs $q_{i-1/2,\mathrm{L/R}}$ from the values $q_{i-2}$, $q_{i-1}$, $q_i$, and $q_{i+1}$. This algorithm correctly takes into account nonuniform spacing, as well as the difference between the volume-averaged cell center coordinates (``centroids of volume'' in that work) and arithmetic means of surface coordinates (``geometrical cell centers''). The slope limiter has been proved to be total variation diminishing (TVD) for all orthogonal coordinate systems. While this guarantee has not been extended to non-orthogonal coordinate systems, to our knowledge this has not been done with any other method, and indeed a number of methods in use are not even TVD with orthogonal but non-Cartesian or non-uniform grids.

For completeness, we summarize the \citeauthor{Mignone2014}\ algorithm here. Define
\begin{subequations} \begin{align}
  \Delta_\mathrm{L} & = \frac{q_{i-1}-q_{i-2}}{x^1_{i-1}-x^1_{i-2}}, \\
  \Delta_\mathrm{C} & = \frac{q_i-q_{i-1}}{x^1_i-x^1_{i-1}}, \\
  \Delta_\mathrm{R} & = \frac{q_{i+1}-q_i}{x^1_{i+1}-x^1_i},
\end{align} \end{subequations}
Define also
\begin{subequations} \begin{align}
  \Delta x_\mathrm{L} & = x^1_{i-1/2} - x^1_{i-1}, & \Delta x_\mathrm{R} & = x^1_i - x^1_{i-1/2}, \\
  C^\mathrm{F}_\mathrm{L} & = \frac{x^1_i-x^1_{i-1}}{\Delta x_\mathrm{L}}, & C^\mathrm{F}_\mathrm{R} & = \frac{x^1_{i+1}-x^1_i}{x^1_{i+1/2}-x^1_i}, \\
  C^\mathrm{B}_\mathrm{L} & = \frac{x^1_{i-1}-x^1_{i-2}}{x^1_{i-1}-x^1_{i-3/2}}, & C^\mathrm{B}_\mathrm{R} & = \frac{x^1_i-x^1_{i-1}}{\Delta x_\mathrm{R}}.
\end{align} \end{subequations}
The left and right quantities are then given by
\begin{subequations} \begin{align}
  q_{i-1/2,\mathrm{L}} & =
  \begin{cases}
    \displaystyle q_{i-1}, & \Delta_\mathrm{L} \Delta_\mathrm{C} \leq 0, \\
    \displaystyle q_{i-1} + \frac{\Delta x_\mathrm{L}\Delta_\mathrm{L}\Delta_\mathrm{C}(C^\mathrm{F}_\mathrm{L}\Delta_\mathrm{L}+C^\mathrm{B}_\mathrm{L}\Delta_\mathrm{C})}{\Delta_\mathrm{L}^2+(C^\mathrm{F}_\mathrm{L}+C^\mathrm{B}_\mathrm{L}-2)\Delta_\mathrm{L}\Delta_\mathrm{C}+\Delta_\mathrm{C}^2}, & \Delta_\mathrm{L} \Delta_\mathrm{C} > 0;
  \end{cases} \\
  q_{i-1/2,\mathrm{R}} & =
  \begin{cases}
    \displaystyle q_i, & \Delta_\mathrm{C} \Delta_\mathrm{R} \leq 0, \\
    \displaystyle q_i - \frac{\Delta x_\mathrm{R}\Delta_\mathrm{C}\Delta_\mathrm{R}(C^\mathrm{F}_\mathrm{R}\Delta_\mathrm{C}+C^\mathrm{B}_\mathrm{R}\Delta_\mathrm{R})}{\Delta_\mathrm{C}^2+(C^\mathrm{F}_\mathrm{L}+C^\mathrm{B}_\mathrm{L}-2)\Delta_\mathrm{C}\Delta_\mathrm{R}+\Delta_\mathrm{R}^2}, & \Delta_\mathrm{C} \Delta_\mathrm{R} > 0.
  \end{cases}
\end{align} \end{subequations}

In the case of transverse magnetic field $q \in \{B^2, B^3\}$, the cell-centered values are used in the reconstruction algorithm. The longitudinal field $B^1$ is already defined at $x^1_{i-1/2}$, and so no reconstruction is needed for this quantity. Reconstruction for $x^2$- and $x^3$-surfaces proceeds analogously.

\subsection{Frame Transformation}
\label{sec:method:algorithm:transformation}

Rather than try to solve a Riemann problem directly in general coordinates, we choose to explicitly transform the left and right states at an interface to a locally Minkowski frame. In this way, we can leverage the advances made in special-relativistic Riemann problems, as suggested by \citet{Pons1998} and \citet{Anton2006}. While the scalar quantities $\rho$ and $\pgas$ transform between frames trivially, vector quantities rely on constructing a new basis $\{\vec{e}_{(\hat{\mu})}\}$ from the old coordinate basis $\{\vec{\pp}_{(\mu)}\}$, where the inner product with respect to the metric yields $\vec{\pp}_{(\mu)} \cdot \vec{\pp}_{(\nu)} = g_{\mu\nu}$ but $\vec{e}_{(\hat{\mu})} \cdot \vec{e}_{(\hat{\nu})} = \eta_{\mu\nu}$.

For a surface of constant $x^i$, we define the new basis to have the following properties:
\begin{enumerate}
  \item $\vec{e}_{(\hat{\mu})}$ must be orthogonal to $\vec{e}_{(\hat{\nu})}$ for all $\mu \neq \nu$.
  \item Each $\vec{e}_{(\hat{\mu})}$ must be normalized to have an inner product of $\pm1$ with itself, with $\vec{e}_{(\hat{t})}$ being timelike and $\vec{e}_{(\hat{\jmath})}$ being spacelike.
  \item $\vec{e}_{(\hat{t})}$ must be orthogonal to surfaces of constant $x^0$.
  \item The projection of $\vec{e}_{(\hat{x})}$ onto the surface of constant $x^0$ must be orthogonal to the surface of constant $x^i$ within that submanifold.
\end{enumerate}
The first two properties are a restatement of the condition $\vec{e}_{(\hat{\mu})} \cdot \vec{e}_{(\hat{\nu})} = \eta_{\mu\nu}$. Without the third property, the fluxes returned by the Riemann solver would not necessarily correspond to time evolution in the understood $x^0$-direction. The fourth property allows us to ignore $y$- and $z$-fluxes in the Riemann problem, using only the $x$-fluxes returned by a $1$-dimensional Riemann solver to infer the $x^1$-fluxes.

The details of the frame transformation are worked out in Appendix~\ref{chap:transformation}. Here we quote the results. For an $x^1$-interface, define the components
\begin{equation} \label{eq:to_global}
  \tensor{M}{^\mu_{\hat{\nu}}} =
  \begin{pmatrix}
    A g^{00} & 0                               & 0                   & 0 \\
    A g^{01} & B (g^{01}g^{01} - g^{00}g^{11}) & 0                   & 0 \\
    A g^{02} & B (g^{01}g^{02} - g^{00}g^{12}) & \phantom{-}D g_{33} & 0 \\
    A g^{03} & B (g^{01}g^{03} - g^{00}g^{13}) & -D g_{23}           & C
  \end{pmatrix},
\end{equation}
where we have the shorthands
\begin{subequations} \begin{align}
  A & = -\paren[\big]{-g^{00}}^{-1/2}, \\
  B & = \paren[\Big]{g^{00} \paren[\big]{g^{00}g^{11} - g^{01}g^{01}}}^{-1/2}, \\
  C & = (g_{33})^{-1/2}, \\
  D & = \paren[\big]{g_{33} (g_{22}g_{33} - g_{23}g_{23})}^{-1/2}.
\end{align} \end{subequations}
This matrix transforms contravariant tensors in the locally Minkowski basis to the coordinate basis, so for example $\tensor{T}{^1_\mu} = g_{\mu\nu} \tensor{M}{^1_{\hat{\mu}}} \tensor{M}{^\nu_{\hat{\nu}}} T^{\hat{\mu}\hat{\nu}}$ are the $x^1$-fluxes of $4$-momentum in the coordinate basis if $T^{\hat{\mu}\hat{\nu}}$ are the stress-energy tensor components in the orthonormal basis. The inverse transformation takes the form
\begin{equation} \label{eq:to_local}
  \tensor{M}{^{\hat{\mu}}_\nu} =
  \begin{pmatrix}
    -A                     & 0                      & 0                     & 0   \\
    B g^{01}               & -B g^{00}              & 0                     & 0   \\
    B^2E g^{00} / D g_{33} & B^2F g^{00} / D g_{33} & 1 / D g_{33}          & 0   \\
    I                      & J                      & (1/C) g_{23} / g_{33} & 1/C
  \end{pmatrix},
\end{equation}
where we define
\begin{subequations} \begin{align}
  E & = g^{01} g^{12} - g^{11} g^{02}, \\
  F & = g^{01} g^{02} - g^{00} g^{12}, \\
  G & = g^{01} g^{13} - g^{11} g^{03}, \\
  H & = g^{01} g^{03} - g^{00} g^{13}, \\
  I & = \frac{B^2g^{00}}{C} \paren[\bigg]{G + \frac{Eg_{23}}{g_{33}}}, \\
  J & = \frac{B^2g^{00}}{C} \paren[\bigg]{H + \frac{Fg_{23}}{g_{33}}},
\end{align} \end{subequations}
With this matrix we can for example write the orthonormal frame fluid $4$-velocity components in terms of the coordinate basis components:\ $u^{\hat{\mu}} = \tensor{M}{^{\hat{\mu}}_\nu} u^\nu$.

These formulas hold for the interfaces of constant $x^2$ (respectively $x^3$) under one (respectively two) applications of the cyclic permutation $1 \mapsto 2 \mapsto 3$ in all indices, including the rows of $\tensor{M}{^\mu_{\hat{\nu}}}$ and the columns of $\tensor{M}{^{\hat{\mu}}_\nu}$.

One important detail to be noted is that the constant-coordinate interface across which we desire fluxes will generally appear to be moving in the orthonormal coordinates. To see this, consider an interface of constant $x^1$ and how it appears in orthonormal coordinates $(\hat{t},\hat{x},\hat{y},\hat{z})$. According to \eqref{eq:to_global} we know
\begin{equation} \label{eq:interface_coordinates}
  x^1 = A g^{01} \hat{t} + B (g^{01} g^{01} - g^{00} g^{11}) \hat{x}.
\end{equation}
Given that $x^1$ is constant, differentiation of \eqref{eq:interface_coordinates} tells us the interface velocity can be calculated as
\begin{equation} \label{eq:interface_velocity}
  v^{\hat{x}} \equiv \frac{\dd\hat{x}}{\dd\hat{t}} = -\frac{Ag^{01}}{B(g^{01}g^{01}-g^{00}g^{11})} = \frac{g^{01}}{\sqrt{g^{01}g^{01}-g^{00}g^{11}}}.
\end{equation}
In the language of the $3{+}1$ formalism with lapse $\alpha$, shift components $\beta^i$, and contravariant $3$-metric components $\gamma^{ij}$, we can write $v^{\hat{x}} = \beta^1\! / \alpha \sqrt{\gamma^{11}}$, in agreement with \citet{Pons1998}. That is, a nonzero shift in the direction of interest results in a nonzero interface velocity in the chosen orthonormal frame.

As this transformation yields physical quantities as measured by a normal observer, it can be applied even inside the ergosphere or, in the case of horizon-penetrating coordinates, inside the event horizon. This would not be the case if for example we attempted to transform into the frame of a stationary observer, since the spacetime might not everywhere admit timelike stationary worldlines. The transformation can be singular at coordinate singularities such as a polar axis, but such cases must be handled carefully no matter what method is being employed.

\subsection{Riemann solver}
\label{sec:method:algorithm:riemann}

The Riemann problem is that of determining time-averaged fluxes across an interface given constant left and right states. A wide variety of exact and approximate solvers have been developed. For example, Roe solvers \citep{Roe1981} solve a full linearized set of equations, obtaining high accuracy at the cost of performance.

An alternative is the HLL family of Riemann solvers (\S\ref{sec:intro:numerics:riemann}), based on the work of \citet{Harten1983a}, which have the advantage of being computationally simple while guaranteeing positivity (i.e.\ returning physically admissible intermediate states given physically admissible inputs). All HLL Riemann solvers take as input left and right states as well as leftgoing and rightgoing signal speeds, which we calculate as either the sound speeds in hydrodynamics or the fast magnetosonic speeds in MHD. The calculation of special-relativistic MHD wavespeeds, which are needed even in general relativity in the transformed frame, is detailed in Appendix~\ref{chap:wavespeeds}.

\begin{figure}
  \centering
  \includegraphics[width=6in]{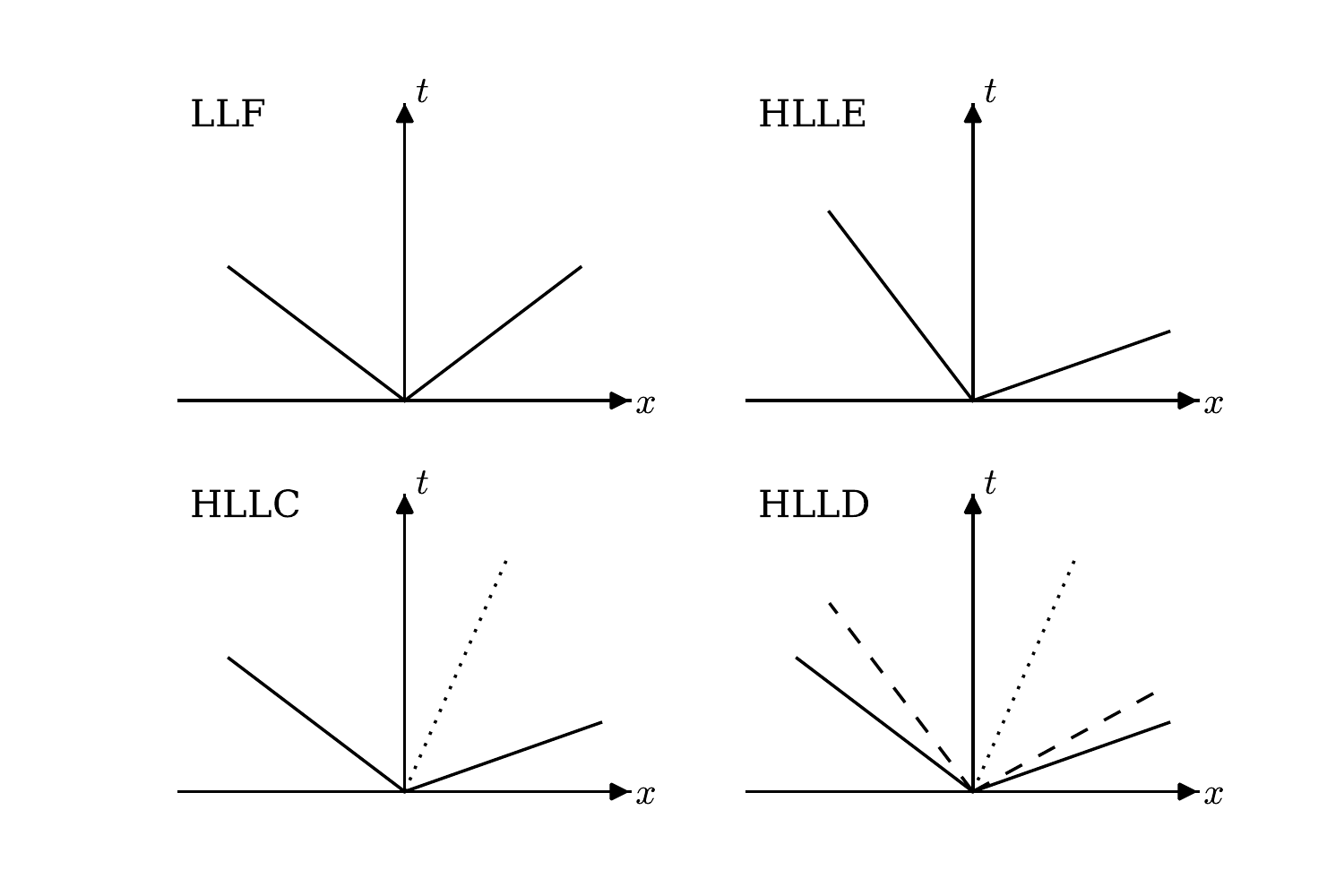}
  \caption{Schematic $1{+}1$ spacetime diagrams of different HLL Riemann solvers' wavefans. The solid diagonal lines indicate the fastest moving waves in either direction, with LLF assuming the speeds are the same. The dotted lines in HLLC and HLLD represent the contact discontinuity, and the dashed lines in HLLD represent the Alfv\'en waves. The ordering of the waves shown is accurate, though waves can be degenerate. Fluxes are determined along the $x = 0$ surface for interfaces at rest, which can be in any of the $3$, $4$, or $6$ regions of the wavefan. Slopes are qualitative; in particular all wavespeeds are subluminal. \label{fig:riemann}}
\end{figure}

The HLL solvers we consider are introduced in \S\ref{sec:intro:numerics:riemann}. Figure~\ref{fig:riemann} illustrates how the different solvers treat the internal shock structure.

Owing to the frame transformation, the same Riemann solvers can be used in both special and general relativity. As noted in \S\ref{sec:method:algorithm:transformation}, however, one must be careful to consider the Riemann problem with a moving interface. This is done by constructing the waves and intermediate states as usual and then determining the relevant portion of the wavefan according to where the generally nonzero interface velocity \eqref{eq:interface_velocity} falls relative to the wavespeeds. In the schematic sense of Figure~\ref{fig:riemann}, the fluxes sought are not necessarily along the vertical lines, but could be along angled lines also passing through the origin.

\subsection{Inverse Frame Transformation}
\label{sec:method:algorithm:inverse_transformation}

The transformation \eqref{eq:to_global} back to global coordinates will have a nonzero component $\tensor{M}{^1_{\hat{t}}}$, in addition to the ever-present $\tensor{M}{^1_{\hat{x}}}$, whenever there is a moving interface. This necessitates having not just the components $T^{\hat{x}\hat{\nu}}$ but also $T^{\hat{t}\hat{\nu}}$ in the orthonormal frame. That is, we need the conserved state in the appropriate region of the wavefan, not just the associated fluxes. In practice even the more complicated Riemann solvers solve for the intermediate states by finding primitive variables, from which one can deduce consistent conserved states as well as fluxes.

The general procedure for obtaining fluxes proceeds as follows. First, the Riemann problem consisting of left and right states, extremal wavespeeds, and an interface velocity is transformed into the appropriate locally Minkowski frame. Next, the Riemann solver is used to solve for the wavefan, including the speeds of internal waves as well as the conserved quantities and their associated fluxes normal to the interface in each region separated by the waves. The appropriate region is then selected based on where the interface velocity falls inside the wavefan. Finally, the fluxes and conserved quantities in that region are transformed back to fluxes in the original coordinate system.

\subsection{Constrained Transport}
\label{sec:method:algorithm:ct}

We employ a CT update of the face-centered magnetic fields in order to maintain the divergence-free constraint without resorting to divergence cleaning. The fundamentals of CT were in fact developed within the context of general relativity \citep{Evans1988}. Our implementation differs slightly from that original description, for example by having cell-centered rather than face-centered velocities. Ours is a simple extension of the algorithm detailed in \citet{Gardiner2005} to relativity, and we summarize it here.

\begin{figure}
  \centering
  \includegraphics[width=5in]{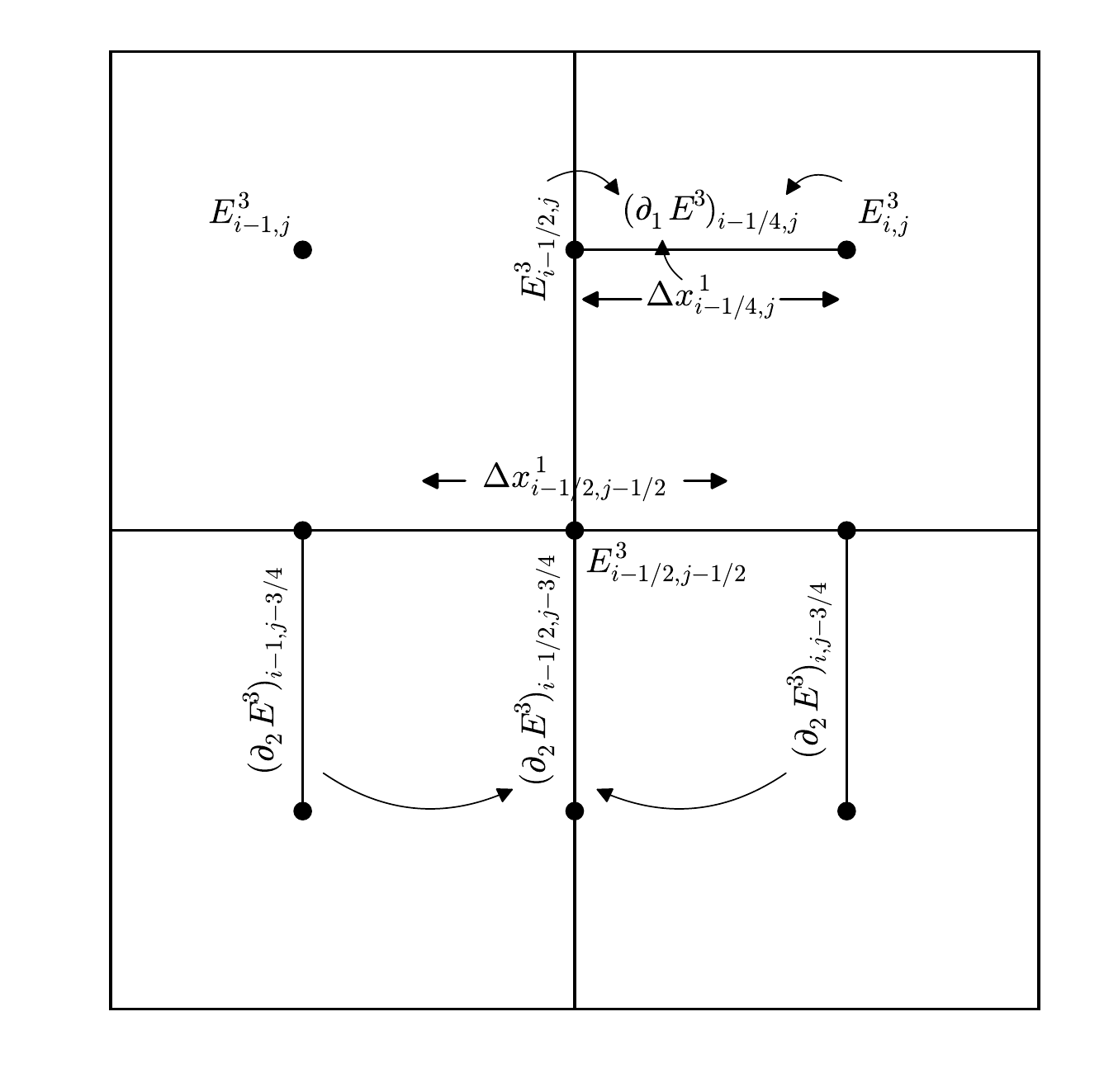}
  \caption{$2{+}0$ spacetime diagram showing a timeslice of four cells. The $x^3$ direction is out of the plane of the figure, and indices $k$ have been suppressed from all quantities. The top left cell shows what is known at the end of the Riemann problem:\ electric fields on the faces, as well as easily deducible cell-centered electric fields. The top right cell illustrates how a gradient is determined. The bottom two cells show gradients being upwinded back to the faces. Four face-centered fields, four upwinded gradients, and two edge-centered widths are combined to yield the edge-centered electric field $E^3_{i-1/2,j-1/2}$. \label{fig:ct_diagram}}
\end{figure}

The goal of the algorithm is to determine edge-centered electric fields to use in the update \eqref{eq:b_update} consistent with the fluxes returned by the Riemann solver. For concreteness, we will show how we calculate $E^3_{i-1/2,j-1/2}$, dropping all time superscripts. The algorithm proceeds in five steps:
\begin{enumerate}
  \item Obtain the four face-centered electric fields $E^3_{i-1/2,j(-1)}$ and $E^3_{i(-1),j-1/2}$ from the Riemann solver.
  \item Calculate the four cell-centered electric fields $E_{i(-1),j(-1)}$.
  \item Find the eight electric field gradients
    \begin{equation*}
      (\pp_1 E^3)_{i-1/4(-1/2),j(-1)} \text{ and } (\pp_2 E^3)_{i(-1),j-1/4(-1/2)}.
    \end{equation*}
  \item Upwind the electric field gradients to obtain the four gradients
    \begin{equation*}
      (\pp_1 E^3)_{i-1/4(-1/2),j-1/2} \text{ and } (\pp_2 E^3)_{i-1/2,j-1/4(-1/2)}.
    \end{equation*}
  \item Combine the face-centered fields and upwinded gradients in a specific way to form the edge-centered electric field $E^3_{i-1/2,j-1/2}$.
\end{enumerate}
The relevant quantities are laid out schematically in Figure~\ref{fig:ct_diagram}.

For the first step, we simply note that the electric fields in the $3$-direction are nothing more than the $1,2$-components of the dual of the electromagnetic field tensor \eqref{eq:electromagnetic_tensor_upper} and are thus the fluxes returned by the Riemann solver in the $x^1$- and $x^2$-directions (after transforming back to global coordinates). In particular, the electric fields $E^3_{i-1/2,j(-1)} = -(\ast F)^{21}_{i-1/2,j(-1)}$ are the negatives of the $x^1$-fluxes of $B^2$. Similarly, we use the positive $x^2$-fluxes of $B^1$ to obtain $E^3_{i(-1),j-1/2} = (\ast F)^{12}_{i(-1),j-1/2}$.

For the second step (which does not depend on the first), we calculate the cell-centered electric fields from the cell-centered velocities and the interpolated magnetic fields (e.g.\ \eqref{eq:b_interpolation}). This amounts to simply constructing the appropriate $4$-vector components $u^\mu$ and $b^\mu$ and using \eqref{eq:electromagnetic_tensor} and \eqref{eq:electromagnetic_tensor_upper}.

In the third step we take gradients of electric field components from cell centers to faces, similar to \extref{45} of \citeauthor{Gardiner2005}. In order to calculate such gradients, we require cell widths, though as we shall see these can be canceled from the final expression to a good approximation. Consider the gradient $\pp_1 E^3$ located between cell center $i,j$ and face center $i-1/2,j$, as in the top right cell of Figure~\ref{fig:ct_diagram}. If we have
\begin{equation}
  \Delta x^1_{i-1/4,j} = \int\limits_{C} \sqrt{g_{11}} \, \dx^1
\end{equation}
with $C$ the curve of constant $x^0$, $x^2$, and $x^3$ running from the cell center to the face, then we can take
\begin{equation} \label{eq:e_grad}
  (\pp_1 E^3)_{i-1/4,j} = \frac{E^3_{i,j}-E^3_{i-1/2,j}}{\Delta x^1_{i-1/4,j}}.
\end{equation}
Similarly we can calculate the other seven such gradients.

Fourth, we shift the gradients onto the faces via upwinding. For example, as illustrated in Figure~\ref{fig:ct_diagram}, we calculate
\begin{equation}
  (\pp_2 E^3)_{i-1/2,j-3/4} =
  \begin{cases}
    (\pp_2 E^3)_{i-1,j-3/4}, & u^1_{i-1/2,j-1} > 0, \\
    (\pp_2 E^3)_{i,j-3/4}, & u^1_{i-1/2,j-1} < 0,
  \end{cases}
\end{equation}
according to the sign of the mass flux returned by the Riemann solver. This is \extref{50} of \citeauthor{Gardiner2005}.

Finally, we compute the edge-centered value as is done in \extref{41} of \citeauthor{Gardiner2005}. If we have an appropriate width $\Delta x^1_{i-1/2,j-1/2}$ defined similarly to $\Delta x^1_{i-1/4,j}$ -- that is, by integrating $\sqrt{g_{11}}$ as $x^1$ varies over half a cell, as shown in Figure~\ref{fig:ct_diagram} -- then the value we seek is
\begin{equation} \begin{split} \label{eq:e_combined}
  E^3_{i-1/2,j-1/2} & = \frac{1}{4} \paren[\Big]{E^3_{i-1/2,j} + E^3_{i-1/2,j-1} + E^3_{i,j-1/2} + E^3_{i-1,j-1/2}} \\
  & \quad \qquad + \frac{\Delta x^1_{i-1/2,j-1/2}}{4} \paren[\Big]{E^3_{i-3/4,j-1/2} - E^3_{i-1/4,j-1/2}} \\
  & \quad \qquad + \frac{\Delta x^2_{i-1/2,j-1/2}}{4} \paren[\Big]{E^3_{i-1/2,j-3/4} - E^3_{i-1/2,j-1/4}}.
\end{split} \end{equation}
However, if the metric and grid spacing are varying smoothly, we can approximately cancel the $\Delta x$ factors in \eqref{eq:e_combined} with those in \eqref{eq:e_grad}.

\subsection{Source Terms}
\label{sec:method:algorithm:sources}

In non-Cartesian coordinates one must consider not only fluxes but also geometric source terms when updating conserved quantities according to \eqref{eq:finite_volume}. As can be seen in \eqref{eq:differential}, the continuity and magnetic field equations never have source terms but the energy-momentum equation generally does.

As noted in \citet{Gammie2003}, with the adopted index placement on the stress-energy tensor the source terms for $\tensor{T}{^0_\mu}$ will vanish when the metric does not depend on $x^\mu$. In particular, since the metric is stationary we never have geometric source terms in energy.

For the remaining momentum equations we evaluate the stress-energy components using the primitive variables at the appropriate timestep, taking the metric to have its cell-centered values. We then contract these components with the connection coefficients according to \eqref{eq:source_decomposition}, resulting in the appropriate quantities for adding to the conserved variables.

Source terms for additional physics (e.g.\ heating and cooling functions) can be added to the conserved quantities along with the geometric terms.

\subsection{Time Integrator}
\label{sec:method:algorithm:integrator}

At the heart of the numerical implementation is the time integrator. We employ a temporally second-order van~Leer integrator for all quantities. Given the conserved variables at time step $n$, we first calculate the associated primitive variables (\S\ref{sec:method:algorithm:inversion}). Next, Riemann problems are set up (\S\ref{sec:method:algorithm:reconstruction} and \S\ref{sec:method:algorithm:transformation}) and solved (\S\ref{sec:method:algorithm:riemann}), yielding fluxes. We also compute any necessary source terms from the primitives (\S\ref{sec:method:algorithm:sources}).

The fluxes and source terms are used to update all conserved variables by half a timestep, including the magnetic fields (\S\ref{sec:method:algorithm:ct}). The same procedure is repeated at time step $n{+}1/2$, except the resulting fluxes and source terms are used to update the step $n$ state to step $n+1$. Schematically, we advance the grid by a single timestep via
\begin{subequations} \begin{align}
  C^{n+1/2} & = C^n + \frac{\Delta t}{2} S^n + \frac{\Delta t}{2\Delta V} \sum_\mathrm{faces} F^n \Delta A, \\
  C^{n+1} & = C^n + \Delta t S^{n+1/2} + \frac{\Delta t}{\Delta V} \sum_\mathrm{faces} F^{n+1/2} \Delta A.
\end{align} \end{subequations}
The van~Leer integrator is TVD, and so it will not introduce spurious extrema.

Athena++ implements a variety of other temporal integration algorithms, including the TVD second-order and third-order Runge--Kutta methods (RK2 and RK3) of \citet{Shu1988}. The RK2 integrator is the same as Heun's method and we have found is no better than the van~Leer method we implement. It suffers from the drawback that second-order spatial reconstruction in both substeps, whereas we can use first-order reconstruction for the first half step with the van~Leer scheme and still obtain second-order convergence. RK3 and other higher-order schemes generally require extra storage of intermediate results, and for our present purposes we find second order in time to be sufficient. CT corner transport upwind (CTU) integrators have also been developed for MHD \citep{Gardiner2005,Gardiner2008}, but CTU methods require time-advanced variable estimates that are difficult to generalize to arbitrary coordinate systems.

\subsection{Storage Requirements}
\label{sec:method:algorithm:storage}

At each cell, we must store $5$ primitive variables and $5$ conserved variables for each half step as required by the van~Leer integrator. For MHD in $3$ spatial dimensions we also store $3$ each of face-centered and cell-centered magnetic fields and edge-centered, face-centered, and cell-centered electric fields, all at each half step. This amounts to $20$ numbers per cell in hydrodynamics, and $50$ numbers per cell in MHD.

Any finite volume code must have access to cell volumes, interface areas, edge lengths (in the case of MHD), and cell widths (in order to determine maximum stable timesteps). In general relativity, the values of $g_{\mu\nu}$, $g^{\mu\nu}$, $\Gamma^\sigma_{\mu\nu}$, and the transformation matrices \eqref{eq:to_global} and \eqref{eq:to_local} are also required at various stages of the integration. Fortunately the assumption of a stationary metric means these terms can be precomputed. Moreover, symmetries of the metric combined with the restriction that cells be divided by interfaces of constant coordinate values often results in separable formulas for these values, obviating the need for 3D storage arrays.

For example, in the Kerr--Schild coordinates used in Chapters~\ref{chap:test} and~\ref{chap:application} and defined in \S\ref{sec:examples:kerr:kerr-schild}, the volume \eqref{eq:volume} of a cell bounded by coordinates $r_\pm$, $\theta_\pm$, and $\phi_\pm$ is
\begin{equation}
  \Delta V = \frac{1}{3} (r_+-r_-) (\cos\theta_--\cos\theta_+) (\phi_+-\phi_-) \paren[\big]{f(r_-,r_+) + a^2 f(\cos\theta_-,\cos\theta_+)},
\end{equation}
where $f(x,y) = x^2 + xy + y^2$. The three parenthesized prefactors and the two $f$ terms can each be stored in 1D arrays, where $\Delta V$ is computed using just a small handful of additions and multiplications when needed. In fact, no more than 2D storage is required for any metric discussed in this work. For Kerr--Schild, the most complicated metric discussed here, we store $22 N_r + 32 N_\theta + 9 N_\phi + 20 N_r N_\theta$ values on an $N_r \times N_\theta \times N_\phi$ grid. This may seem substantial in comparison to the $4 N_x + 4 N_y + 4 N_z$ coordinate-related values stored when using the Minkowski metric, but still pales in comparison to the $(20\text{--}50) N_1 N_2 N_3$ values holding evolving variables. Thus the memory footprint of our code is substantially smaller than that of codes that store all geometric factors without regard for symmetry or separability.

All the necessary expressions are cataloged in Appendix~\ref{chap:examples} for a number of different coordinate systems.


\chapter{Testing Athena++}
\label{chap:test}

As with any large code, there are many places where errors can be introduced, either in the underlying algorithm or the actual implementation. While some errors may lead to crashes or other such obvious symptoms, there exists a worrying class of issues wherein the code returns reasonable results that are nonetheless wrong. A battery of quantitative tests can be used to combat such insidious bugs.

The strategy, then, is to have the code solve problems that altogether exercise all of the components of the procedure. For example, small perturbations (small enough to be able to neglect nonlinear terms in the evolution) can be initialized matching the eigenfunctions of the linearized equations. These linear waves should hold their shape over time, with the errors decreasing with increasing resolution. Success indicates reconstruction and temporal integration are almost certainly operating correctly. Similarly, the nonlinear terms are important in the Riemann solver and variable inversion, and running shock tubes should detect problems with these parts of the code.

A suite of selected tests is described in the following sections. Because much of the machinery for our code is the same in both general and special relativity, several of our tests are special relativistic in nature. Some are run in both special- and general-relativistic settings, and others are purely general relativistic.

\section{Linear Waves}
\label{sec:test:linear}

Linear wave convergence is a strong test of a code, and so we present convergence results for special-relativistic hydrodynamics and magnetohydrodynamics (MHD).

Choose a background constant state and consider only perturbations in the $x$-direction. If we write the time evolution equations in terms of a matrix $A$ depending on the background and the vector of primitives $\altvec{P}$,
\begin{equation}
  \pp_t \altvec{P} + A \pp_x \altvec{P} = 0,
\end{equation}
then the linear waves are perturbations $\delta\altvec{P}$ that are eigenvectors of $A$.

In order to quantify the error, we evolve a wave for exactly one period and calculate the error for each primitive quantity $q$ as the $L^1$ norm of the difference between the initial and final states:
\begin{equation} \label{eq:l1_error}
  \epsilon_q = \frac{1}{N} \sum_{i=1}^N \abs{q^\mathrm{init}_i-q^\mathrm{fin}_i}.
\end{equation}
We then take the overall error to be the root-mean-square value of the set of errors $\epsilon_q$.

\begin{figure}
  \centering
  \includegraphics[width=\textwidth]{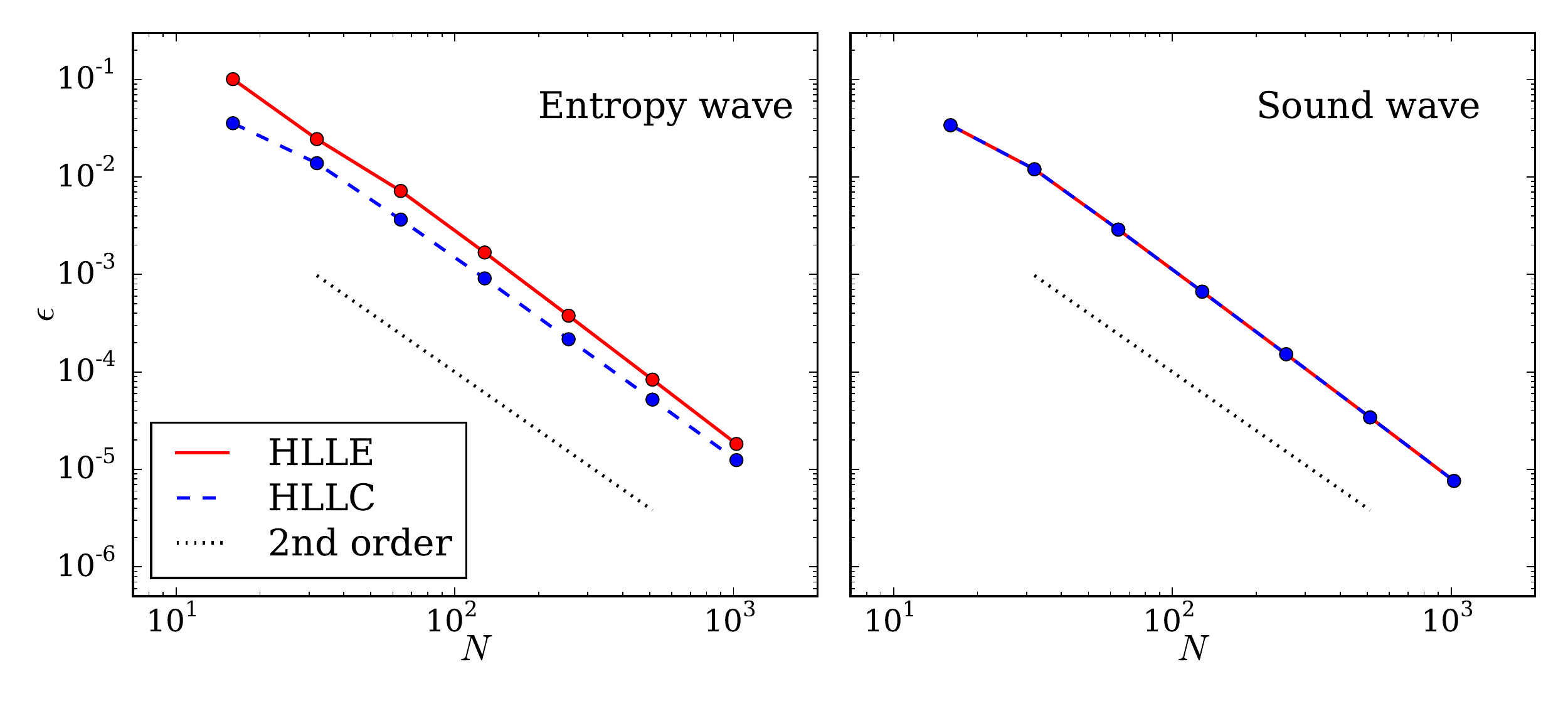}
  \caption{Convergence tests for special-relativistic hydrodynamics. Both Riemann solvers converge at second order, with HLLC having a smaller overall error compared to HLLE for the entropy wave. LLF results are indistinguishable from HLLE. \label{fig:linear_sr_hydro}}
\end{figure}

The eigenvectors for hydrodynamics are given in \citet[their Appendix~A]{Falle1996}. We choose a background state consisting of $\rho = 4$, $\pgas = 1$, and $v^i = (0.1, 0.3, -0.05)$, with ratio of specific heats $\Gamma = 4/3$. The resulting wavespeeds are approximately $-0.30$, $0.1$, and $0.47$. We evolve the entropy and rightgoing sound wave, using both HLLE and HLLC Riemann solvers. The results are shown in Figure~\ref{fig:linear_sr_hydro}. Both methods converge at second order in the number of cells, as is expected since the reconstruction and time integration are both second order. Note that the error in the entropy wave is $1.7$ times larger using HLLE compared to HLLC for these parameters, while the solvers have the same error for the sound wave. This reflects the fact that when inside the wavefan, using just the sound speeds leads to averaging over too large a domain of dependence, diffusing the solution.

\begin{figure}
  \centering
  \includegraphics[width=\textwidth]{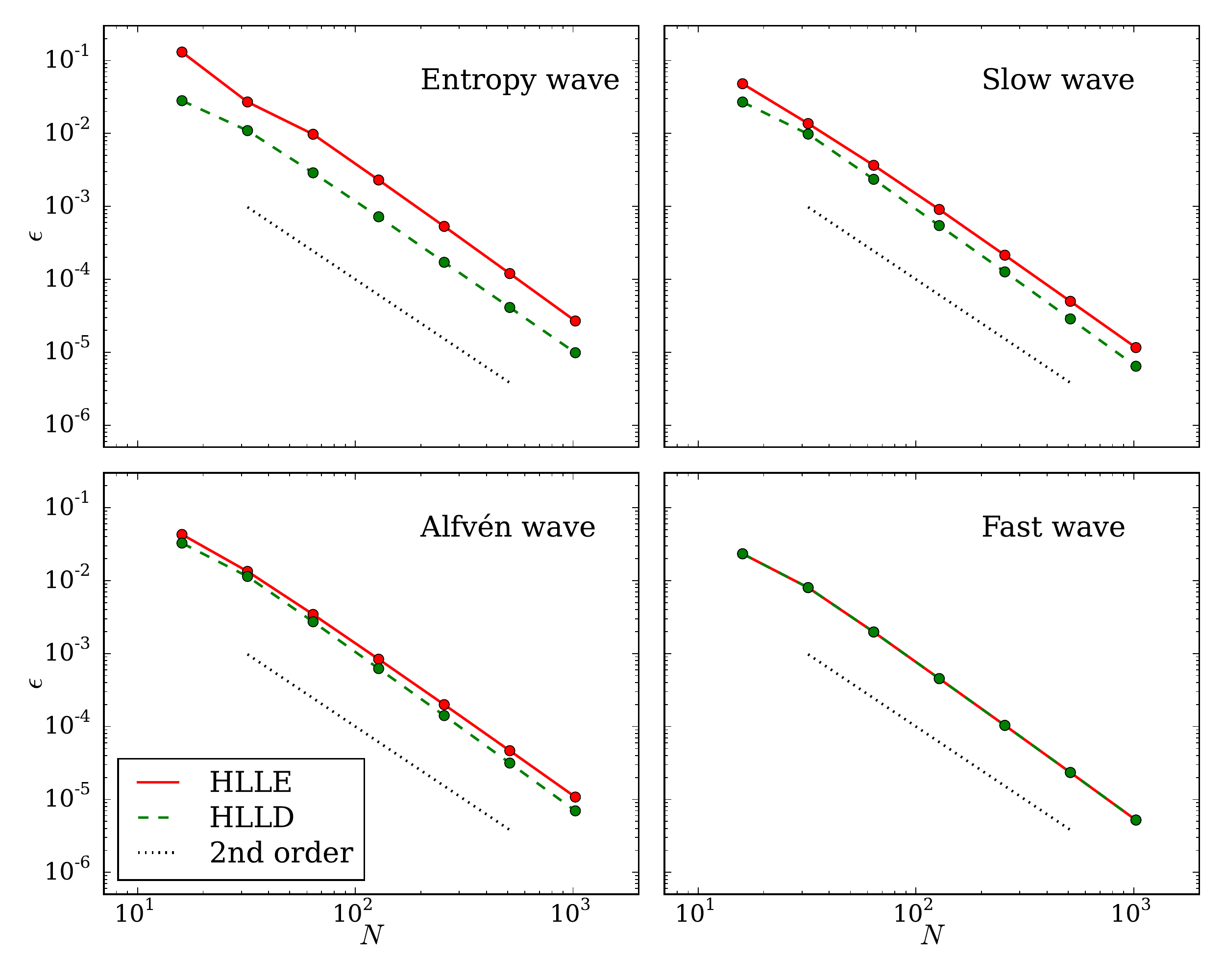}
  \caption{Convergence tests for special-relativistic MHD. Both Riemann solvers converge at second order, with HLLD having a smaller overall error compared to HLLE for all but the fast wave. LLF results are indistinguishable from HLLE. \label{fig:linear_sr_mhd}}
\end{figure}

For MHD we take the same hydrodynamic background state and add a magnetic field with components $B^i = (2.5, 1.8, -1.2)$, resulting in all wavespeeds (approximately $-0.71$, $-0.55$, $-0.21$, $0.1$, $0.35$, $0.54$, and $0.81$) being well separated. The eigenvectors are given in \citet{Anton2010} as their \extref{46} (entropy), \extref{65} (Alfv\'en), and \extref{71} (magnetosonic). The results for the four rightmost waves are given in Figure~\ref{fig:linear_sr_mhd}. Again we have second order convergence, and again the more advanced Riemann solver has smaller errors for the internal waves of the Riemann fan. Here HLLD's error is a factor of $3.1$, $1.7$, and $1.4$ times lower than that of HLLE for the entropy, slow magnetosonic, and Alfv\'en waves, respectively.

We can also perform these same tests in a way that exercises the general-relativistic portions of the code. Consider the ``tilted'' coordinates $(T, X, Y, Z)$ related to Minkowski coordinates $(t, x, y, z)$ by
\begin{subequations} \begin{align}
  T & = \frac{t+ax}{\sqrt{1+a^2}}, \\
  X & = \frac{x-at}{\sqrt{1+a^2}}, \\
  Y & = y, \\
  Z & = z
\end{align} \end{subequations}
for some parameter $a$, $\abs{a} < 1$. This coordinate system has its time axis corresponding to an observer moving with velocity $a$ with respect to the Minkowski frame, while the surface of constant time corresponds to that of an observer with velocity $-a$. Visually this corresponds to rigidly rotating the $tx$-axes into the $TX$-axes. While the connection coefficients and thus source terms vanish in this coordinate system, the $TX$ metric coefficients are nonvanishing and so there are off-diagonal components in the transformation matrices \eqref{eq:to_global} and \eqref{eq:to_local}. Additionally, the interface velocity \eqref{eq:interface_velocity} is nonzero. (See \S\ref{sec:examples:minkowski:tilted} for the details of this coordinate system.)

We can rerun the previous convergence tests in this new coordinate system. We use the same background states, with the given numbers understood to be in Minkowski coordinates. The linear solution is known throughout spacetime, and we initialize the perturbation over a single wavelength on the surface $T = 0$. This is evolved until $T = (1 + a\lambda) / \lvert \lambda-a \rvert$, which corresponds to one wavelength crossing the domain at Minkowski $3$-velocity $\lambda$. For most waves we choose $a = 0.1$. For the entropy waves we choose $a = 0.05$, since if $a$ matches $\lambda$ the HLLC and HLLD solvers will exactly capture the wave, placing our comparison \eqref{eq:l1_error} at the machine precision noise floor even for low resolutions.

\begin{figure}
  \centering
  \includegraphics[width=\textwidth]{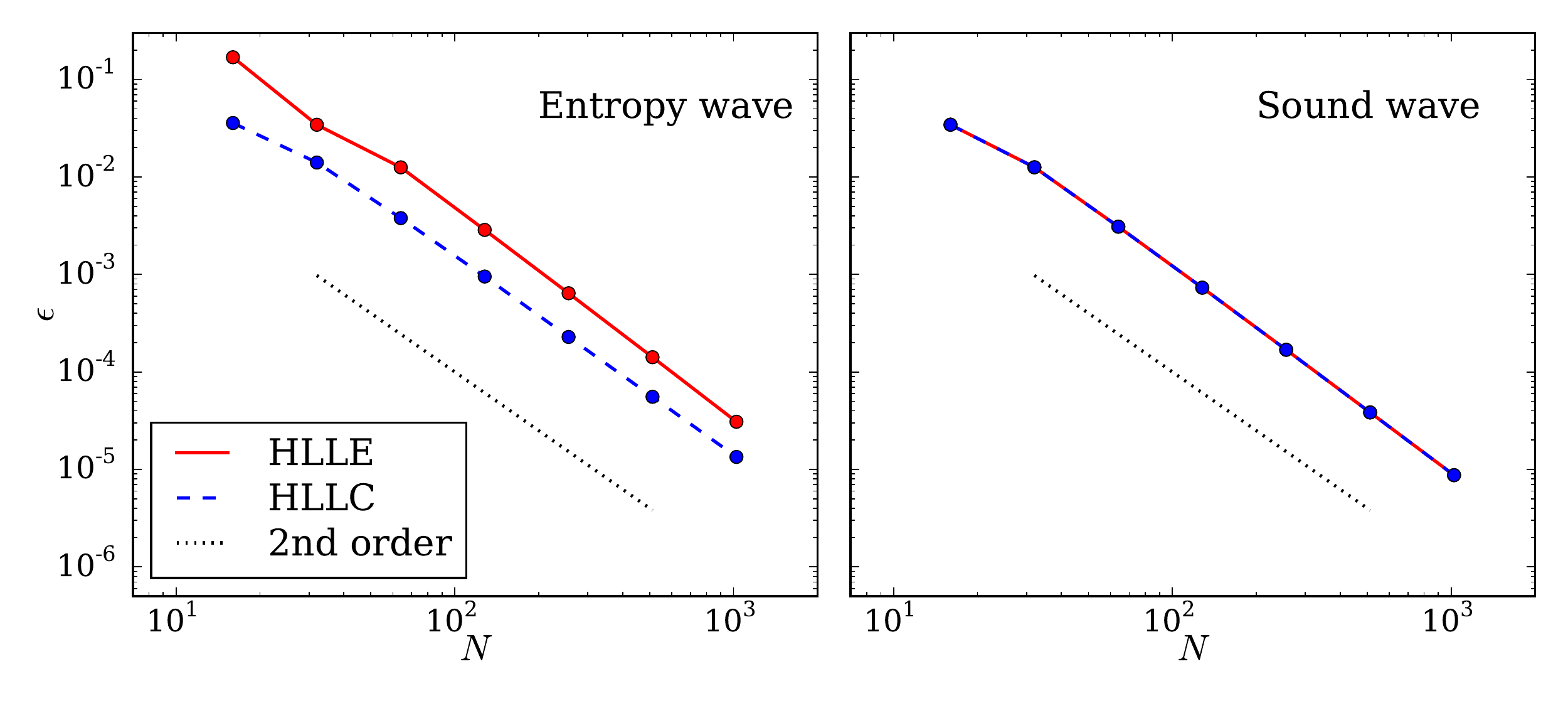}
  \caption{Convergence tests for general-relativistic hydrodynamics. Both Riemann solvers converge at second order, with HLLC improving upon HLLE for the entropy wave, as in Figure~\ref{fig:linear_sr_hydro}. \label{fig:linear_gr_hydro}}
\end{figure}

\begin{figure}
  \centering
  \includegraphics[width=\textwidth]{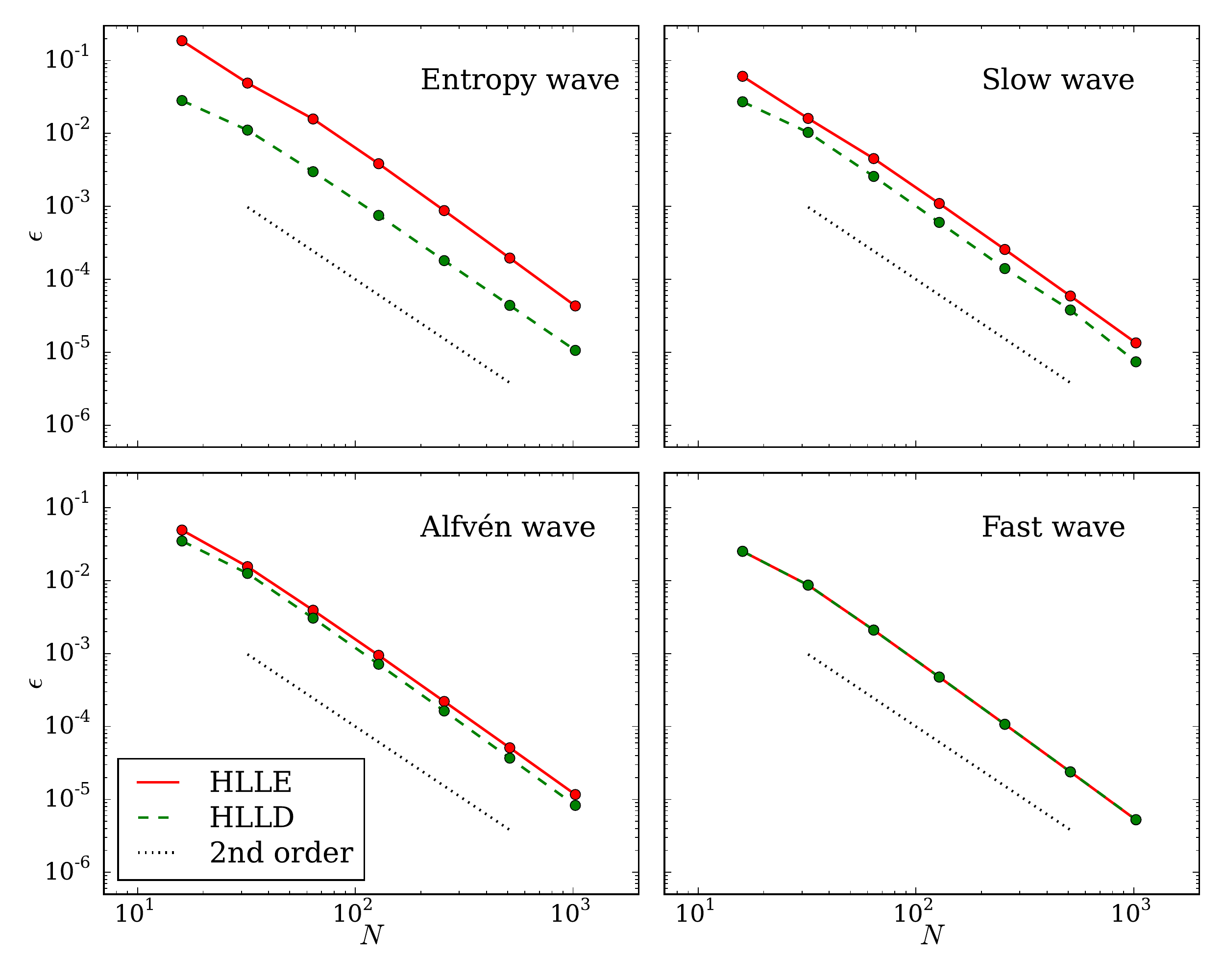}
  \caption{Convergence tests for general-relativistic MHD. Both Riemann solvers converge at second order, with HLLD improving upon HLLE for all but the fast wave, as in Figure~\ref{fig:linear_sr_mhd}. \label{fig:linear_gr_mhd}}
\end{figure}

The results for hydrodynamics and MHD are given in Figures~\ref{fig:linear_gr_hydro} and~\ref{fig:linear_gr_mhd}. For the hydrodynamic entropy wave, the HLLC solver's error is lower than that of HLLE by a factor of $2.8$. In MHD, the HLLD solver's error beats that of HLLE by $4.8$, $1.8$, and $1.4$ in the entropy, slow, and Alfv\'en waves. We note that in these particular tests the advanced Riemann solvers outperform HLLE by larger margins in general relativity compared to special relativity.

\section{Shock Tubes}
\label{sec:test:shocks}

Another class of one-dimensional special-relativistic tests is that of shock tubes. While the above wave tests focus on linear perturbations, shock tubes emphasize the nonlinear behavior of the equations, as well as the ability to resolve discontinuities in the solutions.

\begin{figure}
  \centering
  \includegraphics[width=\textwidth]{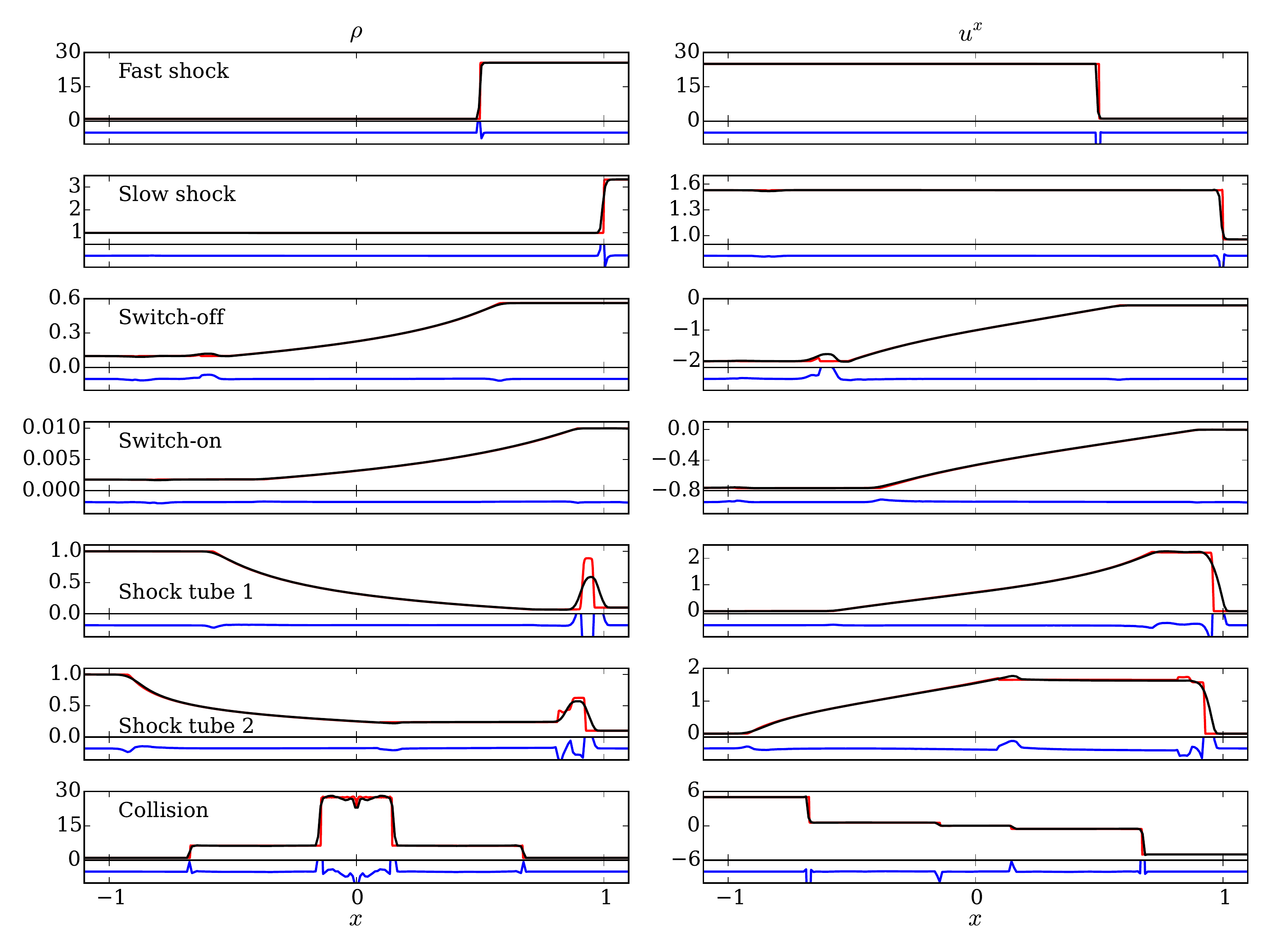}
  \caption{\Citeauthor{Komissarov1999}\ shock tubes for special-relativistic MHD. Results are computed with HLLD on a grid of $220$ (black) and $2200$ (red) cells. The blue lines show the relative error between the two resolutions normalized by the maximum value attained by the variable over the domain, with the plot limits being $\pm10\%$ in all cases. \label{fig:nonlinear_waves}}
\end{figure}

We show the results for seven shock tubes from \citet{Komissarov1999} \citep[cf.][their Figures~4 and~5]{Gammie2003} in Figure~\ref{fig:nonlinear_waves}. These results in black all use the HLLD solver with $220$ grid cells in the interval $[-1.1,1.1]$, corresponding to the same fixed resolution used by \citeauthor{Gammie2003}\ in all cases. The underplotted red line is a reference solution computed with ten times the resolution. The Courant--Friedrichs--Lewy (CFL) number is $0.8$ except in the cases of the fast shock ($0.2$) and the collision ($0.5$).

Some of the same numerical artifacts observed by \citeauthor{Komissarov1999}\ can be seen in our results. For example, we see ringing upstream of the slow shock, a spurious bump at the low-density end of the switch-off rarefaction wave, and overshooting at the low-density end of the shock tube~2 rarefaction wave. In the case of shock tube~1 we do not observe the overshooting in density at the base of the density rarefaction wave, and we have greatly diminished overshooting at the top of the velocity discontinuity, despite using the same resolution as \citeauthor{Komissarov1999}. That HLLD in particular does well at capturing shocks is not surprising; \citet{Mignone2009} note this in comparing different Riemann solvers. We conclude that we are correctly capturing the nonlinear behavior expected in this special-relativistic test suite.

\section{Spherical Blast Wave}
\label{sec:test:blast}

As a two-dimensional test of our general-relativistic formulation, we evolve a spherical blast wave. The initial conditions are uniform density ($\rho = 1$) with a central overpressure ($\pgas = 2.5$ inside a radius of $0.5$; $\pgas = 0.1$ outside). We enforce periodic boundary conditions on a rectangular grid of width $4$ and height $6$. By time $t = 15$, the blast wave has intersected itself a number of times, inducing Richtmeyer--Meshkov instability features. The shock velocities here are not so high as to be in the regime where Richtmeyer--Meshkov instability is strongly suppressed, as discussed in \citet{Mohseni2014} and \citet{Zanotti2015}.

In order to test nontrivial spatial geometry, we evolve the system not only in Minkowski coordinates but also in ``snake'' coordinates that vary sinusoidally with Minkowski position. Explicitly, given Minkowski $(t, x, y, z)$, define
\begin{subequations} \begin{align}
  x^0 & = t, \\
  x^1 & = x, \\
  x^2 & = y + a \sin(kx), \\
  x^3 & = z,
\end{align} \end{subequations}
where $a$ and $k$ are free parameters. If we define the parameter
\begin{equation}
  \delta = a k \cos(kx^1),
\end{equation}
the metric becomes
\begin{equation}
  g_{\mu\nu} =
  \begin{pmatrix}
    -1 & 0                 & 0       & 0 \\
    0  & \sqrt{1+\delta^2} & -\delta & 0 \\
    0  & -\delta           & 1       & 0 \\
    0  & 0                 & 0       & 1
  \end{pmatrix}.
\end{equation}
The transformations \eqref{eq:to_global} and \eqref{eq:to_local} differ from the identity via $\tensor{M}{^2_{\hat{x}}} = -\tensor{M}{^{\hat{y}}_1} = \delta$, and there is a source term due to $\Gamma^2_{11}$. (More details can be found in \S\ref{sec:examples:minkowski:sinusoidal}.)

\begin{figure}
  \centering
  \includegraphics[width=\textwidth]{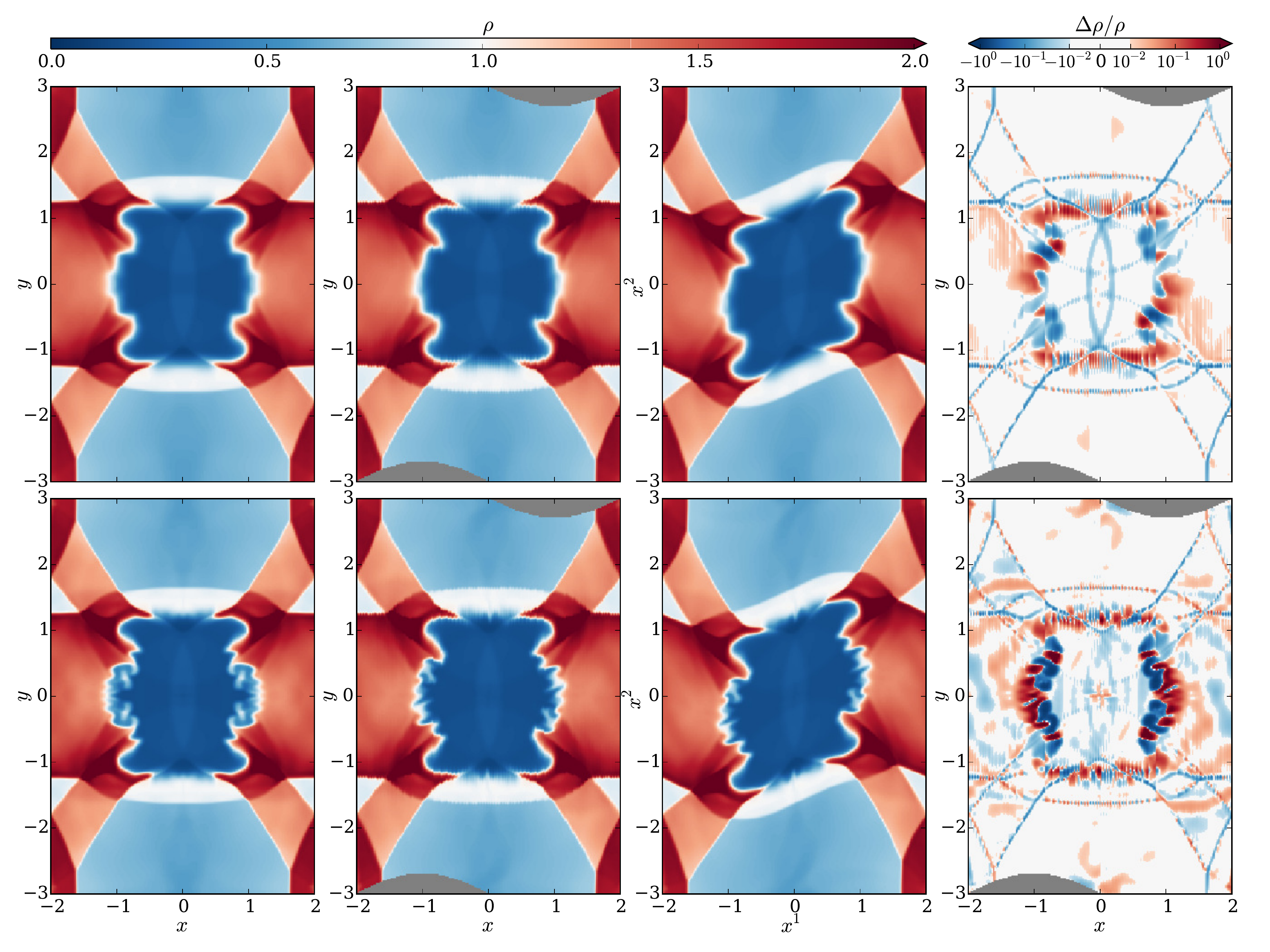}
  \caption{Density $\rho$ for the spherical blast wave at $t = 15$ in Minkowski coordinates (left), snake coordinates transformed back to Minkowski (center left), and snake coordinates (center right). The right panels show the fractional difference $\Delta\rho/\rho$ between the first and second columns, where $\Delta\rho = \rho_\mathrm{snake} - \rho_\mathrm{Minkowski}$ and white regions have a fractional difference of less than $1\%$. The top panels use HLLE, while the bottom use HLLC. \label{fig:blast_hydro}}
\end{figure}

The hydrodynamical blast wave density is shown in Figure~\ref{fig:blast_hydro}. The top row displays results using the HLLE Riemann solver, while the bottom row uses HLLC. The first column shows the results at $t = 15$ as computed in Minkowski coordinates on a $200\times300$ grid with a CFL number of $0.4$. The second column shows the same results as calculated in snake coordinates and transformed back to Minkowski at the end of the calculation. Here we also use a $200\times300$ grid, with $a = 0.3$ and $k = \pi/2$. Lines of constant $x^2$ in snake coordinates parallel the sinusoidal boundary of the shaded region.

One can see that the solutions remain qualitatively unchanged when the coordinates are varied, despite the fact that the fluxes, the source terms, and even the conserved quantities themselves vary between coordinate systems. To give a sense of how the internal representation of the fluid changes between coordinate systems, the third column of Figure~\ref{fig:blast_hydro} plots density in snake coordinates $(x^1,x^2)$.

For a more quantitative comparison, the last column shows the fractional difference in density between the Minkowski solution and the snake solution transformed to Minkowski coordinates. Colored areas indicate regions where the fractional difference exceeds $1\%$. Differences are generally confined to cell-level variations in the locations of shock fronts, or else to the details of the Richtmeyer--Meshkov fingering. This latter effect is to be expected with an instability seeded with grid noise. We also note that the HLLC Riemann solver leads to more sharply defined fingers, whereas using just HLLE the structures are notably more diffuse.

As a further test, we add an initially uniform magnetic field ($B^i = (1, 1, 0)$) and rerun the blast wave, this time with the HLLD solver. In this strongly magnetized regime, the blast wave should be guided by the magnetic field lines, breaking its initial symmetry.

\begin{figure}
  \centering
  \includegraphics[width=\textwidth]{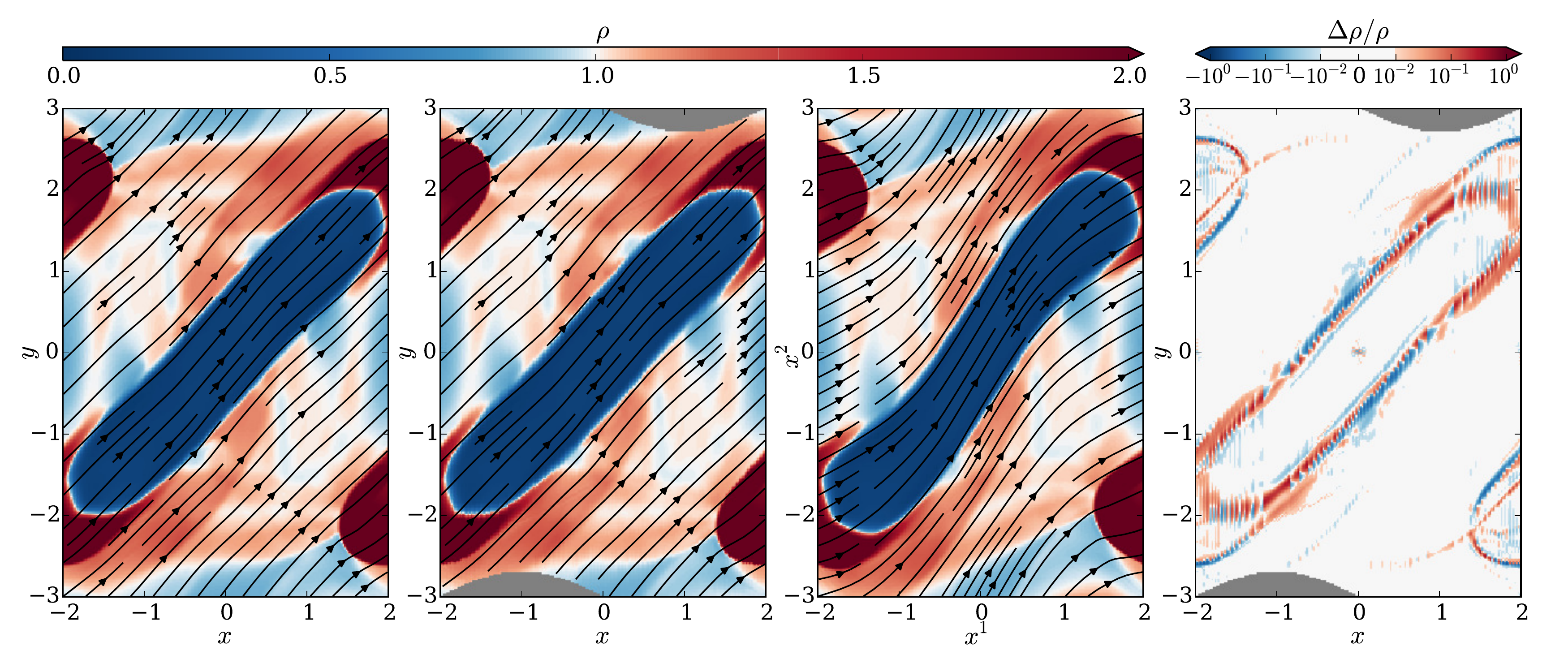}
  \caption{Similar to Figure~\ref{fig:blast_hydro} with MHD added. Density $\rho$ (color) and magnetic field $B^i$ (streamlines) are shown for the strongly magnetized blast wave at $t = 5$ in Minkowski coordinates (left), snake coordinates transformed back to Minkowski (center left), and snake coordinates (center right). The right panel shows the fractional difference between the first and second columns as in Figure~\ref{fig:blast_hydro}. All panels use HLLD. \label{fig:blast_mhd}}
\end{figure}

The MHD results are shown in Figure~\ref{fig:blast_mhd}, with color denoting $\rho$ and streamlines showing the magnetic field. The same layout is used here as in both rows of Figure~\ref{fig:blast_hydro}. Both MHD calculations are done with HLLD on a $200\times300$ periodic grid with a CFL number of $0.4$.

As expected, the field guides the blast wave and suppress the Richtmeyer--Meshkov instability. Moreover, the same results are found in both coordinate systems. The self-interaction of the wave after wrapping around the boundaries induces a slight curvature in the boundary of the rarefaction region and in the magnetic field lines, and even this is the same in the two coordinate systems.

\section{Bondi Accretion}
\label{sec:test:bondi}

As a test of the code in a nontrivial spacetime, we model Bondi accretion onto a Schwarzschild black hole. That is, we seek steady-state solutions to radial flow onto a black hole, where pressure can be nonnegligible.

The Bondi solution is derived in \citet{Hawley1984}, and we summarize the pertinent formulas here. Begin with a black hole mass $M$, an adiabatic index $\Gamma$, an adiabat $K$, and a critical radius $\rcrit$. Define the polytropic index $n = 1/(\Gamma-1)$. The critical velocity and temperature are then
\begin{gather}
  \ucrit^1 = -\sqrt{\frac{M}{2\rcrit}}, \\
  \tcrit = \frac{n}{n+1} \cdot \frac{\ucrit^1\ucrit^1}{1-(n+3)\ucrit^1\ucrit^1}.
\end{gather}
Next we calculate the constants
\begin{align}
  C_1 & = \tcrit^n \ucrit^1 \rcrit^2, \\
  C_2 & = \paren[\big]{1 + (n+1)\tcrit}^2 \paren[\bigg]{1 - \frac{2M}{\rcrit} + \ucrit^2\ucrit^1}.
\end{align}
The run of temperature with radius is given by solving the equation
\begin{equation}
  \paren[\big]{1 + (n+1)T}^2 \paren[\bigg]{1 - \frac{2M}{r} + \frac{C_1^2}{r^4T^{2n}}} = C_2.
\end{equation}
There will generally be two roots to this equation. The lesser root is taken for $r < \rcrit$, while the greater root is taken outside the critical point. We can then find the radial velocity according to
\begin{equation}
  u^1 = \frac{C_1}{r^2T^n}.
\end{equation}
Finally, density and pressure are given by
\begin{align}
  \rho & = \paren[\bigg]{\frac{T}{K}}^n, \\
  \pgas & = T \rho.
\end{align}
Note these formulas assume Schwarzschild coordinates.

We choose parameters $M = 1$, $\Gamma = 4/3$, $K = 1$, and $\rcrit = 8$, and we simulate the region $3 < r < 10$, $\pi/4 < \theta < 3\pi/4$. In order to test MHD as well as pure hydrodynamics, we optionally include a purely radial monopole field with $B^r \propto 1/r^2$, normalized such that $b_\lambda b^\lambda / \rho = 10$ at the inner radius. This corresponds to plasma $\beta = 0.246$ at the critical point. The 2D domain is divided into $N$ equally spaced cells in each direction. We initialize the Bondi solution and run the simulation until time $t = 10$, computing the $L^1$ difference in $\pgas$ between the beginning and end. We use the inner $3/4$ of the grid in each direction, and normalize the errors $\epsilon$ by the $L^1$ norm of the initial $\pgas$.

\begin{figure}
  \centering
  \includegraphics[width=6in]{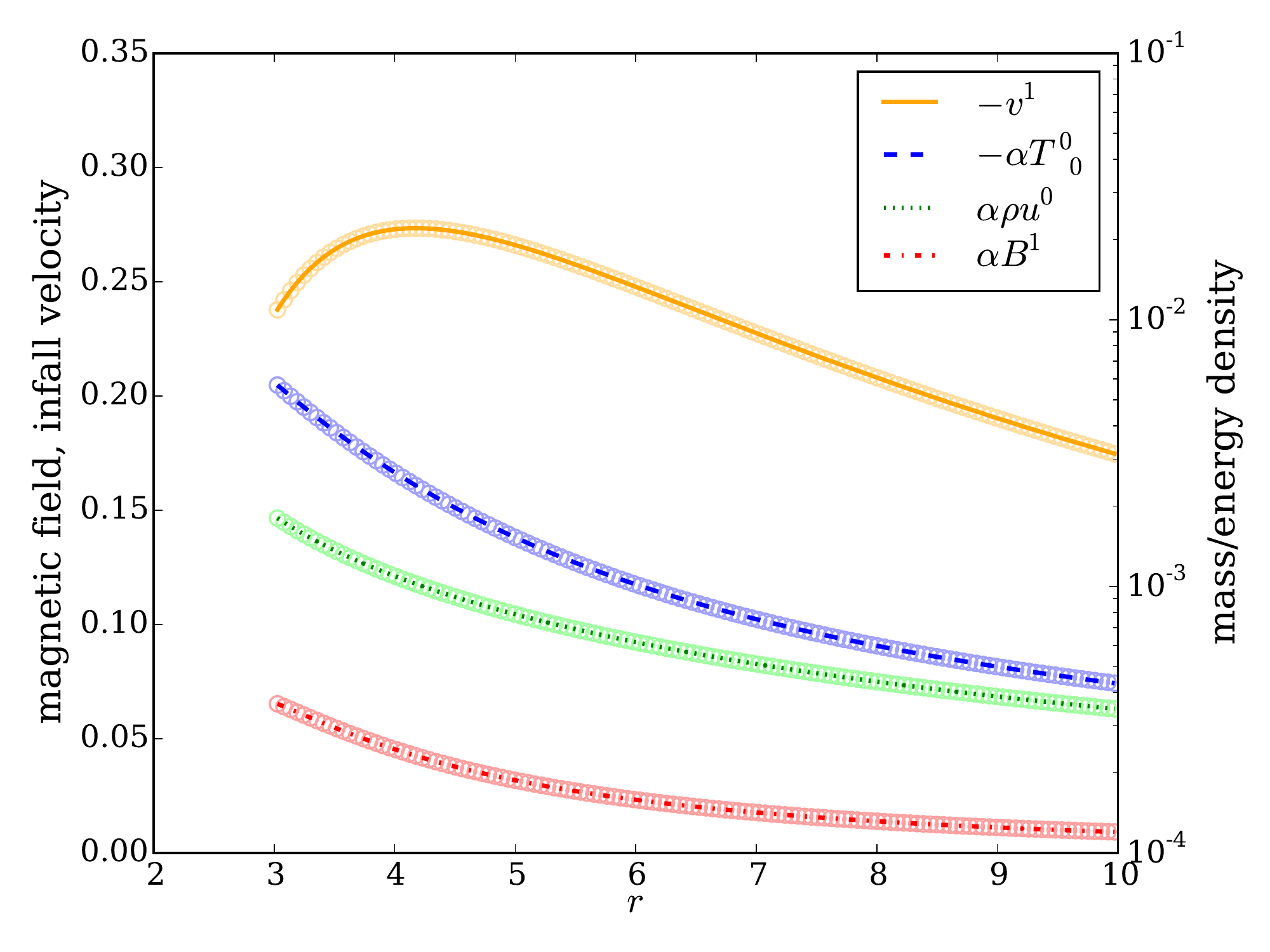}
  \caption{Magnetized Bondi flow in Schwarzschild coordinates with parameters as described in the text. Lines indicate the initial solution, while the circles indicate the values at each cell at $t = 10$. The conserved variables are scaled by the lapse $\alpha$. \label{fig:bondi_solution}}
\end{figure}

\begin{figure}
  \centering
  \includegraphics[width=0.45\textwidth]{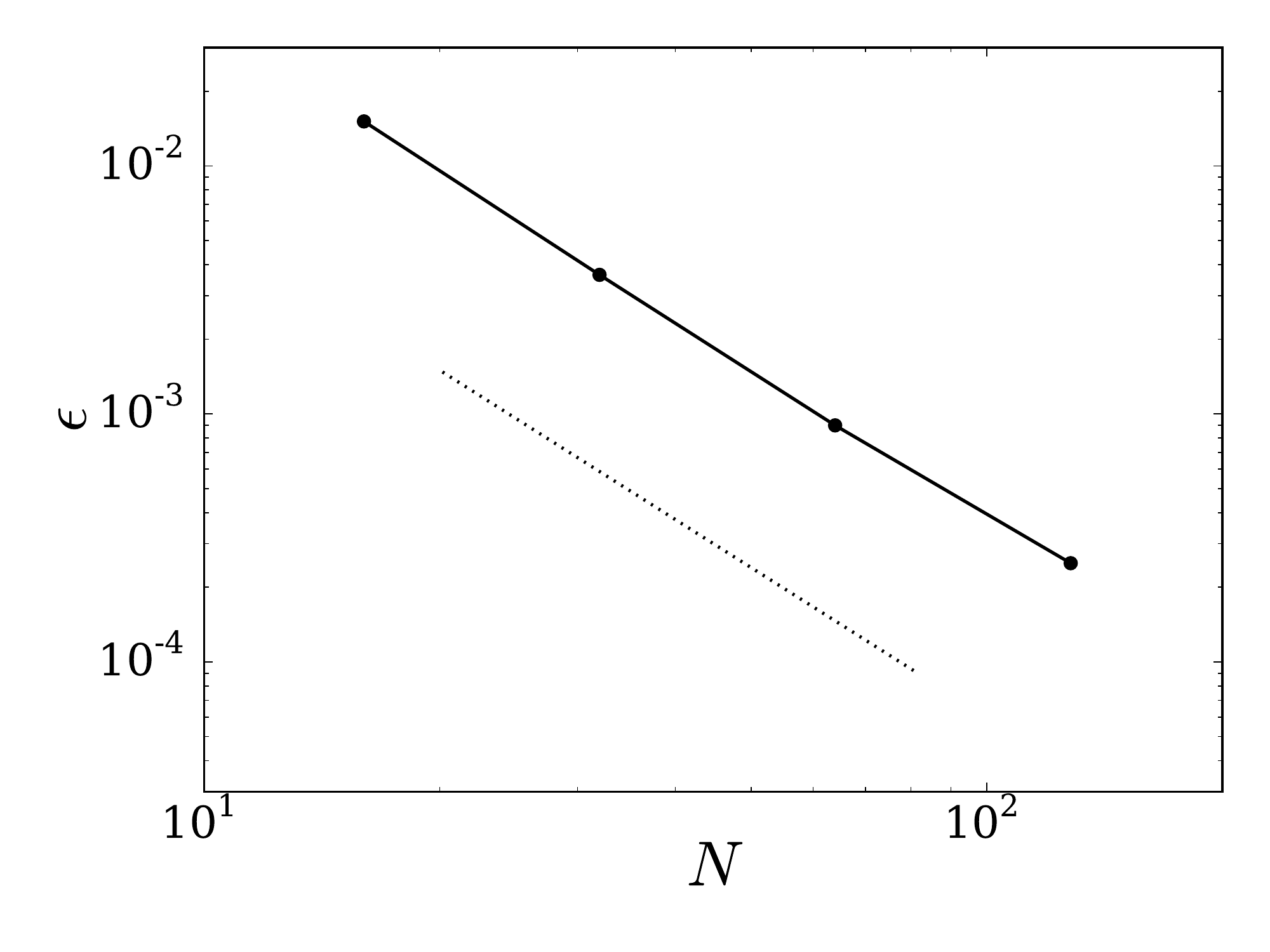}
  \includegraphics[width=0.45\textwidth]{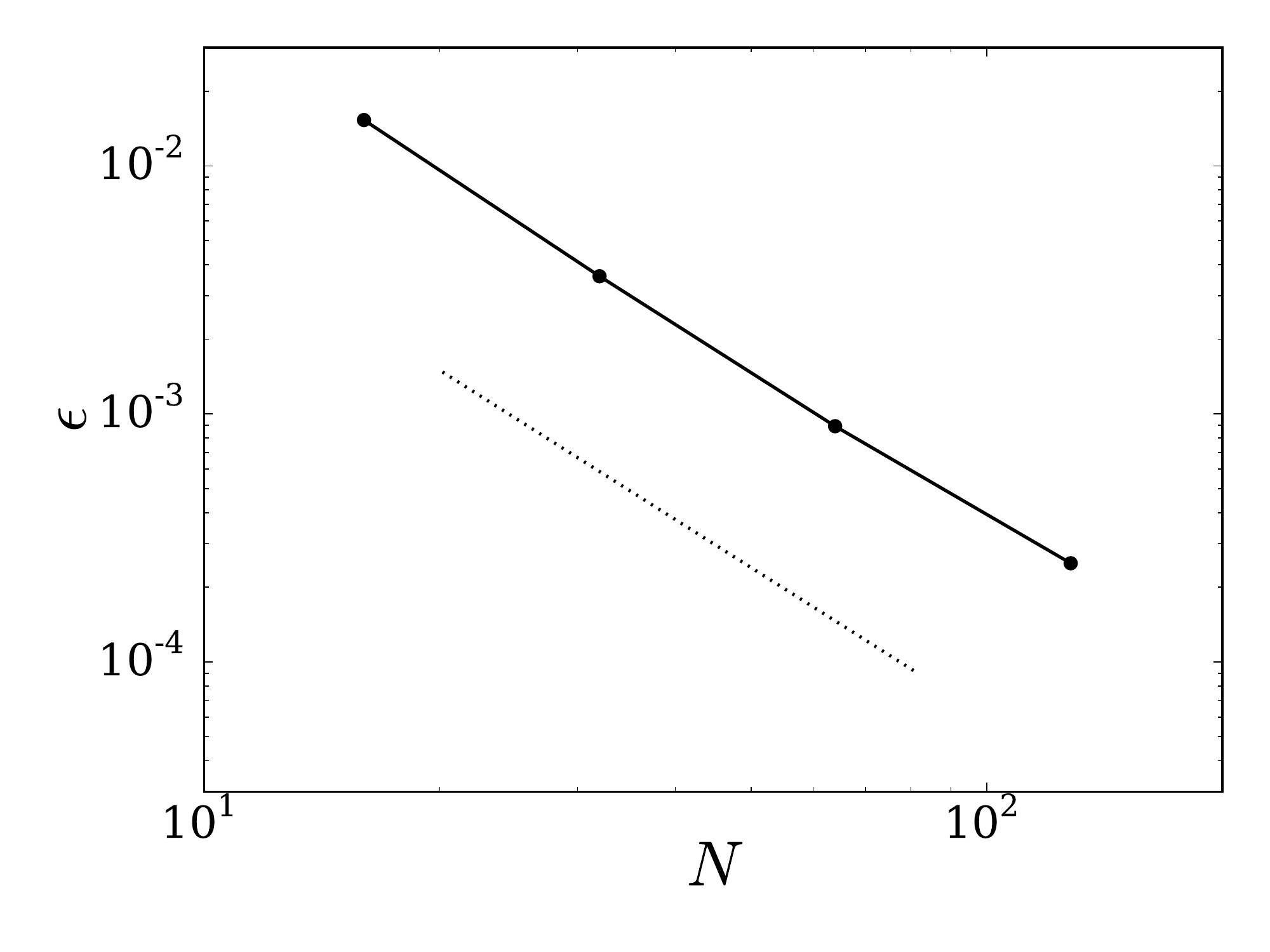}
  \caption{Errors in gas pressure for hydrodynamical (left) and MHD (right) Bondi flow on various $N \times N$ grids. In both cases we achieve second-order convergence, as indicated by the dotted lines. \label{fig:bondi_error}}
\end{figure}

The steady-state solution is shown in Figure~\ref{fig:bondi_solution}. The convergence is second order for both hydrodynamics using HLLC and MHD using HLLD, as shown in Figure~\ref{fig:bondi_error}. Since this is a stationary problem, convergence is not affected by the order of the integrator. However, it is limited by the order of reconstruction, and so these tests show we are correctly getting the full second-order convergence in space expected, even in a curved spacetime.

\section{Magnetized Equatorial Inflow}
\label{sec:test:inflow}

For a final test problem, we simulate the plunging region of a magnetized thin disk around a spinning black hole, as described in \citet{Gammie1999}. We choose the same values for the free parameters as in \citet{Gammie2003}:\ $M = 1$, $a/M = 0.5$, $F_M = -1$, and $F_{\theta\phi} = 0.5$, with the flow matching onto the innermost stable circular orbit. The simulation covers the range $2 < r < 4$ (where the horizon is at $r = 1.86603$ and the innermost stable circular orbit is at $r = 4.23300$). We use the same Kerr--Schild coordinates $\{t, r, \theta, \phi\}$ as \citeauthor{Gammie2003}, defined in \S\ref{sec:examples:kerr:kerr-schild}. Note that Kerr--Schild $r$ and $\theta$ are the same as Boyer--Lindquist $r$ and $\theta$.

\begin{figure}
  \centering
  \includegraphics[width=6in]{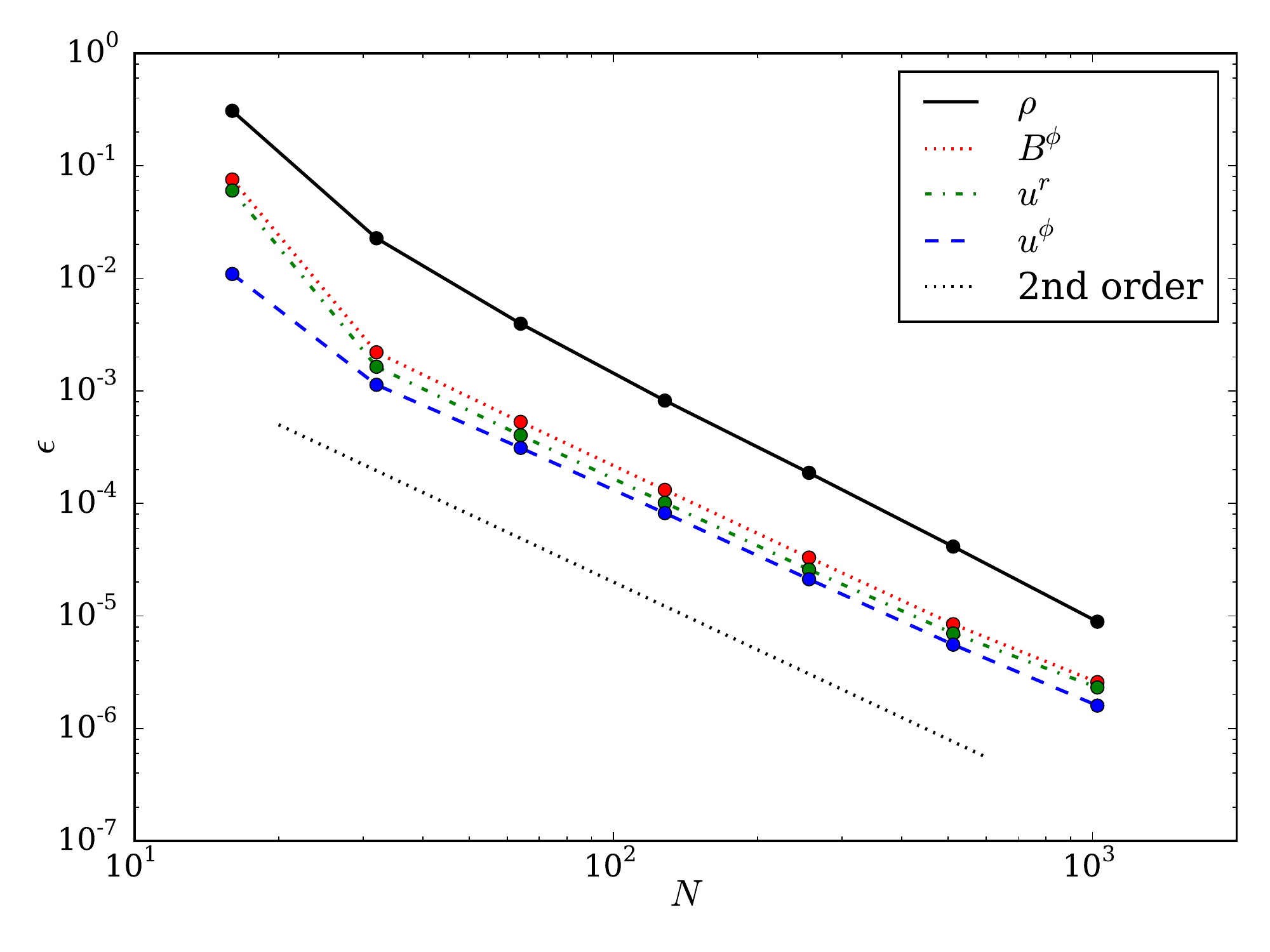}
  \caption{Errors in density, azimuthal magnetic field, and radial and azimuthal velocities as functions of the number of radial grid cells in the magnetized inflow test. As expected, all quantities converge at second order. \label{fig:magnetized_inflow}}
\end{figure}

The flow is initialized to the steady-state solution, with gas pressure set to $10^{-10}$ times the density, and is run to time $t = 15$. A plot of the $L^1$ error in four quantities is shown in Figure~\ref{fig:magnetized_inflow}. We find second-order convergence, demonstrating that the code is solving the correct equations even in a highly nontrivial spacetime. In particular, this test is performed with a frame-transforming HLLE Riemann solver, showing that the frame transformations work even when there is a large shift between the frames.

\section{Torus Simulations}
\label{sec:test:torus}

As a full demonstration of the code in a science application setting, we simulate a fluid torus around a spinning black hole. As proven in \citet{Fishbone1976}, there are steady-state solutions for hydrodynamical tori of an isentropic fluid with constant angular momentum orbiting in Kerr spacetime. Given a black hole mass and spin and a fluid equation of state, the solutions are defined by choosing the radius of the inner edge of the torus, $\redge$, and either the radius of the pressure maximum, $\rpeak$, or the angular momentum per unit mass, $\ell = u^t u_\phi$ (in Boyer--Lindquist coordinates).

A general-relativistic code should be able to maintain such a torus, balancing gravity, centrifugal force, and pressure gradients. We take a black hole with mass $M = 1$ and spin $a/M = 0.95$ and initialize a $\Gamma = 13/9$ fluid torus with $\redge = 3.7$ and $\ell = 3.85$ (so $\rpeak = 7.82$). The torus is evolved on a 2D grid with $N_r$ cells in the radial direction and $N_r/2$ cells in the polar direction using the aforementioned Kerr--Schild coordinates (see \S\ref{sec:examples:kerr:kerr-schild}). Our grid extends radially from $0.98$ times the outer horizon radius $M + \sqrt{M^2-a^2}$ to $r = 20$, with geometric grid spacing such that each cell is $1.025$ times as wide as its inner neighbor. We limit the polar angle to $\pi/4 < \theta < 3\pi/4$, excising parts of the necessarily non-stationary atmosphere that should not impact the torus proper. This simulation employs uniform spacing in $\theta$. We impose outflow conditions on the inner and outer boundaries and reflecting boundary conditions on the constant-$\theta$ boundaries.

\begin{figure}
  \centering
  \includegraphics[width=0.45\textwidth]{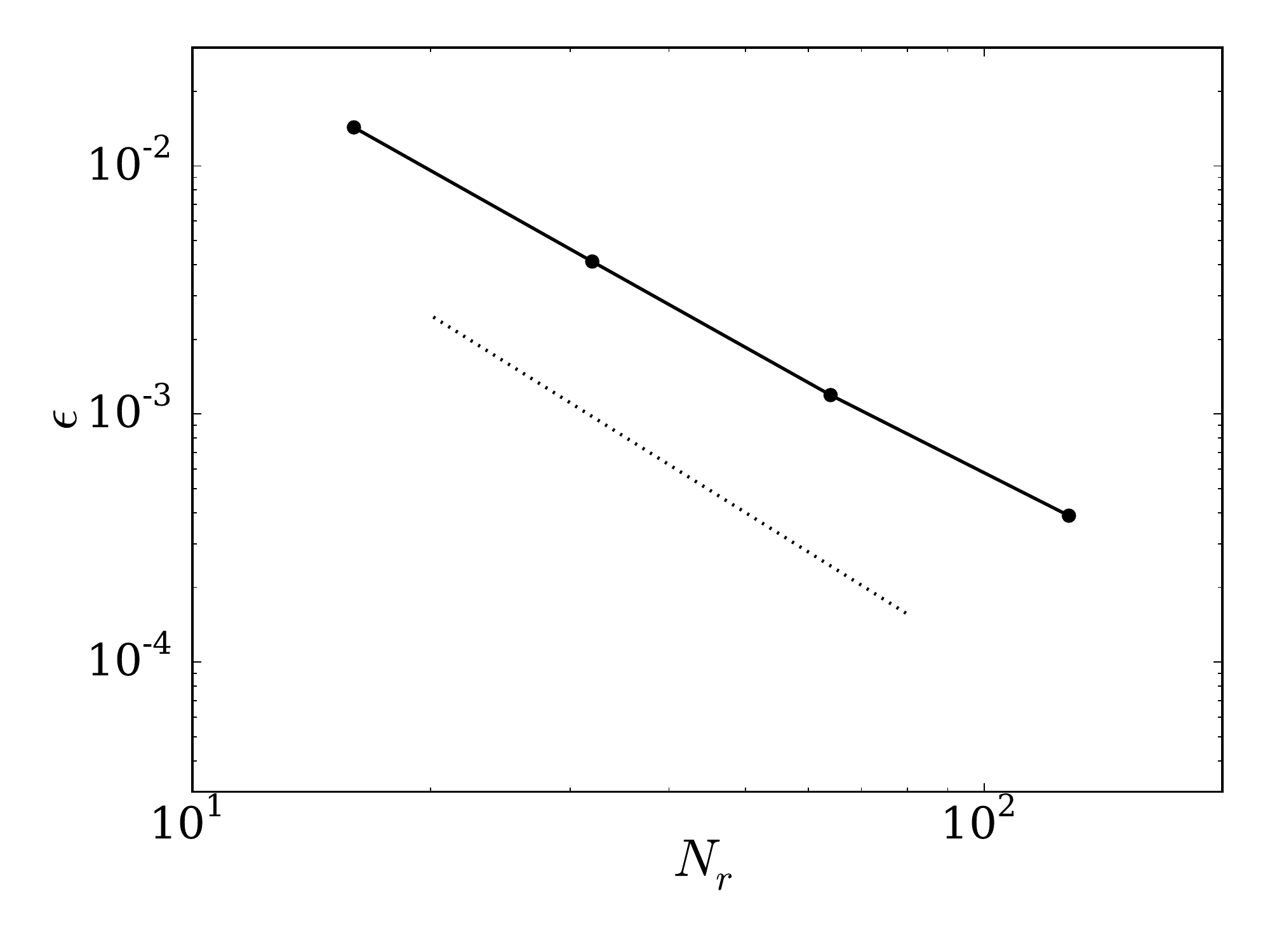}
  \caption{Errors in density for 2D Fishbone--Moncrief hydrodynamical tori on various $N_r \times N_r/2$ grids. Errors are defined by \eqref{eq:torus_error}. The dotted line indicates second-order convergence. \label{fig:torus_hydro_convergence}}
\end{figure}

The simulation is evolved using the HLLC solver with a CFL number of $0.2$ until a time $t = 1$, at which point the error accrued in $\rho$ is calculated. Define the region $T$ to be those cells for which $\rho$ is initially at least $0.02$ its peak value. We calculate the error $\epsilon$ over the region $T$ (thus avoiding edge effects) in the $L^1$ sense:
\begin{equation} \label{eq:torus_error}
  \epsilon = \frac{\displaystyle\int_T\abs{\rho_\mathrm{fin}-\rho_\mathrm{init}}\sqrt{-g}\,\dr\,\dth}{\displaystyle\int_T\abs{\rho_\mathrm{init}}\sqrt{-g}\,\dr\,\dth}.
\end{equation}
The errors are shown in Figure~\ref{fig:torus_hydro_convergence}, where one can see that they are second order. This is the same test problem as done in \citet[cf.\ their Figure~15]{Gammie2003} with the exception that we use a slightly different grid with no equatorial compression of the constant-$\theta$ surfaces.

Our torus is indeed able to maintain equilibrium, even over longer times. To demonstrate this we continue the $64\times32$ simulation to $t = 430$, corresponding to three orbital periods at the pressure maximum, using both HLLE and HLLC solvers. With the HLLE solver the error is $\epsilon = 0.030$, while it is $\epsilon = 0.021$ with HLLC.

We next consider a 3D Fishbone--Moncrief torus with an added magnetic field, upsetting equilibrium via the magnetorotational instability (MRI). Here we choose $M = 1$, $a/M = 0.9375$, $\Gamma = 13/9$, $\redge = 6$, and $\rpeak = 12$ (so $\ell = 4.28$). The grid is expanded to $0.98 (M + \sqrt{M^2-a^2}) < r < 40$ and $0 < \theta < \pi$ (with no artificial boundary at the poles), and it covers the entire azimuthal range. In one case we use the equivalent of $144\times128\times96$ cells in the radial, polar, and azimuthal directions, with geometric radial spacing (ratio $1.02397$) and uniform spacing in the angles. Static mesh refinement is used to de-refine the polar regions outside the torus by a factor of $4$ in each dimension, thus preventing the timestep from being limited by those regions. Explicitly, there are $80$ cells equally spaced in the range $3\pi/16 < \theta < 13\pi/16$. For a higher resolution grid, we further double the number of cells in each dimension in the region $7\pi/32 < \theta < 25\pi/32$. This forms an effective $288\times256\times192$ grid.

To this torus we add a magnetic field with purely azimuthal vector potential $A^\phi \propto \max(\rho-0.2, 0)$, calculated from a density normalized to unity at its maximum. The magnetic field is normalized such that the ratio of maximum gas pressure to maximum magnetic pressure (necessarily located at different points) is $100$. These parameters are chosen to roughly match those of \citet{Shiokawa2012} except in not having an equatorially compressed grid like theirs. In particular, our low and high resolution runs have $\Delta\theta = 0.0245$ and $\Delta\theta = 0.0123$ in their respective most refined regions, while the $96\times96\times64$ and $144\times144\times96$ grids from \citeauthor{Shiokawa2012}\ have $\Delta\theta = 0.00983$ and $\Delta\theta = 0.0065$ at the midplane.

\begin{figure}
  \centering
  \includegraphics[width=\textwidth]{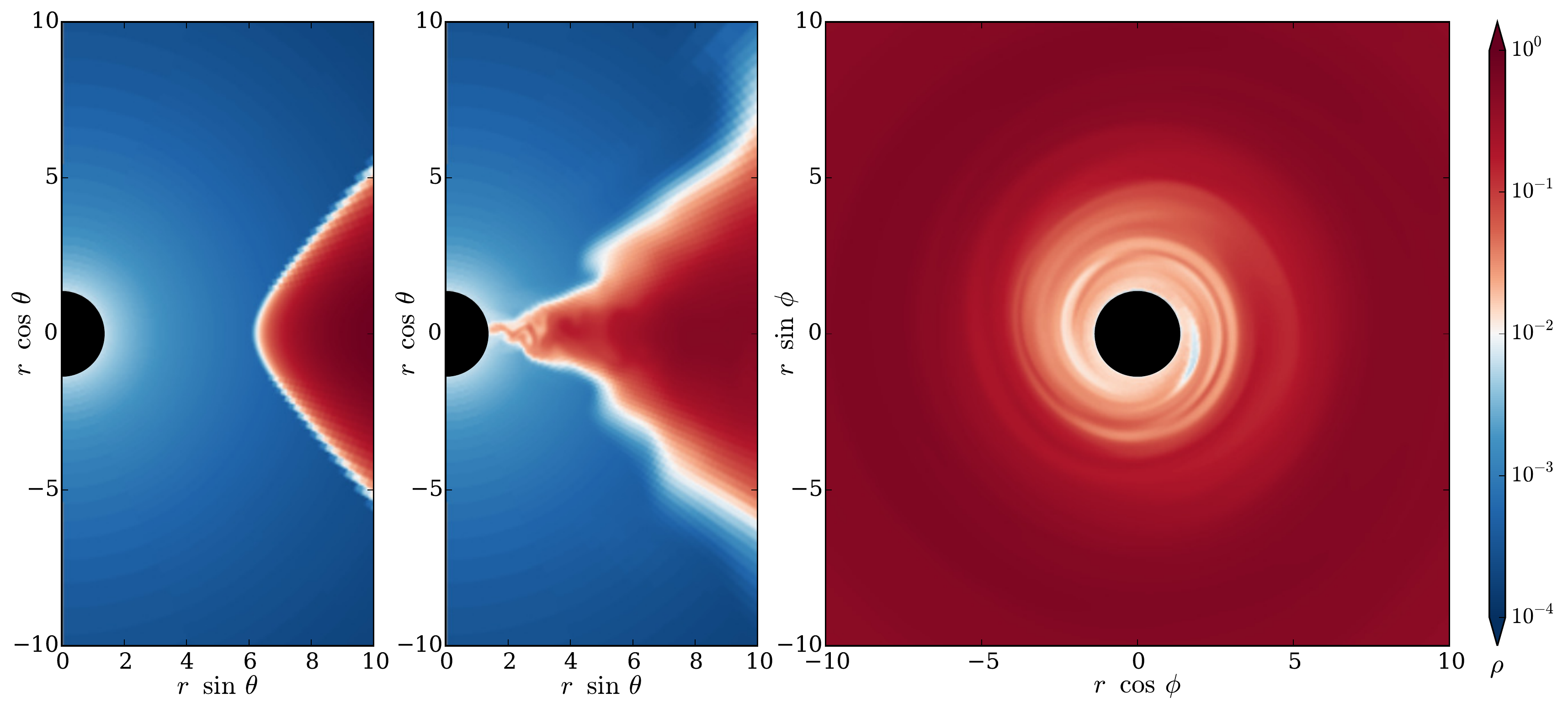}
  \caption{Density $\rho$ of a 3D Fishbone--Moncrief MHD torus with effective resolution $288\times256\times192$. Shown are the initial conditions in the $r\theta$-plane (left), the final state at $t = 12{,}000$ in the same plane (center), and the final state in the $r\phi$-plane (right). \label{fig:torus_slices}}
\end{figure}

We evolve the system using an LLF Riemann solver just as is used in Harm \citep{Gammie2003}. The simulation is run until a time $t = 12{,}000$. For comparison, a geodesic circular orbit at the pressure maximum has a period of $t = 267$. After several orbits the MRI sets in, causing turbulence that eventually disrupts the entire torus. The initial and final densities for the run are illustrated in Figure~\ref{fig:torus_slices}.

\begin{figure}
  \centering
  \includegraphics[width=\textwidth]{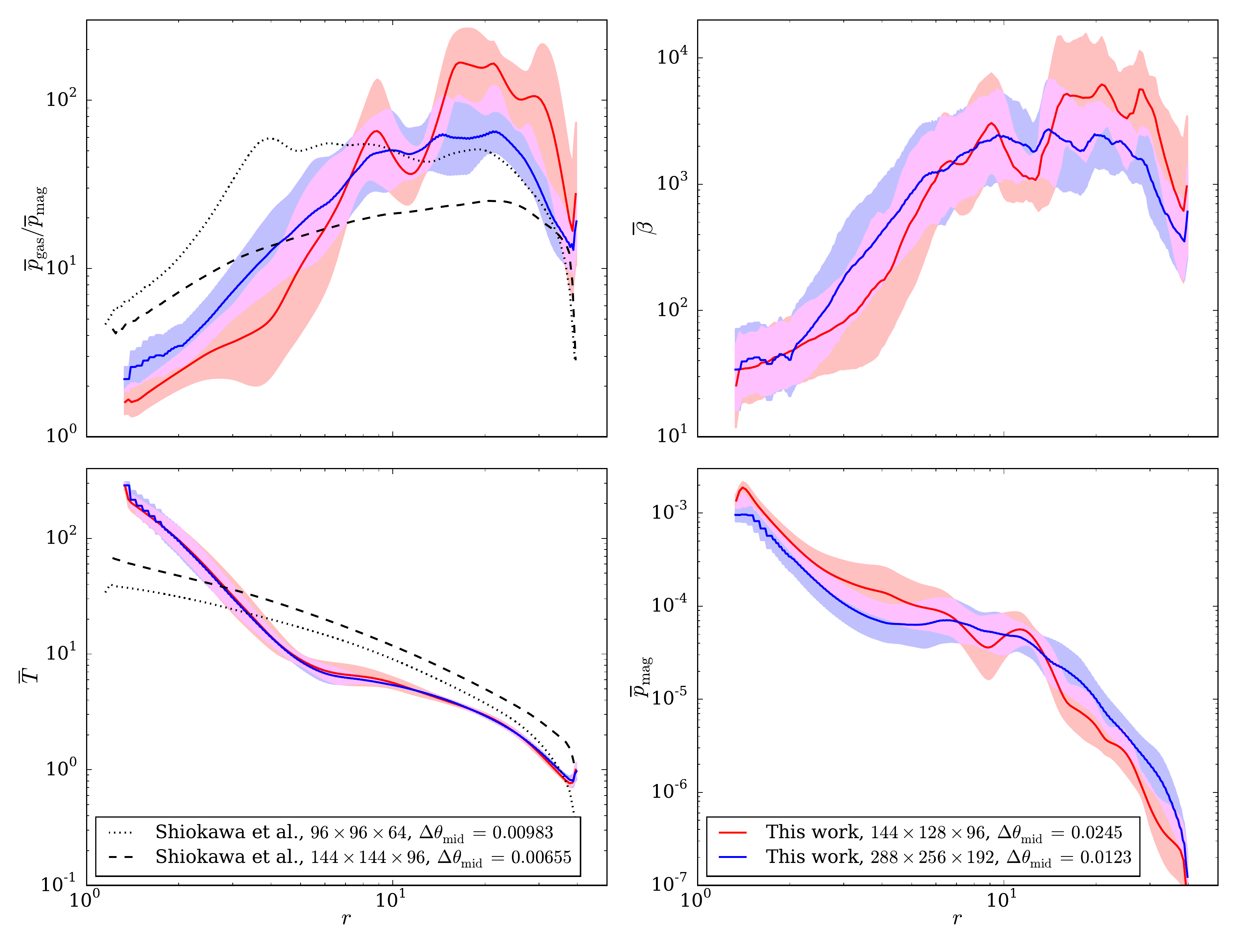}
  \caption{Spherically averaged, time-averaged profiles of various quantities in the 3D torus. These are the gas-to-magnetic pressure ratio (spherical averages taken before ratio, time averages take after; top left), plasma $\beta$ (top right), electron temperature (bottom left), and magnetic pressure (bottom right). The left two panels show the same quantities as plotted in Figure~2 of \citet{Shiokawa2012}, corresponding to their ``soft'' polar boundary condition. In all cases shaded regions indicate $1\sigma$ standard deviations from $800$ samples at different times. \label{fig:torus_profile}}
\end{figure}

For a more quantitative analysis, we construct radial profiles of physical quantities. Following \citet[cf.\ their \extref{9} and \extref{10}]{Shiokawa2012}, we define a spherical shell average of a quantity $q$ at a Kerr--Schild time $t$ according to
\begin{equation}
  \bar{q} = \frac{\displaystyle\int_0^{2\pi}\!\!\int_{\theta_\mathrm{min}}^{\theta_\mathrm{max}}q\rho\sqrt{-g}\,\dth\,\dph}{\displaystyle\int_0^{2\pi}\!\!\int_{\theta_\mathrm{min}}^{\theta_\mathrm{max}}\rho\sqrt{-g}\,\dth\,\dph},
\end{equation}
where in our case we take $(\theta_\mathrm{min},\theta_\mathrm{max}) = (\pi/4,3\pi/4)$ in order to exclude any effects of the atmosphere. We then average $\bar{q}$ over the time range $4000$ to $12{,}000$, noting the mean and standard deviation of values at each radius. This is done for plasma $\beta$, electron temperature $T = \mprot\pgas/2\melec\rho$ (called $\theta_\mathrm{e}$ by \citeauthor{Shiokawa2012}), and magnetic pressure $\pmag = b^\lambda b_\lambda/2$ in Figure~\ref{fig:torus_profile}. For a more direct comparison to \citeauthor{Shiokawa2012}\ we also show the result of time-averaging $\bar{p}_\mathrm{gas}/\bar{p}_\mathrm{mag}$, the result of which is simply called $\beta$ in that work.

The electron temperature roughly matches what \citeauthor{Shiokawa2012}\ find, both in magnitude and in trend with radius, though they do not have the same kink at around $r = 5\text{--}6$ and our disk gets notably hotter inside the innermost stable circular orbit at $r = 2.04420$. We get the same shape as \citeauthor{Shiokawa2012}\ for $\pgas/\pmag$, especially their low resolution ``soft'' polar run (cf.\ the upper left panel of their Figure~2; data reproduced in Figure~\ref{fig:torus_profile}). They find $\pgas/\pmag$ to be less than about $60$ in all cases, whereas our simulations get up to about $200$ in the outer regions. Our ratio also approaches unity in the innermost regions, while theirs never drops below $4$. The same shape is found when we perform spherical averages of $\beta$ itself, rather than $\pgas$ and $\pmag$ separately, as shown in the upper right of Figure~\ref{fig:torus_profile}. While there are some differences, we also note that as we increase resolution our $\pgas/\pmag$ profile approaches theirs, and we expect even better agreement if we were to have midplane $\Delta\theta$ values as small as theirs.

\Citeauthor{Shiokawa2012}\ also examine azimuthal correlation lengths in the equatorial plane. We use the same definitions as they do (their \extref{13} through \extref{17}), which we summarize here. For a vector quantity with components $q^\mu$ define the time-dependent correlation function
\begin{equation}
  R(r,\phi,t) = \frac{1}{r\Delta r\Delta\theta} \int_{\theta_\mathrm{min}}^{\theta_\mathrm{max}} \int_{r_\mathrm{min}}^{r_\mathrm{max}} \int_0^{2\pi} \delta q^\mu(r',\theta',\phi_0,t) \delta q_\mu(r',\theta',\phi_0+\phi,t) \, \dph \, r' \, \dr'\,\dth'.
\end{equation}
The integrals run over a region of radial width $\Delta r = r_\mathrm{max} - r_\mathrm{min}$ of a single cell and polar width $\Delta \theta = \theta_\mathrm{max} - \theta_\mathrm{min}$ of two cells bordering the equatorial plane. The quantities $\delta q^\mu$ and $\delta q_\mu$ are deviations from averages of $q^\mu$ and $q_\mu$ over the domains of integration. For scalar quantities we simply use the integrand $\delta q(r',\theta',\phi_0,t) \delta q(r',\theta',\phi_0+\phi,t)$ instead. Define the time-averaged correlation function as
\begin{equation}
  \bar{R}(r,\phi) = \int_{t_\mathrm{min}}^{t_\mathrm{max}} \frac{R(r,\phi,t)}{R(r,0,t)} \, \dt,
\end{equation}
where we average from $4000$ to $12{,}000$. Then the correlation length $\lambda$ at a given radius $r$ is the value for which $\bar{R}(r,\lambda) = \bar{R}(r,0)/\ee$.

\begin{figure}
  \centering
  \includegraphics[width=\textwidth]{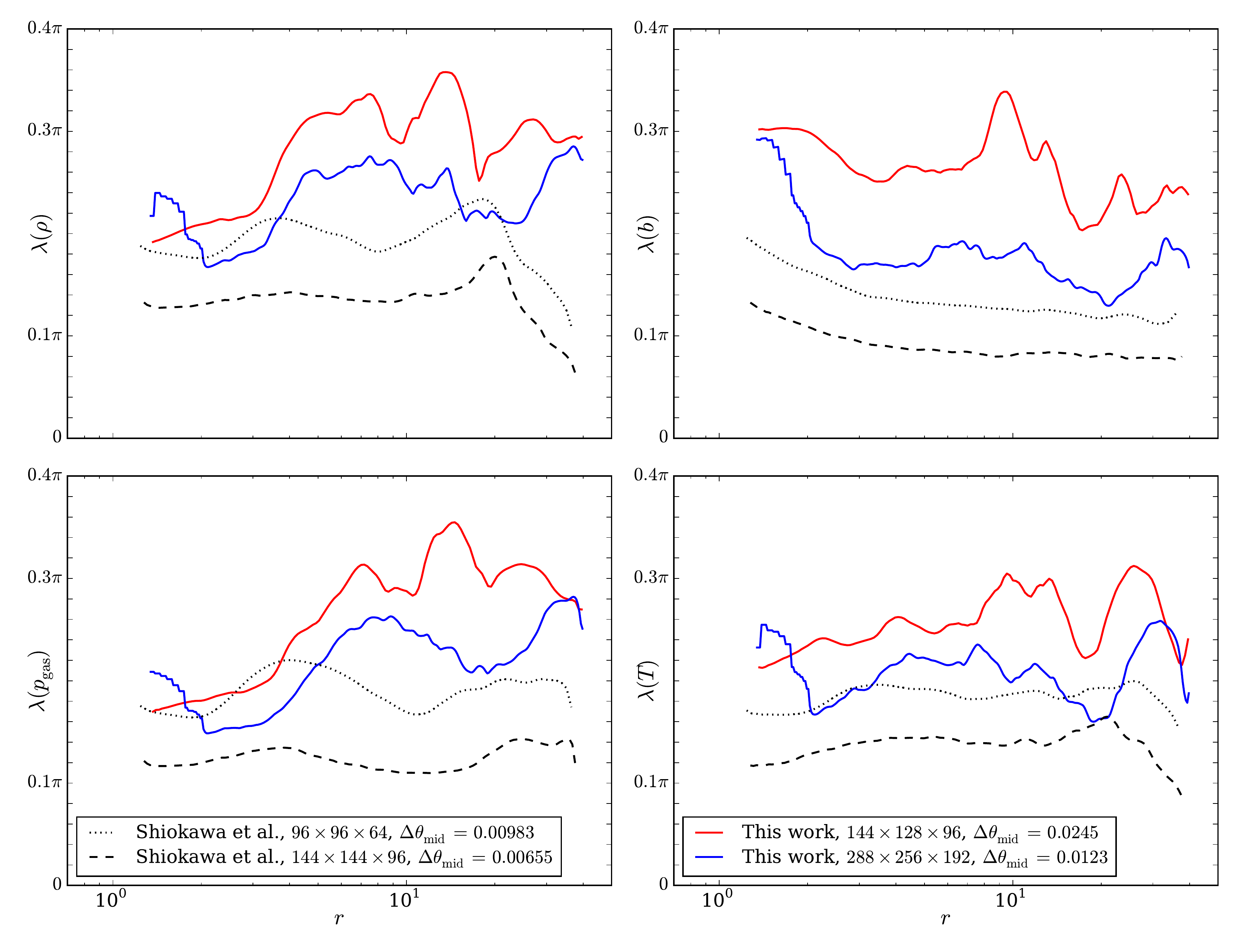}
  \caption{Azimuthal correlation lengths as functions of radius in the 3D torus runs. These are for density (top left), magnetic field (top right), gas pressure (equivalently internal energy; bottom left), and electron temperature (bottom right). The black lines show two select datasets from \citet{Shiokawa2012}, both using the ``soft'' polar boundary condition. \label{fig:torus_correlation}}
\end{figure}

We plot the runs of four correlation lengths in Figure~\ref{fig:torus_correlation}, to be compared to Figure~4 of \citet{Shiokawa2012}, from which select data has been replotted here. Compared to \citeauthor{Shiokawa2012}\ our disks have larger correlation lengths, especially in the outer regions. This is a sign that we have not yet reached the fully saturated, turbulent state in that region. It is worth repeating, however, that our midplane resolution is not as high as in \citeauthor{Shiokawa2012}, and moreover our correlation lengths decrease as we increase resolution.

\chapter[Application to MAD]{Application to Magnetically Arrested Disks}
\label{chap:application}

In selecting a first scientific application of Athena++, two considerations come to the forefront:\ an existing body of literature and room for improvement.

If we were to apply a new code to a situation previously lacking in simulations, results would be subject to a degree of doubt as to whether they are physical or peculiarities of the implementation at hand. Such forays into the unknown of course should be done, but we prefer to save them for later applications (see \S\ref{sec:conclusion:science}). As a result, we refrain from situations where the initial conditions and other parameters are largely unconstrained. Instead, we prefer a setting in which some numerical work has been done.

At the same time, merely reproducing old results, though arguably necessary for parts of the scientific method, is of little interest. It is important that we be able to say more about the situation afterward, whatever the result, than could have been said before our investigation.

We choose therefore to reexamine the situation of magnetically arrested disks (MAD). In particular, the question arises as to whether MAD states and their corresponding spin-powered jets survive in high-resolution simulations, where more instabilities are naturally present.

\section{The Importance and Physics of MAD}
\label{sec:application:mad_importance}

Powerful relativistic jets have long been observed by astronomers. While such jets need not be associated with black holes -- for example, pulsars produce jets -- often there is no other source that can explain a particular jet. The connection between black holes -- whose existence, recall, is only indirectly inferred, though with very strong evidence -- and well-observed jets was solidified by \citet{Blandford1977}, where a mechanism was outlined by which accretion could drive energy extraction from a spinning black hole along its polar axis.

In the Blandford--Znajek (BZ) mechanism, the magnetosphere of a spinning black hole can lend itself to extracting energy from the spin of the black hole itself. Under the assumption that pair production will lead to the presence of a diffuse plasma even if there was no matter originally, they approximate this magnetosphere in the force-free limit. This is what is modeled in our ideal MHD simulations in the low-$\beta$ limit. The spin torques the field in such a way as to produce a net Poynting flux in a method analogous to that of \citet{Goldreich1969} around pulsars. It was noted in \citet{Penrose1969} that energy can be extracted to infinity at the cost of reducing the spin of the black hole using mechanisms that rely on the fact that a particle's energy as seen at infinity changes sign as it crosses into the ergosphere. Indeed the BZ mechanism can be cast in terms of energy-momentum transfer via the electromagnetic field between the ergosphere and distant regions. The key then for the BZ effect to occur is to have a spinning black hole with a low-$\beta$ plasma hosting magnetic field ordered on large scales as was assumed in their model.

Under the right circumstances, one expects to be able to extract more energy from a black hole than is lost to it by infall, corresponding to an efficiency of greater than $100\%$. \Citeauthor{Tchekhovskoy2011}\ claim to see such efficiencies, attributing them to the simulations being in a magnetically arrested disk (MAD). As described in \citet{Narayan2003} and based on the ideas in \citet{Bisnovatyi-Kogan1974}, MAD occurs when enough vertical magnetic flux has accumulated so as to prevent or at least restrict further mass accretion. While a weak field threading a disk will destabilize it to the MRI, a strong enough field will disrupt this instability and thus reduce the effective viscosity. The overall effect is to limit the inflow of matter to a low rate and low velocity.

We know some black holes are launching jets. If MAD is to be part of the explanation for such observations, it must meet two conditions:
\begin{enumerate}
  \item The conditions for the effect -- the initial conditions of the flow, the feeding of the flow, the spin of the black hole, etc.\ -- must be realized often enough in nature. It is conceivable that a mechanism could work but never actually be at play in the systems we observe.
  \item The effect must be generic enough to occur given the right conditions. That is, we should be able to consistently see MAD states in the appropriate simulations, and we should see some of them producing jets as a result of their MAD states. Moreover, we should be confident that what we see in simulations is an accurate portrayal of reality.
\end{enumerate}
We leave the first consideration for a more observationally oriented study. The second, though, is where there is room for improvement upon what is in the computational literature.

\section{The Current State of Simulations}
\label{sec:application:current_state}

As already discussed, \citet{Tchekhovskoy2011} pioneered the study of MAD around black holes as a driving force behind jet launching. Their fiducial model A0.99f is of a black hole with dimensionless spin $0.99$. The size of the simulation is $288 \times 128 \times 64$ in $(r,\theta,\phi)$. The radial coordinate is roughly logarithmically spaced, extending from approximately $1$ (the outer horizon is at $r \approx 1.14$) to $10^5$, with most of the cells inside $10^3$. (All distances here are in terms of the natural scale that is the mass $M$ of the black hole.) The polar coordinate is not uniform but rather concentrated both at the poles and at the equator; it is also deformed to produce a more cylindrical grid near the poles. This results in a cell aspect ratio of $\Delta r : r\Delta\theta : r\Delta\phi \approx 2 : 1 : 7$, with approximately $96$ cells per decade in radius. It should be noted that they have a model (A0.99fh) that doubles the number of cells in the azimuthal direction, though this is not run from the initial conditions but rather starting with a turbulent state produced by the fiducial model.

The initial conditions they choose are those of an equilibrium torus as defined in \citet{Chakrabarti1985}, where the quantity $-u_\phi/u_t$ (Boyer--Lindquist coordinates) is taken to be constant. The inner edge is at $r = 15$, and the pressure maximum is at $r = 34$. For comparison, the innermost stable circular orbit is at $r \approx 1.45$. This equilibrium only holds in pure hydrodynamics, however the torus is seeded with a weak poloidal magnetic field. As described in \citet{Tchekhovskoy2011}, this field is configured so as to make the parameter $\beta = \pgas/\pmag$ reach a minimum of $100$, with a large portion of the inner part of the torus close to this minimum.

The key to the field configuration for the purposes of MAD is that the loops circulate about $r \approx 300$ rather than the pressure maximum. Often in magnetized torus simulations the field is initialized from a purely azimuthal vector potential whose magnitude is proportional to (some power of) density, causing the field lines to match isocontours of density and pressure. However this quickly leads to accretion of oppositely directed field and thus avoids the accumulation of ordered flux that defines MAD.

\section{MAD Simulations in Athena++}
\label{sec:application:simulations}

We now describe the exact details of how we set up our MAD simulations.

\subsection{Grid Parameters}
\label{sec:application:simulations:grid}

We use spherical Kerr--Schild coordinates (see \S\ref{sec:examples:kerr:kerr-schild}) for a black hole of mass $M = 1$ and spin $a = 0.99$. For these values the outer event horizon is at radius $\rhor = M + \sqrt{M^2-a^2} \approx 1.14107$ and the innermost stable circular orbit is at radius $\risco \approx 1.45450$.

The root grid covers $0.98\rhor < r < 1000$ with $96$ cells spaced as close as possible to logarithmically. That is, we have a constant geometric ratio $\Delta r_i/\Delta r_{i-1} \approx 1.07336$ in adjacent cell widths, where this value is chosen such that the ratio of cell width to cell center (taken to be the arithmetic mean of the inner and outer radii) is the same for the innermost and outermost cells. The root grid also covers both angular coordinates in their entirety, $0 < \theta < \pi$ and $0 < \phi < 2\pi$, using $32$ cells for each.

\begin{figure}
  \centering
  \includegraphics[width=\textwidth]{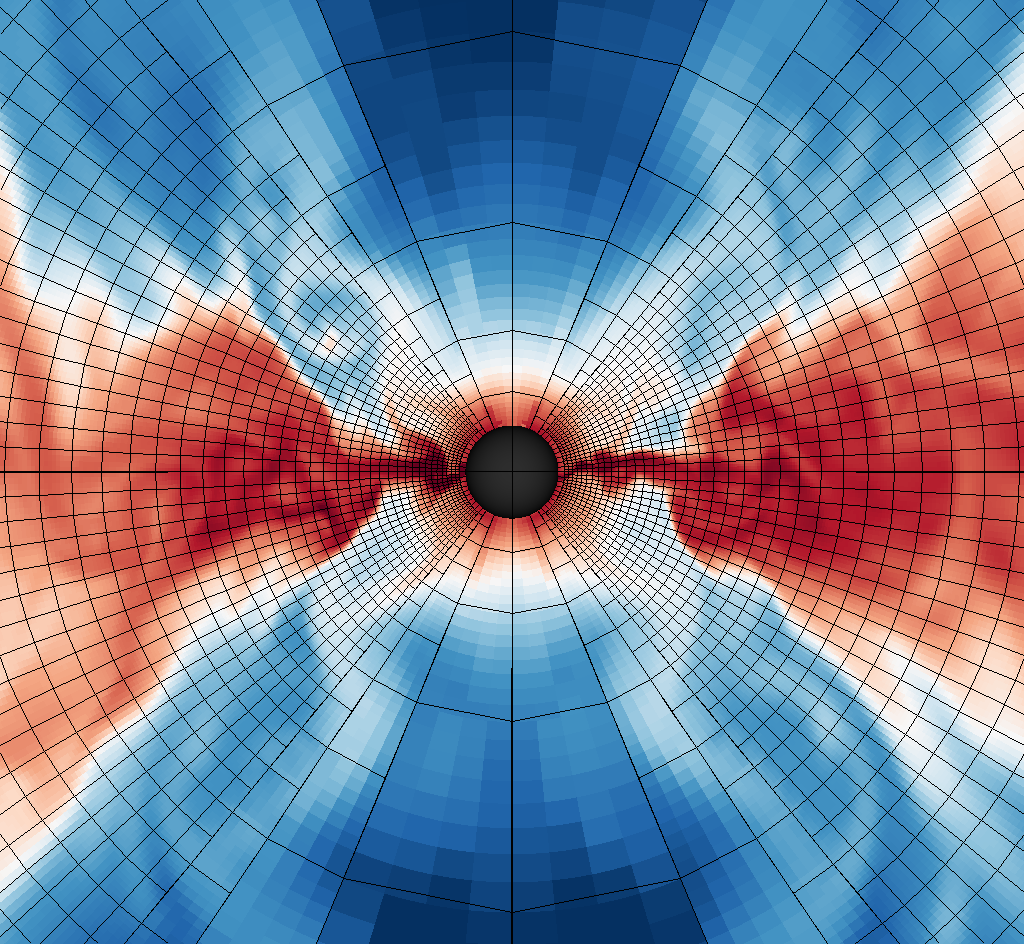}
  \caption{Grid used for MAD simulations, up to level $3$ refinement. In this poloidal slice of the inner $11\ M$, each boxed region corresponds to $8$ cells in $r$ and $4$ in $\theta$. The background is the density $\rho$ at time $5000\ M$, with the color scale running from $10^{-3}$ to $10^1$. \label{fig:mad_grid}}
\end{figure}

We refer to this root grid as level $0$. Level $1$ mesh refinement involves dividing all cells in the range $\pi/8 < \theta < 7\pi/8$ into $8$ new cells, doubling the resolution in each dimension. Note that the radial geometric ratio becomes the square root of its level $0$ value here. Similarly, level $2$ refinement is used in the region $3\pi/16 < \theta < 13\pi/16$, the resolution again doubling in each dimension. Finally, we consider level $3$ refinement to take place in the region $7\pi/32 < \theta < 25\pi/32$, though here we leave the innermost $16$ cells in radius at level $2$ unrefined. That is, our level $3$ grid only extends inward to $r \approx 1.48428 \approx 1.02047\ \risco$. Figure~\ref{fig:mad_grid} illustrates the statically refined grid. With this scheme the timestep at any refinement level is limited by $W_\phi$ (see \eqref{eq:width}) at the poles (at the innermost radii). A result of our refinement scheme and of the code's scaling (\S\ref{sec:intro:new_code:speed}) is that we can increase resolution at the cost of only more computational cores but not wall time.

The outer radial boundary is held fixed, while a strict outflow condition is imposed at the inner boundary (outflow from the grid, which is inflow into the black hole). The grid is of course periodic in $\phi$, and we stitch the grid together along the poles as well (\S\ref{sec:intro:new_code:poles}).

\subsection{Torus Parameters}
\label{sec:application:simulations:torus}

On this grid we initialize an equilibrium hydrodynamical torus in the style of \citet{Fishbone1976}, which takes $\ell = u_\phi u^t$ (Boyer--Lindquist coordinates) to be constant. Such tori are parameterized by an inner edge radius $\redge$ and either a density peak radius $\rpeak$ or a specific angular momentum $\ell$. We choose $\redge = 16.45$ and $\rpeak = 34$, resulting in $l \approx 6.29943$ \citep[\extref{3.8}]{Fishbone1976}. The orbital period at the pressure maximum is $P \approx 1251.88$ \citep[\extref{3.11}]{Fishbone1976}.

\begin{figure}
  \centering
  \includegraphics[width=\textwidth]{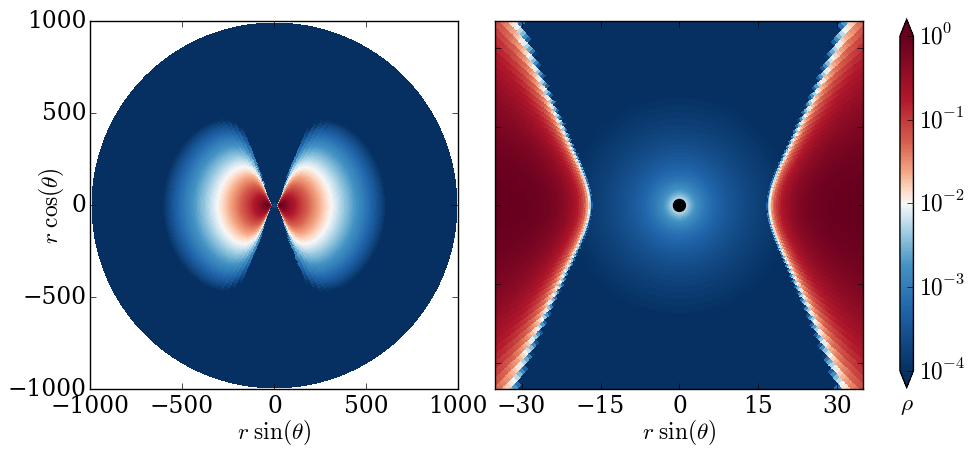}
  \caption{Initial density of MAD torus, shown at level $2$ refinement and two zoom levels. \label{fig:mad_slices_initial_rho}}
\end{figure}

\begin{figure}
  \centering
  \includegraphics[width=\textwidth]{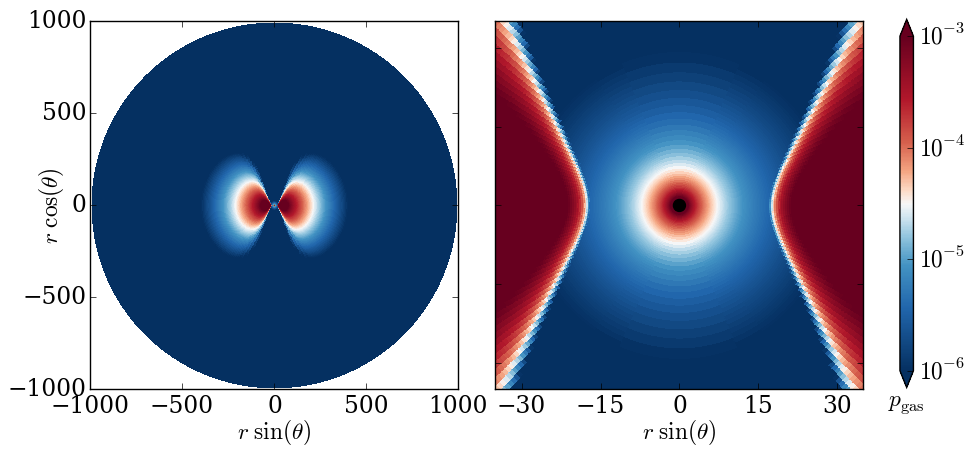}
  \caption{Initial gas pressure of MAD torus, shown at level $2$ refinement and two zoom levels. \label{fig:mad_slices_initial_pgas}}
\end{figure}

\begin{figure}
  \centering
  \includegraphics[width=\textwidth]{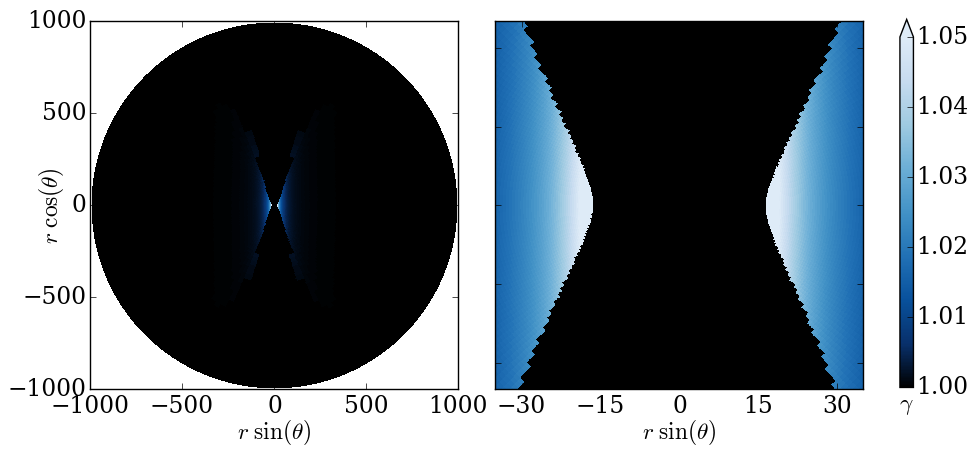}
  \caption{Initial normal observer Lorentz factor of MAD torus, shown at level $2$ refinement and two zoom levels. \label{fig:mad_slices_initial_gamma}}
\end{figure}

The equation of state is taken to be $\pgas = \rho^\Gamma$ for $\Gamma = 13/9$, and the initial density is normalized to unity at its maximum. Rather than have vanishing density and pressure outside the torus, we impose radially varying floors of $\rho \ge \max\{10^{-2} r^{-3/2}, 10^{-6}\}$ and $\pgas \ge \max\{10^{-2} (\Gamma-1) r^{-5/2}, 10^{-8}\}$. We also restrict the Lorentz factor of the fluid as seen by the normal observer to be $\gamma \le \gammamax = 10$ throughout the simulation, where cells found to have higher velocities are adjusted according to $\tilde{u}^i \to \sqrt{(\gammamax^2-1)/(\gamma^2-1)} \tilde{u}^i$. The initial density, gas pressure, and Lorentz factor are shown in Figures~\ref{fig:mad_slices_initial_rho}--\ref{fig:mad_slices_initial_gamma}.

In order to investigate MAD, we seed this torus with a weak magnetic field. Following the procedure given in \citet{Tchekhovskoy2011}\ we define an initial toroidal vector potential $\bar{A}_\phi = r^5 \rho^2$ and from this we derive a poloidal magnetic field with components $\bar{B}^i$. Next we rescale this magnetic field locally to fix $\beta = \pgas/\pmag$ to be constant (thus violating the divergence-free constraint). We proceed to define the true vector potential as $A_\phi = \int_{\theta'=0}^{\theta'=\theta} \bar{B}^r \sqrt{-g} \, \dth'$, and from this we derive the true magnetic field components $B^i$ (which obey the divergence-free constraint by construction). Finally, the field is rescaled globally such that $\beta$ reaches a minimum value of $100$ over the region $\rho > 0.05$. This cutoff is used to make the magnetic field strength minimally dictated by jitter in cell boundaries near the edge of the torus. We also perform the above calculations on a grid with $1536 \times 512$ cells in $(r,\theta)$, interpolating $A_\phi$ onto (the corners of) the actual grid just before differencing to find (the face-centered values of) $B^r$ and $B^\theta$. Thus even at different resolutions we have the same magnetic field.

\begin{figure}
  \centering
  \includegraphics[width=\textwidth]{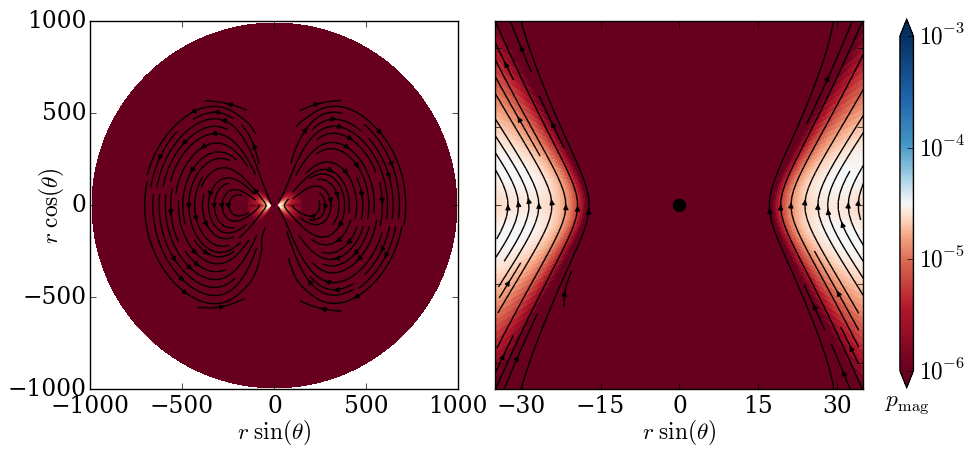}
  \caption{Initial magnetic pressure of MAD torus, shown at level $2$ refinement and two zoom levels. Overlaid are streamlines for the magnetic field. \label{fig:mad_slices_initial_pmag}}
\end{figure}

\begin{figure}
  \centering
  \includegraphics[width=\textwidth]{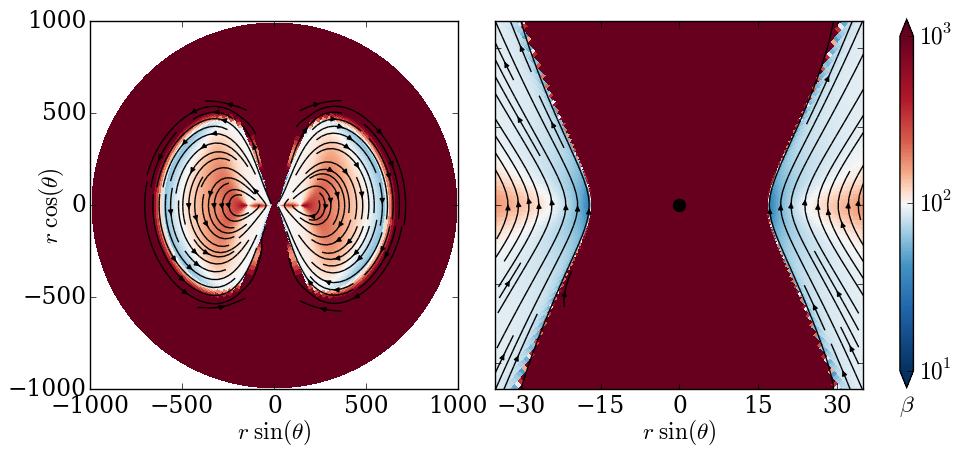}
  \caption{Initial plasma $\beta$ of MAD torus, shown at level $2$ refinement and two zoom levels. Overlaid are streamlines for the magnetic field. \label{fig:mad_slices_initial_beta}}
\end{figure}

The complicated scheme has two effects. First, the scaling of $\bar{A}_\phi$ with $r$ means the poloidal field circulates a ring well outside $\rpeak$. This ensures the flux being accreted is all of one alignment. Second, the local rescaling results in a magnetic field that has roughly constant $\beta$ across the grid. The initial magnetic pressure $\pmag = b_\lambda b^\lambda \! / 2$ and $\beta$ are shown in Figures~\ref{fig:mad_slices_initial_pmag} and~\ref{fig:mad_slices_initial_beta}, with the field configuration illustrated in the latter.

We note that two other limits are imposed on the variables throughout the simulation. The density must obey $\rho \ge \pmag/\gammamax$, and the gas pressure must obey $\pgas \ge (\Gamma-1) \pmag / 10\gammamax$, just as in \citet{Tchekhovskoy2011}. Because we cannot alter the magnetic field locally without violating the divergence-free constraint, these conditions are enforced via floors on $\rho$ and $\pgas$.

\section{Results of New MAD Simulations}
\label{sec:application:results}

We ran three resolutions (refinement levels $0$ through $2$) with a Courant--Friedrichs--Lewy number of $0.25$ out to a time of $t = 30{,}000\ M$. In all cases the flow reached a quasisteady state. We also ran variations on the level $2$ simulation, changing polar resolution, polar boundary conditions, and numerical floors, in order to investigate what effects these have. Furthermore we ran a very high resolution simulation (level $3$) to a time of $t = 4000\ M$ in order to further study convergence.

\subsection{Primary Results at Different Resolutions}
\label{sec:application:results:primary}

\begin{figure}
  \centering
  \includegraphics[width=0.8\textwidth]{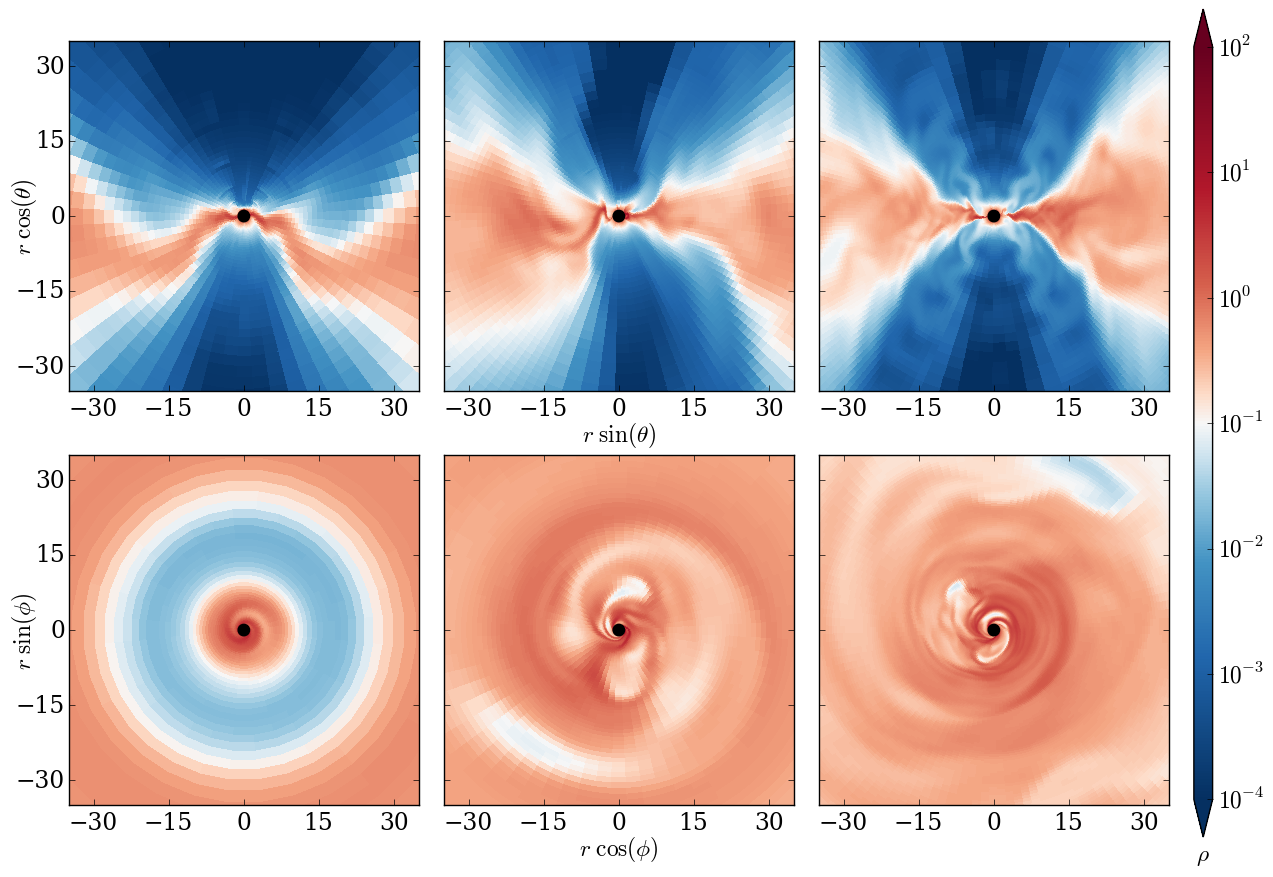}
  \caption{Poloidal (top) and equatorial (bottom) slices of density in MAD torus at time $t = 15{,}000\ M$. Refinement levels $0$, $1$, and $2$ are shown left to right. \label{fig:mad_slices_final_rho}}
\end{figure}

Figure~\ref{fig:mad_slices_final_rho} shows the density in the poloidal ($\phi = 0,\pi$) and equatorial ($\theta = \pi/2$) planes at $t = 15{,}000\ M$. Several features are immediately apparent. First, accretion has set in, with material funneling into the black hole in all cases. However, there is little evidence of turbulence at low resolution. Also, the level $0$ simulation entered a warped, asymmetric state, and it stayed in this shape throughout the simulation. At high resolution, turbulence is visible in both the poloidal and equatorial slices.

\begin{figure}
  \centering
  \includegraphics[width=0.8\textwidth]{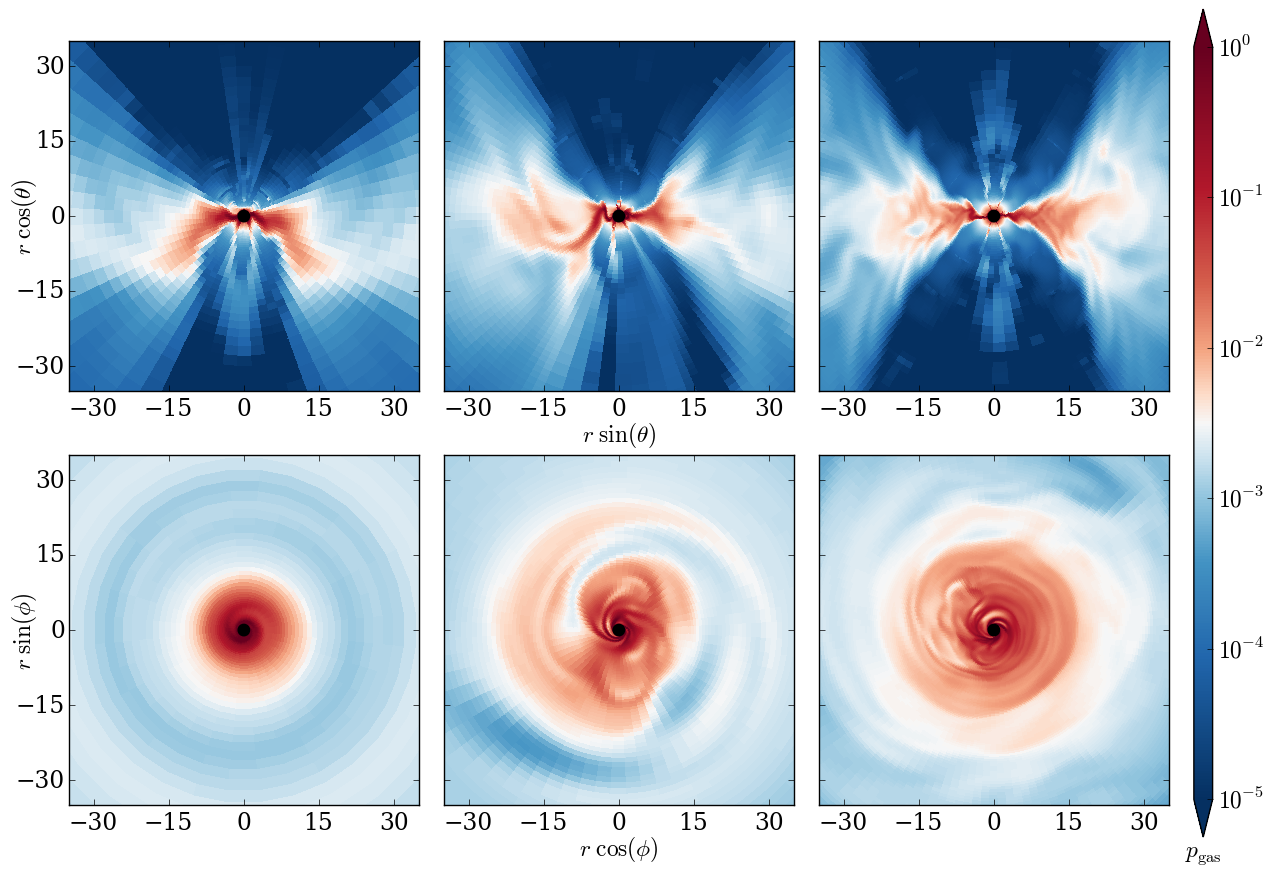}
  \caption{Poloidal (top) and equatorial (bottom) slices of gas pressure in MAD torus at time $t = 15{,}000\ M$. Refinement levels $0$, $1$, and $2$ are shown left to right. \label{fig:mad_slices_final_pgas}}
\end{figure}

\begin{figure}
  \centering
  \includegraphics[width=0.8\textwidth]{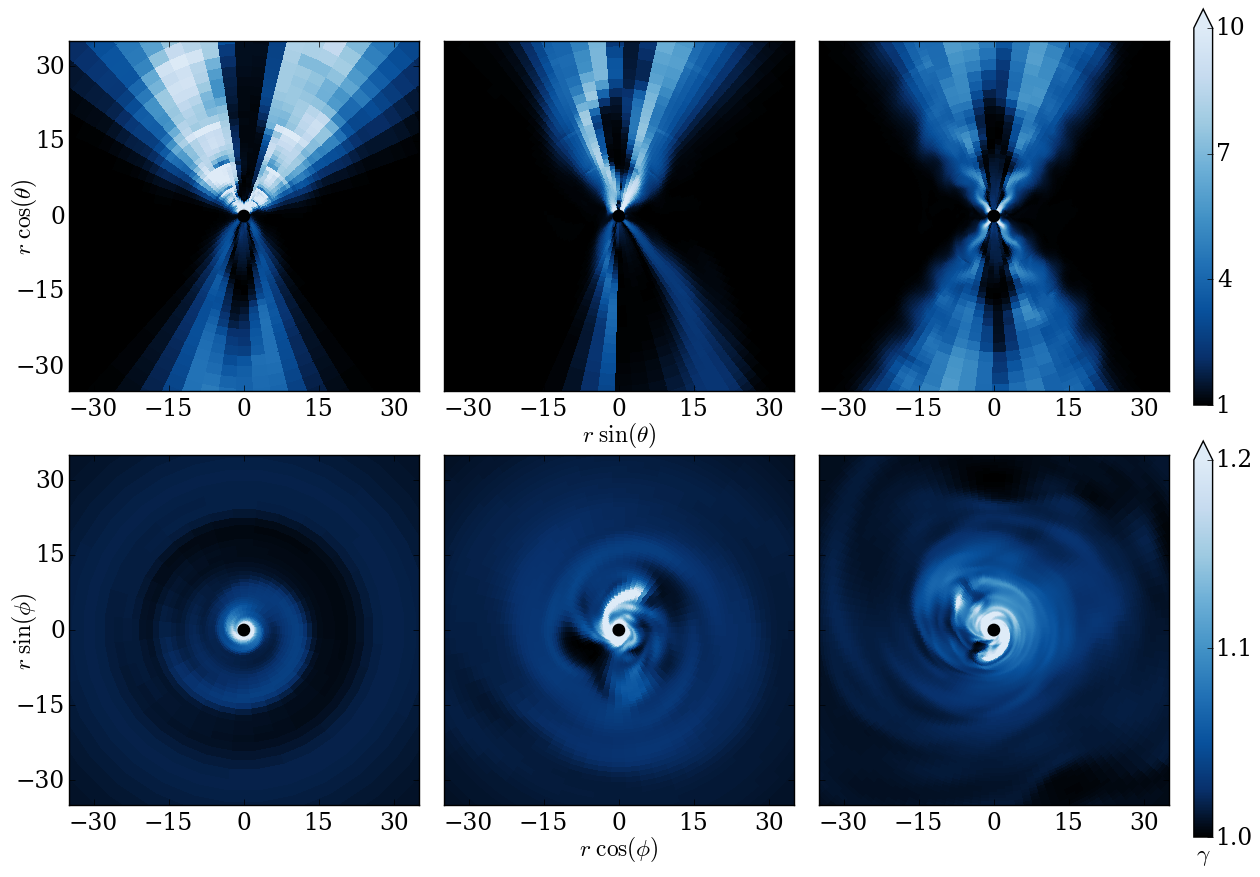}
  \caption{Poloidal (top) and equatorial (bottom) slices of normal observer Lorentz factor in MAD torus at time $t = 15{,}000\ M$. Refinement levels $0$, $1$, and $2$ are shown left to right. \label{fig:mad_slices_final_gamma}}
\end{figure}

Similar features can be seen in the gas pressure shown in Figure~\ref{fig:mad_slices_final_pgas}. In addition, there are regions of high pressure near the poles. These are not visible in the density plot, indicating they are near vacuum but very hot. These jet-like structures are indeed flowing outward from the black hole at high velocity. Figure~\ref{fig:mad_slices_final_gamma} shows the Lorentz factor. From this figure the asymmetry at level $0$ and even level $1$ is quite apparent, with most of the outflow directed upward. This asymmetry is not seen at level $2$, indicating that it probably is an artifact of underresolving the flow.

\begin{figure}
  \centering
  \includegraphics[width=0.8\textwidth]{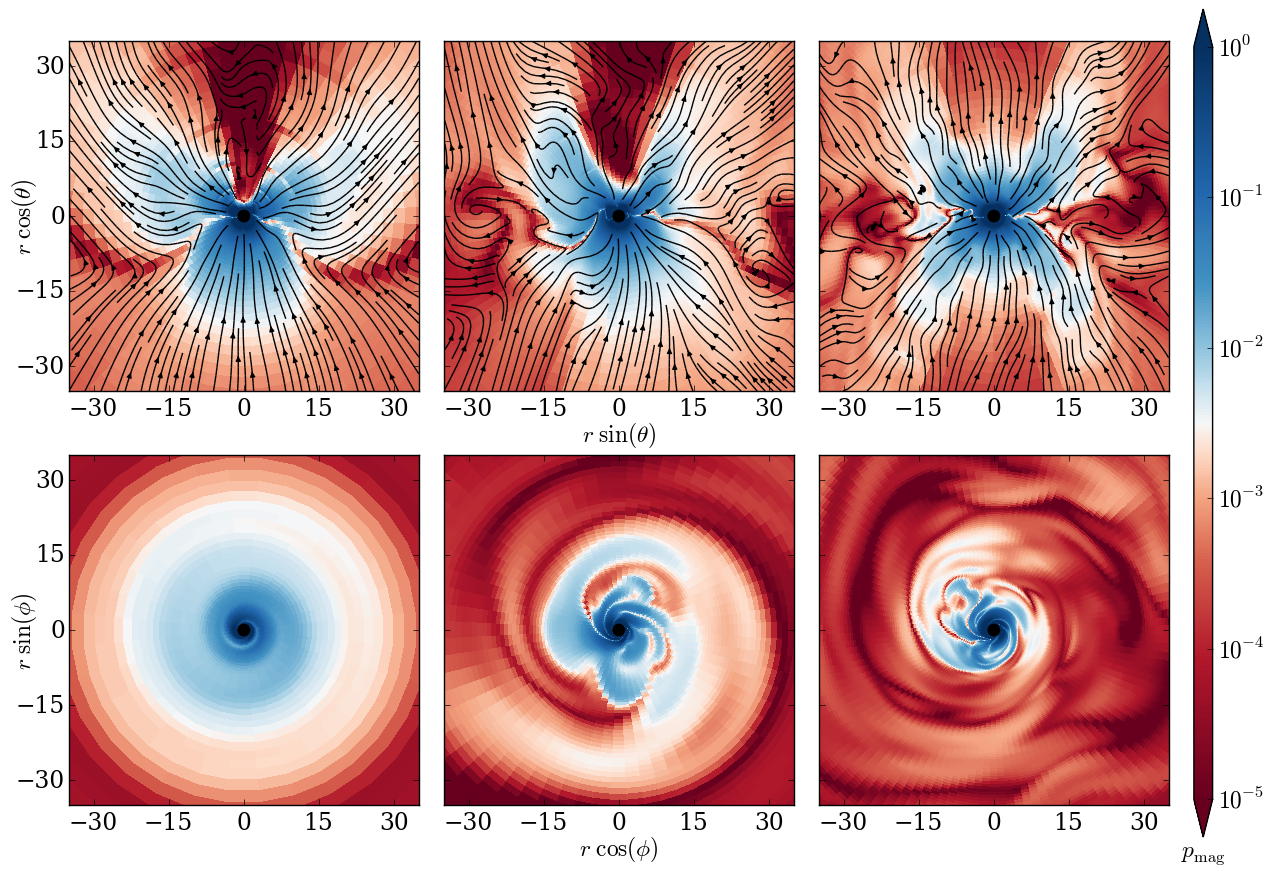}
  \caption{Poloidal (top) and equatorial (bottom) slices of magnetic pressure in MAD torus at time $t = 15{,}000\ M$. Refinement levels $0$, $1$, and $2$ are shown left to right. Overlaid are streamlines for the magnetic field lying on their respective slices through the data. \label{fig:mad_slices_final_pmag}}
\end{figure}

Plots of magnetic pressure are shown in Figure~\ref{fig:mad_slices_final_pmag}. Here we can clearly see the effect of higher resolution on the amount of turbulence present. At level $0$ there is only a single, smooth current sheet in the disk. At level $1$ this sheet is contorted. At level $2$ it is broken into multiple parts. Moreover, the region of high magnetic pressure seen in the equatorial slice becomes fragmented at higher resolution.

This turbulence is reflected in the magnetic field lines, which are more ordered at lower resolution. It is also worth noting that the field lines in the polar funnel region deviate significantly from vertical at lower resolutions, even crossing the polar axis horizontally. The plots show streamlines generated by taking only the slice shown and neglecting the toroidal component. In particular, they are not azimuthally averaged. When averaging is done, these deviations are significantly reduced, showing that even at a single timeslice the azimuthally average field is ordered while there are nonetheless nonaxisymmetric features.

\begin{figure}
  \centering
  \includegraphics[width=0.8\textwidth]{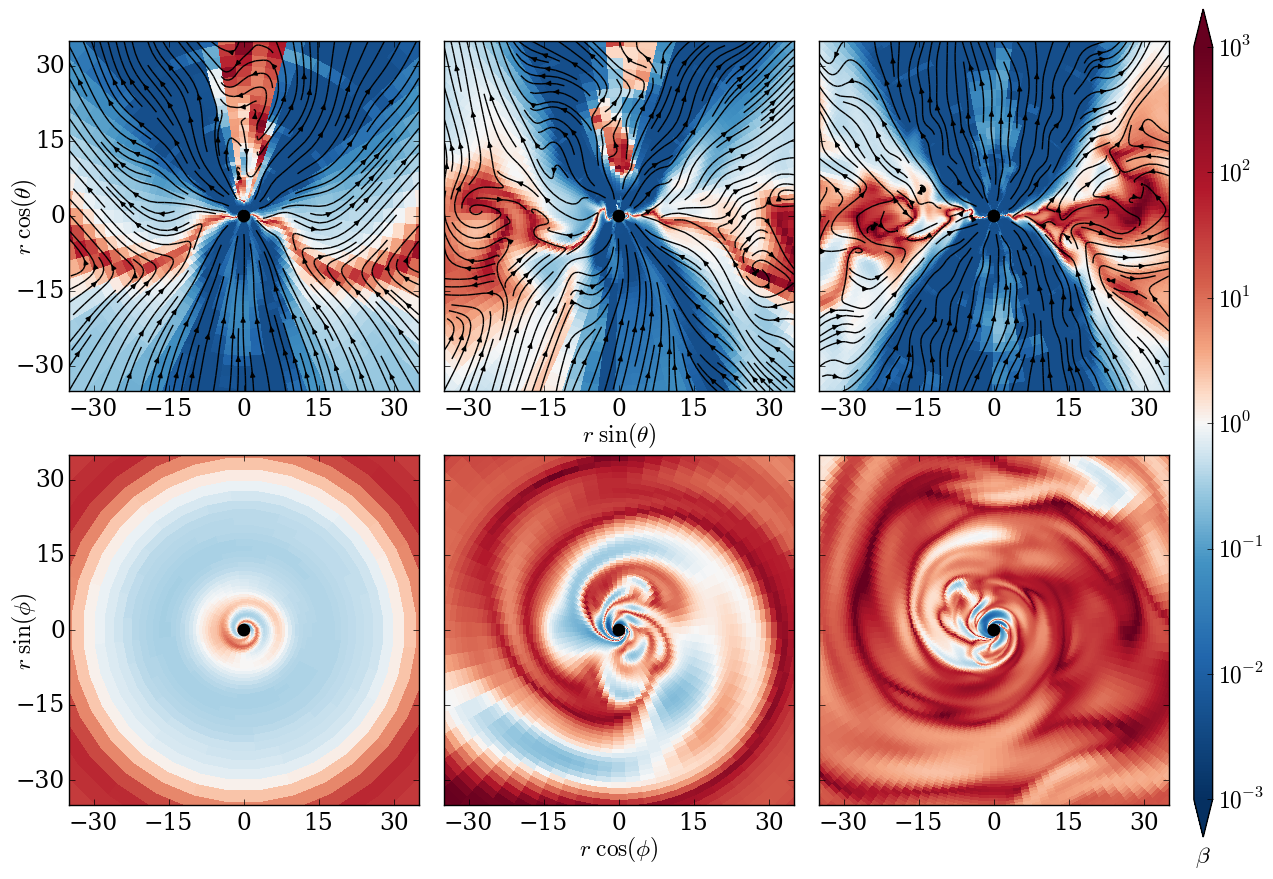}
  \caption{Poloidal (top) and equatorial (bottom) slices of plasma $\beta$ in MAD torus at time $t = 15{,}000\ M$. Refinement levels $0$, $1$, and $2$ are shown left to right. Overlaid are streamlines for the magnetic field lying on their respective slices through the data. \label{fig:mad_slices_final_beta}}
\end{figure}

\begin{figure}
  \centering
  \includegraphics[width=0.8\textwidth]{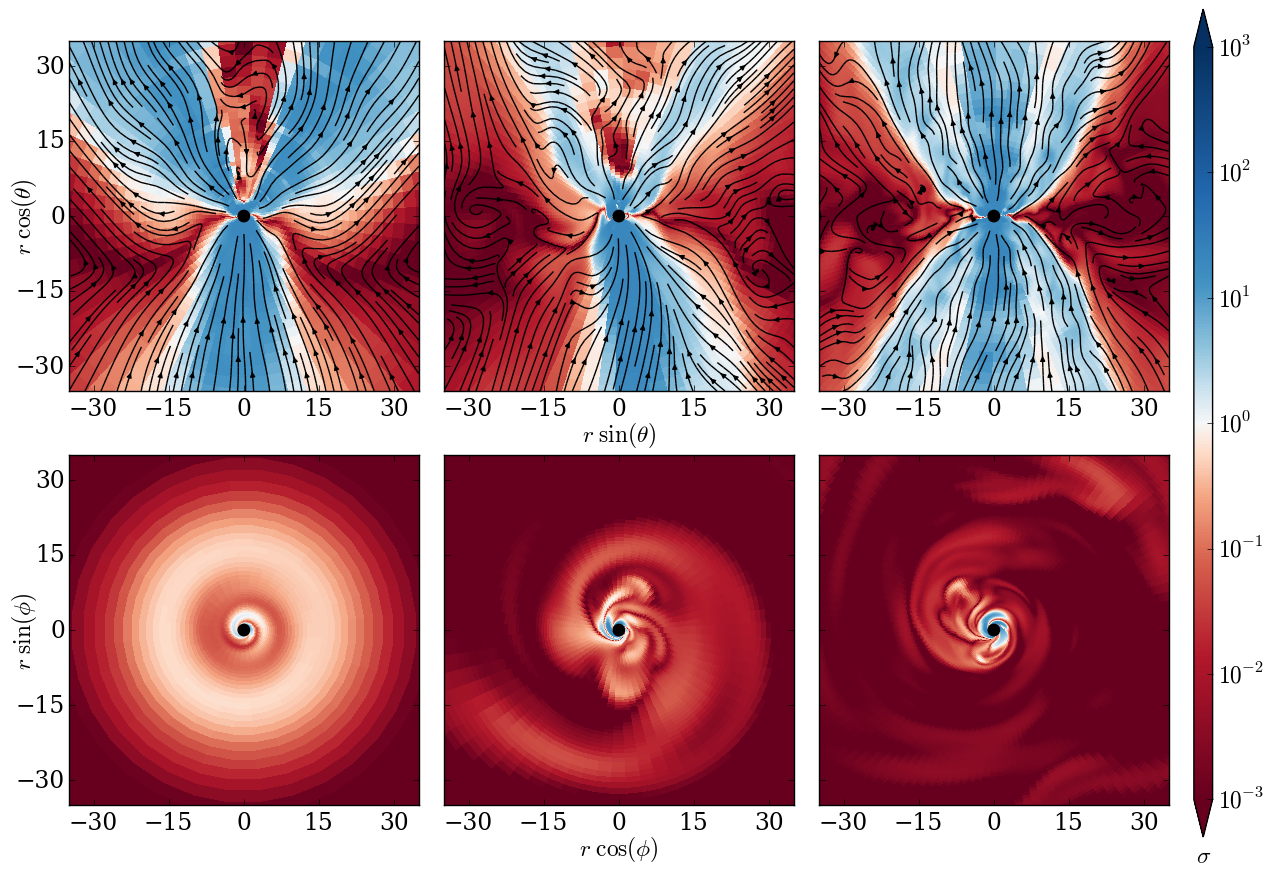}
  \caption{Poloidal (top) and equatorial (bottom) slices of plasma $\sigma$ in MAD torus at time $t = 15{,}000\ M$. Refinement levels $0$, $1$, and $2$ are shown left to right. Overlaid are streamlines for the magnetic field lying on their respective slices through the data. \label{fig:mad_slices_final_sigma}}
\end{figure}

The plots of $\beta$ and $\sigma = b_\lambda b^\lambda\!/\rho$, Figures~\ref{fig:mad_slices_final_beta} and~\ref{fig:mad_slices_final_sigma}, tell a similar story. From them we can also see that the asymmetric jets seen at low resolution have relatively little magnetic pressure (equivalently energy density) given their gas pressure and rest mass density.

\subsection{Variations Near the Poles}
\label{sec:application:results:variations}

In order to run any numerical simulation, a number of choices must be made regarding implementation parameters. Here we take the refinement level $2$ setup and change some of the numerical details in order to see what effect this has on the outcome. We focus on conditions that are likely to most strongly impact the funnel region near the poles.

While we have given much attention to resolving the accretion disk proper, we generally leave the polar regions at level $0$ refinement in all simulations. Low azimuthal resolution here might be forgiven:\ nonaxisymmetric structures are likely to be advected away before significantly affecting the dynamics. However, low resolution in the polar angle could lead to diffusion of the magnetic field or momentum flux away from the pole.

In order to examine the effects of resolution in this region, we run a modified version of the level $2$ grid in which level $1$ refinement goes all the way to the poles (rather than stopping at $\theta = \pi/8$ and $\theta = 7\pi/8$, see \S\ref{sec:application:simulations:grid}). In order to keep the timestep reasonable, we then halve the $\phi$-resolution everywhere, meaning we have $32$ cells in azimuth at the poles in both cases.

\begin{figure}
  \centering
  \includegraphics[width=\textwidth]{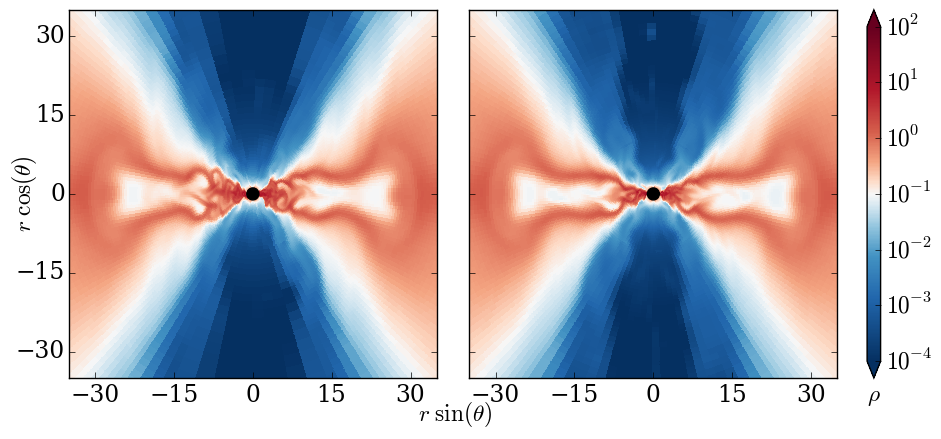}
  \caption{Poloidal slices of density in MAD torus at time $t = 2000\ M$. Shown are the fiducial run (left) and the variation with double resolution near the pole (right). \label{fig:mad_slices_variations_1_rho}}
\end{figure}

\begin{figure}
  \centering
  \includegraphics[width=\textwidth]{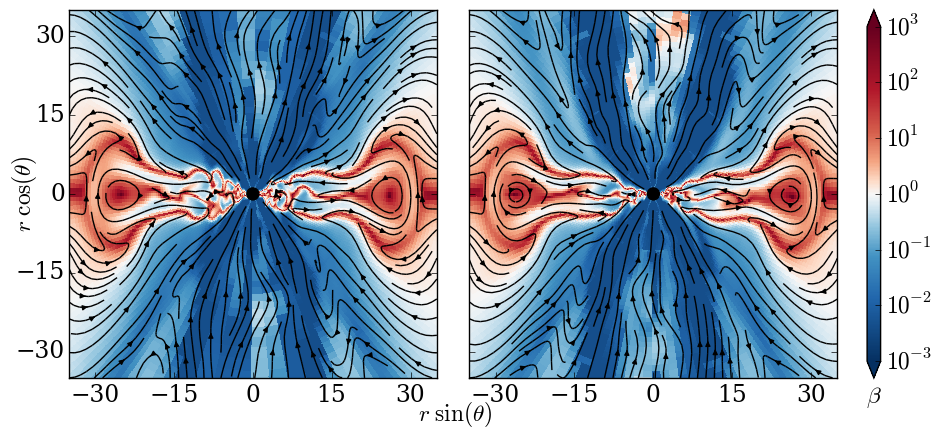}
  \caption{Poloidal slices of plasma $\beta$ in MAD torus at time $t = 2000\ M$. Shown are the fiducial run (left) and the variation with double resolution near the pole (right). Overlaid are streamlines for the magnetic field lying on this slice through the data. \label{fig:mad_slices_variations_1_beta}}
\end{figure}

\begin{figure}
  \centering
  \includegraphics[width=\textwidth]{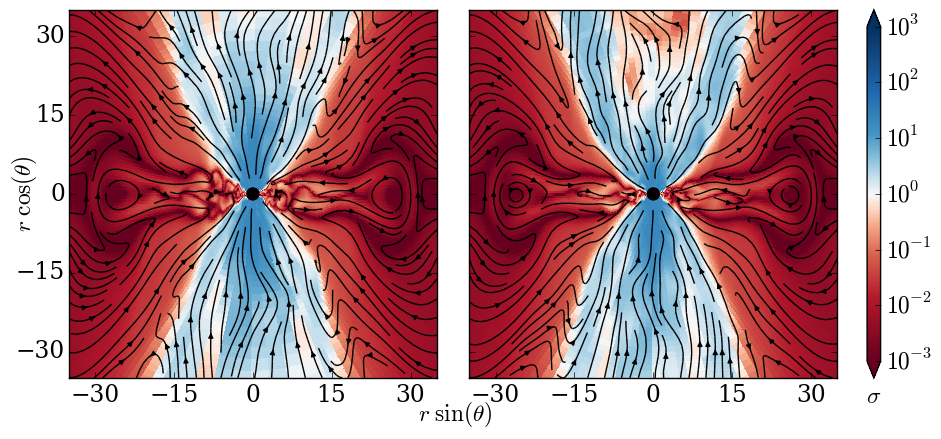}
  \caption{Poloidal slices of plasma $\sigma$ in MAD torus at time $t = 2000\ M$. Shown are the fiducial run (left) and the variation with double resolution near the pole (right). Overlaid are streamlines for the magnetic field lying on this slice through the data. \label{fig:mad_slices_variations_1_sigma}}
\end{figure}

This simulation was run to a time of $t = 2000\ M$, after which numerical issues forced it to stop. Plots of $\rho$, $\beta$, and $\sigma$ for the fiducial run and this variation at this time are shown in Figures~\ref{fig:mad_slices_variations_1_rho} through~\ref{fig:mad_slices_variations_1_sigma}. In all cases the turbulence in the disk is more developed in the fiducial case, however this is a result of the different number of azimuthal cells in the disk ($128$ versus $64$). The jet region in the variation displays higher $\beta$ and lower $\sigma$ values, just as is seen in the fiducial runs at lower resolution (Figures~\ref{fig:mad_slices_final_beta} and~\ref{fig:mad_slices_final_sigma}), as well as hints of the field being less vertically oriented. Thus it appears that azimuthal resolution, even away from the poles, is more important than polar angle resolution for our purposes.

Two further variations were considered:\ imposing a reflecting wall and adjusting the floor values. For the wall we excise the region for which $\theta$ is within $10^{-8}$ of the poles, imposing a reflecting wall boundary condition along the newly created surface. For the new floor values, we change the position-dependent floors to $\rho \ge \max\{10^{-4} r^{-3/2}, 10^{-6}\}$ and $\pgas \ge \max\{10^{-6} (\Gamma-1) r^{-5/2}, 10^{-8}\}$ (see \S\ref{sec:application:simulations:torus} for the original floors). We also increase $\gammamax$ from $10$ to $25$, thus decreasing the limits $\rho \ge \pmag/\gammamax$ and $\pgas \ge (\Gamma-1) \pmag / 10\gammamax$.

\begin{figure}
  \centering
  \includegraphics[width=\textwidth]{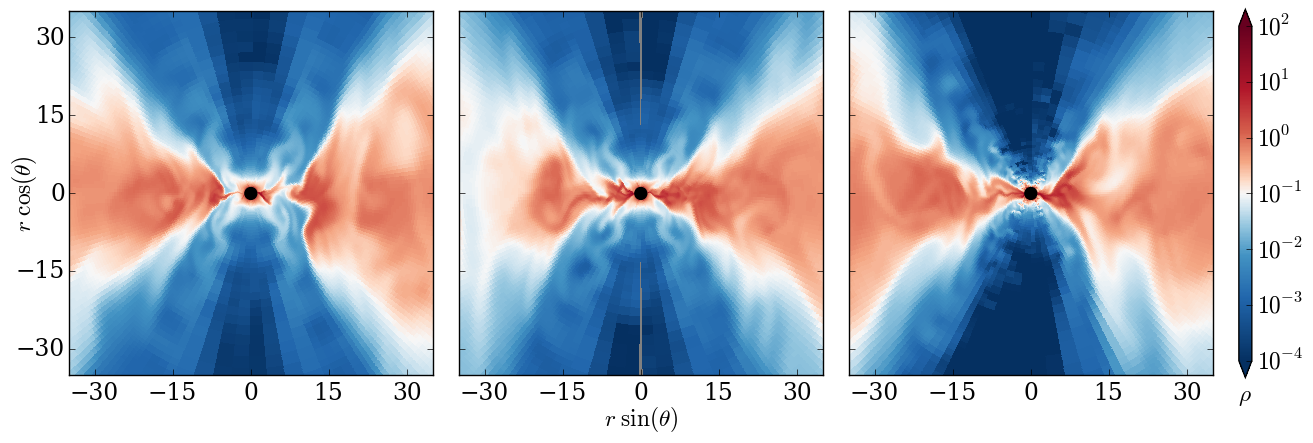}
  \caption{Poloidal slices of density in MAD torus at time $t = 10{,}000\ M$. Shown are the fiducial run (left), the variation with reflecting walls (center), and the variation with looser floors (right). \label{fig:mad_slices_variations_2_rho}}
\end{figure}

\begin{figure}
  \centering
  \includegraphics[width=\textwidth]{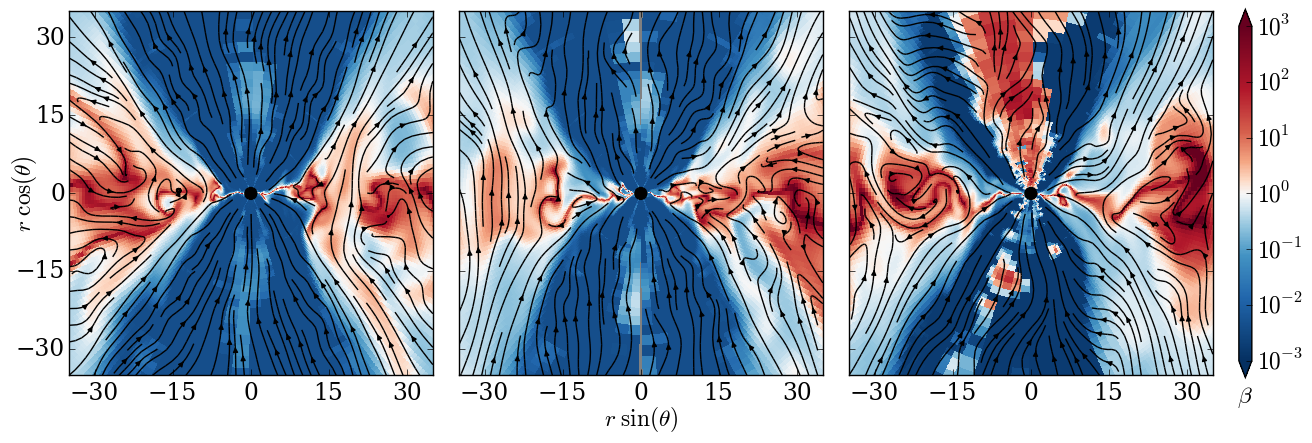}
  \caption{Poloidal slices of plasma $\beta$ in MAD torus at time $t = 10{,}000\ M$. Shown are the fiducial run (left), the variation with reflecting walls (center), and the variation with looser floors (right). Overlaid are streamlines for the magnetic field lying on this slice through the data. \label{fig:mad_slices_variations_2_beta}}
\end{figure}

\begin{figure}
  \centering
  \includegraphics[width=\textwidth]{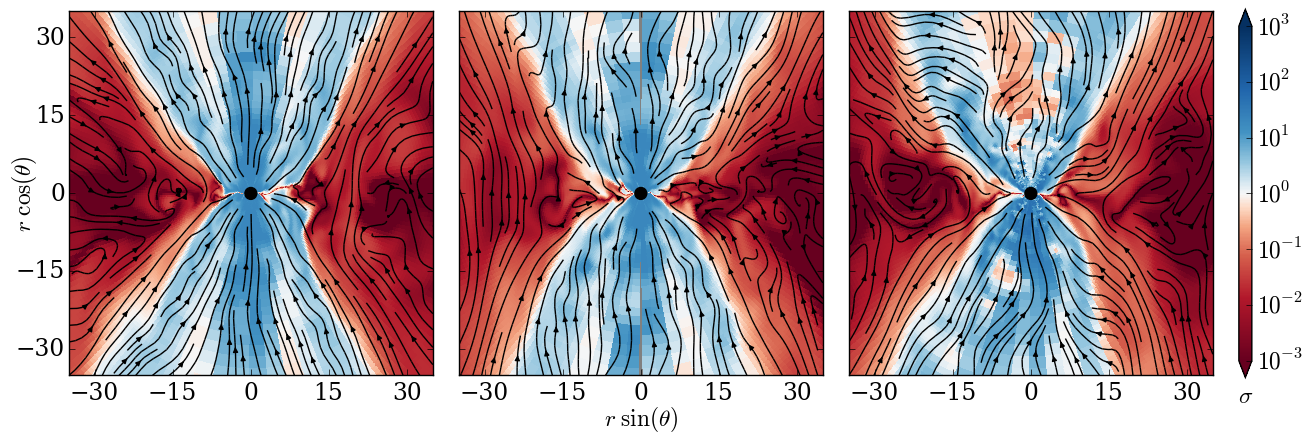}
  \caption{Poloidal slices of plasma $\sigma$ in MAD torus at time $t = 10{,}000\ M$. Shown are the fiducial run (left), the variation with reflecting walls (center), and the variation with looser floors (right). Overlaid are streamlines for the magnetic field lying on this slice through the data. \label{fig:mad_slices_variations_2_sigma}}
\end{figure}

These variations were run to a time of $t = 10{,}000\ M$. Plots of $\rho$, $\beta$, and $\sigma$ at this time are shown in Figures~\ref{fig:mad_slices_variations_2_rho} through~\ref{fig:mad_slices_variations_2_sigma}. The reflecting wall appears little different from the fiducial case. The looser floors, however, led to a significant amount of noise being generated, with large cell-to-cell variations apparent in all three figures. Interestingly, lowering the density a pressure floors, and indeed lowering these floors relative to the magnetic energy density, results in lower magnetization (higher $\beta$, lower $\sigma$) in the core of the funnel region, as is seen in the fiducial runs at lower resolution (Figures~\ref{fig:mad_slices_final_beta} and~\ref{fig:mad_slices_final_sigma}). The direction of the change indicates this likely unphysical demagnetization is not a simple effect of floors being activated in that region, but rather is a complex phenomenon emerging from the accumulation of numerical inaccuracies.

While some of the variations studied show pronounced differences in the outflow region, it is reassuring that they do not appear to affect the bulk of the accretion flow.

\subsection{Very High Resolution}
\label{sec:application:results:high}

In addition to the runs with refinement levels $0$ through $2$, we ran a MAD simulation with refinement level $3$. In the most refined region of this simulation the effective resolution (the resolution if the whole domain were at level $3$ refinement) is $768 \times 256 \times 256$. In the radial direction this is $260$ cells per decade.

\begin{figure}
  \centering
  \includegraphics[width=\textwidth]{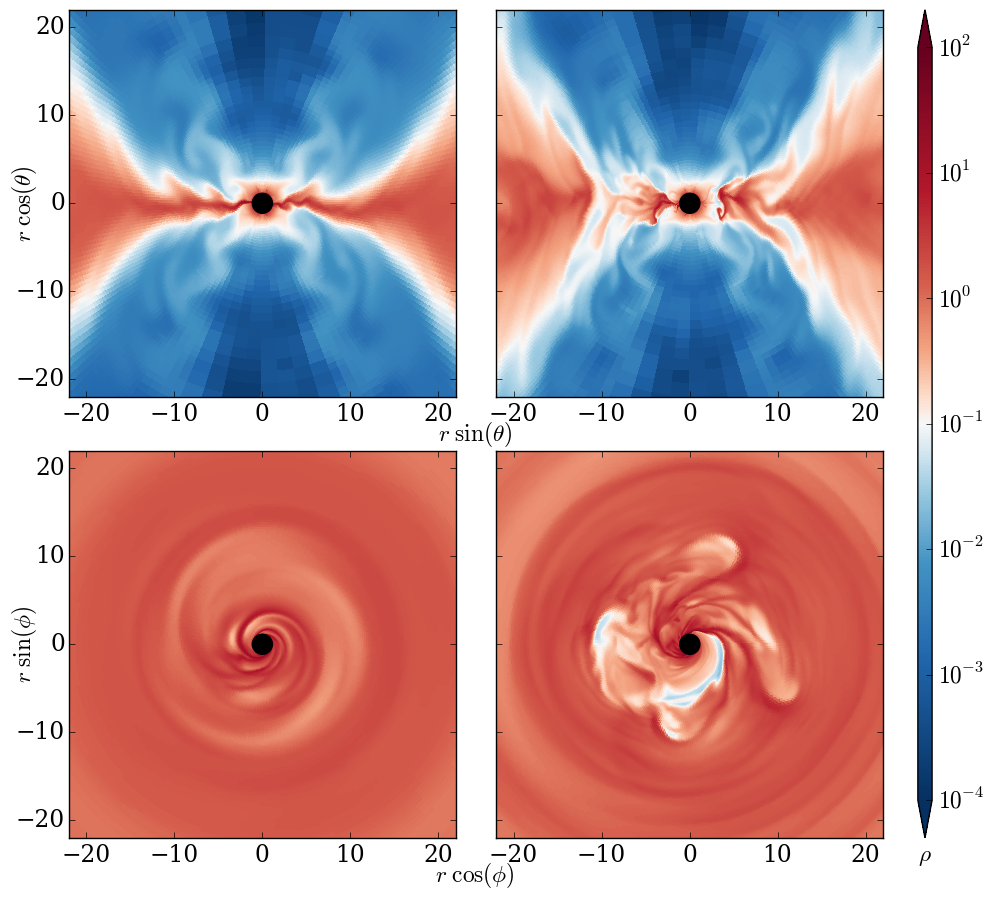}
  \caption{Poloidal (top) and equatorial (bottom) slices of density in MAD torus at time $t = 4000\ M$ with refinement levels $2$ (left) and $3$ (right). See Figure~\ref{fig:mad_slices_high_zoom_rho} for a zoomed in version. \label{fig:mad_slices_high_rho}}
\end{figure}

This costly simulation was only run to $t = 4000\ M$ as of the time of this writing. We compare the fiducial level $2$ simulation at this same time in the evolution. Figure~\ref{fig:mad_slices_high_rho} shows slices of the density in the two simulations. In the poloidal plane the high resolution disk is already disrupted by turbulence, whereas the lower resolution disk only shows mild distortions. It should be noted, though, that the turbulence has not saturated at level $2$ by this point, as can be seen by considering the upper right panel of Figure~\ref{fig:mad_slices_final_rho}, which resembles the upper right panel of Figure~\ref{fig:mad_slices_high_rho}. Indeed turbulence develops more rapidly at higher resolution, as expected.

In the equatorial plane the effects of magnetized, low-density fluid bubbling up are more apparent at higher resolution. Indeed an $m = 4$ Rayleigh--Taylor mode is visually apparent, whereas at lower resolution only $m = 1$ and $m = 2$ spiral patterns are visible. Even at late times the lower resolution resolution simulation does not develop any readily apparent $m = 4$ mode, as can be seen in the lower right panel of Figure~\ref{fig:mad_slices_final_rho}.

\begin{figure}
  \centering
  \includegraphics[width=\textwidth]{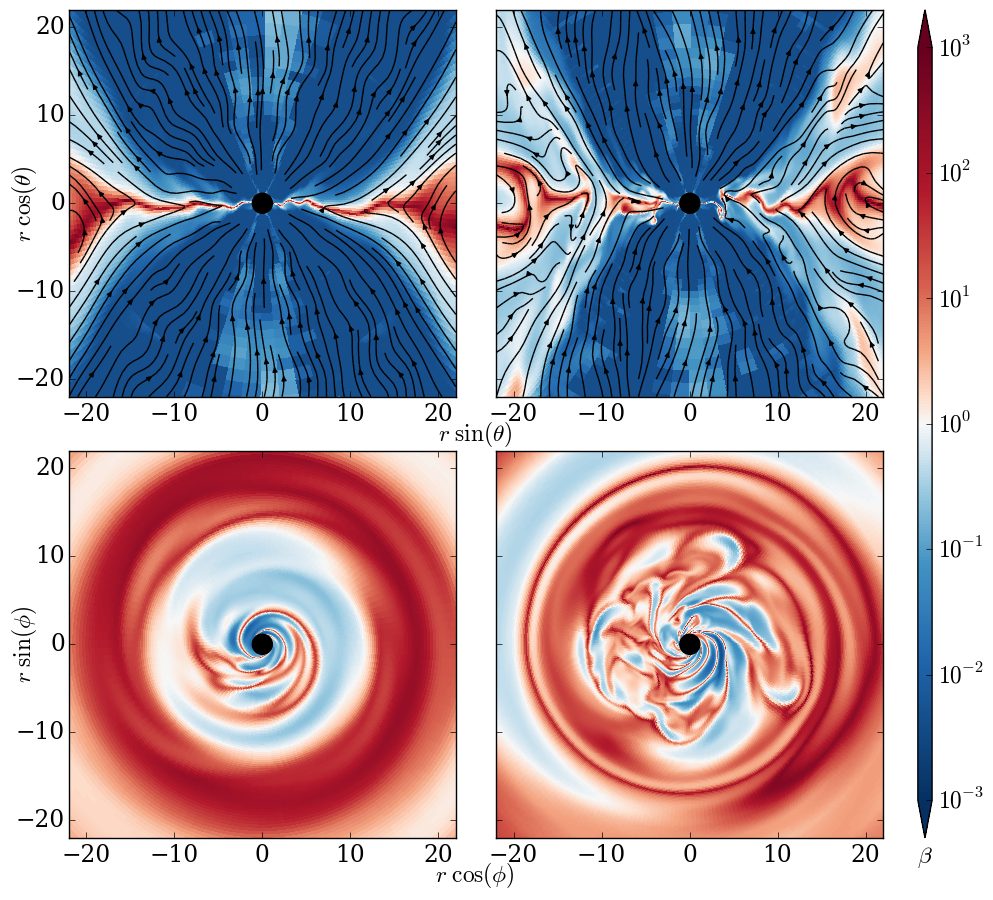}
  \caption{Poloidal (top) and equatorial (bottom) slices of plasma $\beta$ in MAD torus at time $t = 4000\ M$ with refinement levels $2$ (left) and $3$ (right). Overlaid are streamlines for the magnetic field lying on their respective slices through the data. See Figure~\ref{fig:mad_slices_high_zoom_beta} for a zoomed in version. \label{fig:mad_slices_high_beta}}
\end{figure}

\begin{figure}
  \centering
  \includegraphics[width=\textwidth]{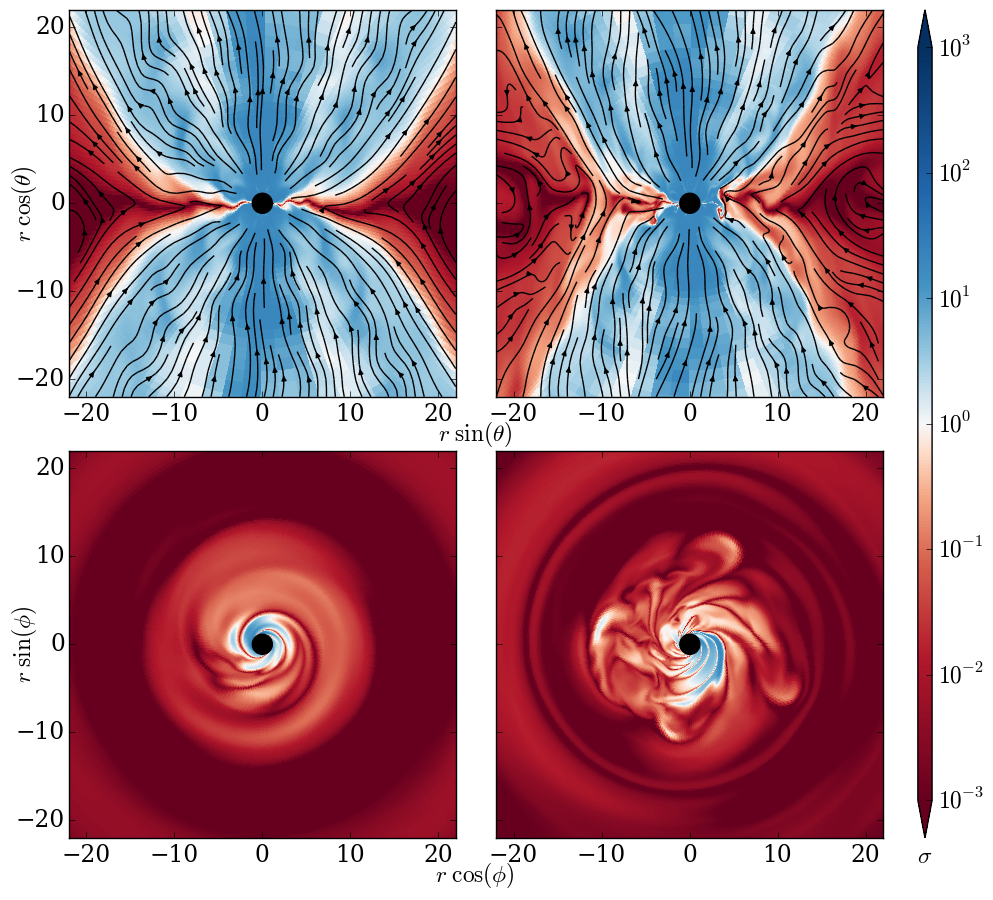}
  \caption{Poloidal (top) and equatorial (bottom) slices of plasma $\sigma$ in MAD torus at time $t = 4000\ M$ with refinement levels $2$ (left) and $3$ (right). Overlaid are streamlines for the magnetic field lying on their respective slices through the data. See Figure~\ref{fig:mad_slices_high_zoom_sigma} for a zoomed in version. \label{fig:mad_slices_high_sigma}}
\end{figure}

Similar patterns can be seen with $\beta$ (Figure~\ref{fig:mad_slices_high_beta}) and $\sigma$ (Figure~\ref{fig:mad_slices_high_sigma}):\ the inner part of the disk is more turbulent at higher resolution. The low-magnetization streams feeding the black hole are thinner and more numerous at higher resolution. This holds even when considering later times at lower resolution, as shown in the lower right panels of Figures~\ref{fig:mad_slices_final_beta} and~\ref{fig:mad_slices_final_sigma}.

\begin{figure}
  \centering
  \includegraphics[width=\textwidth]{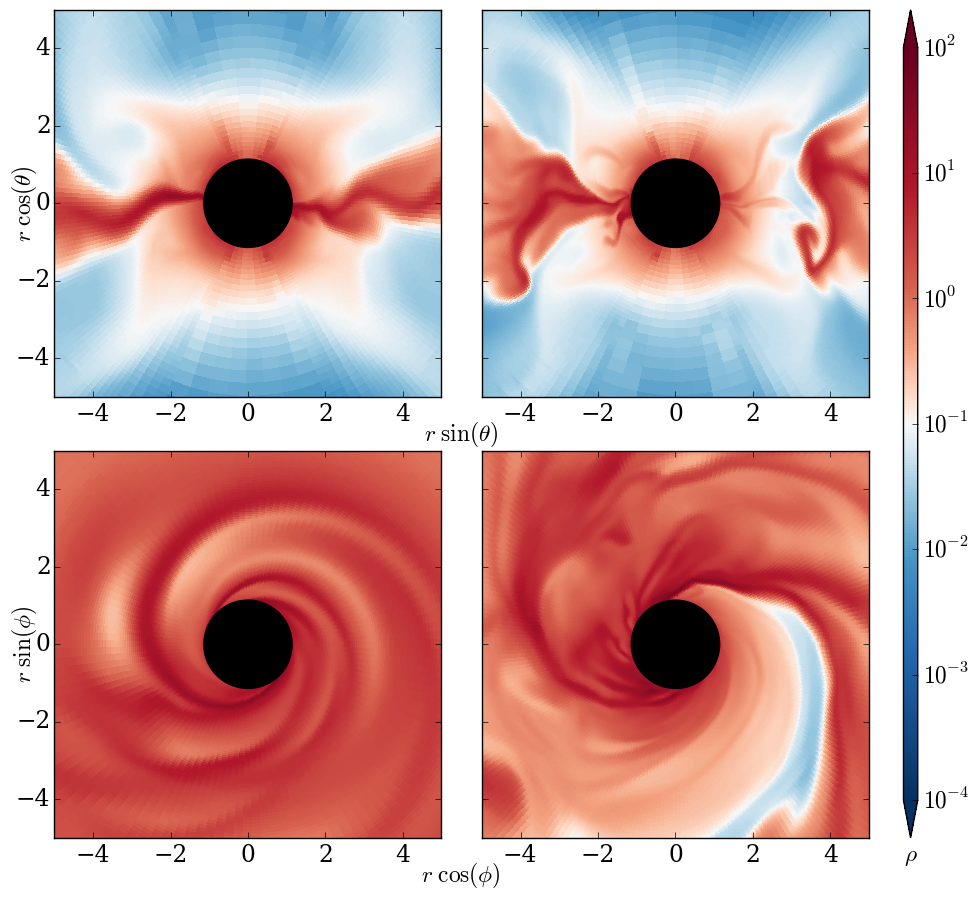}
  \caption{Zoomed in version of Figure~\ref{fig:mad_slices_high_rho}, showing poloidal (top) and equatorial (bottom) slices of density in MAD torus at time $t = 4000\ M$ with refinement levels $2$ (left) and $3$ (right). \label{fig:mad_slices_high_zoom_rho}}
\end{figure}

\begin{figure}
  \centering
  \includegraphics[width=\textwidth]{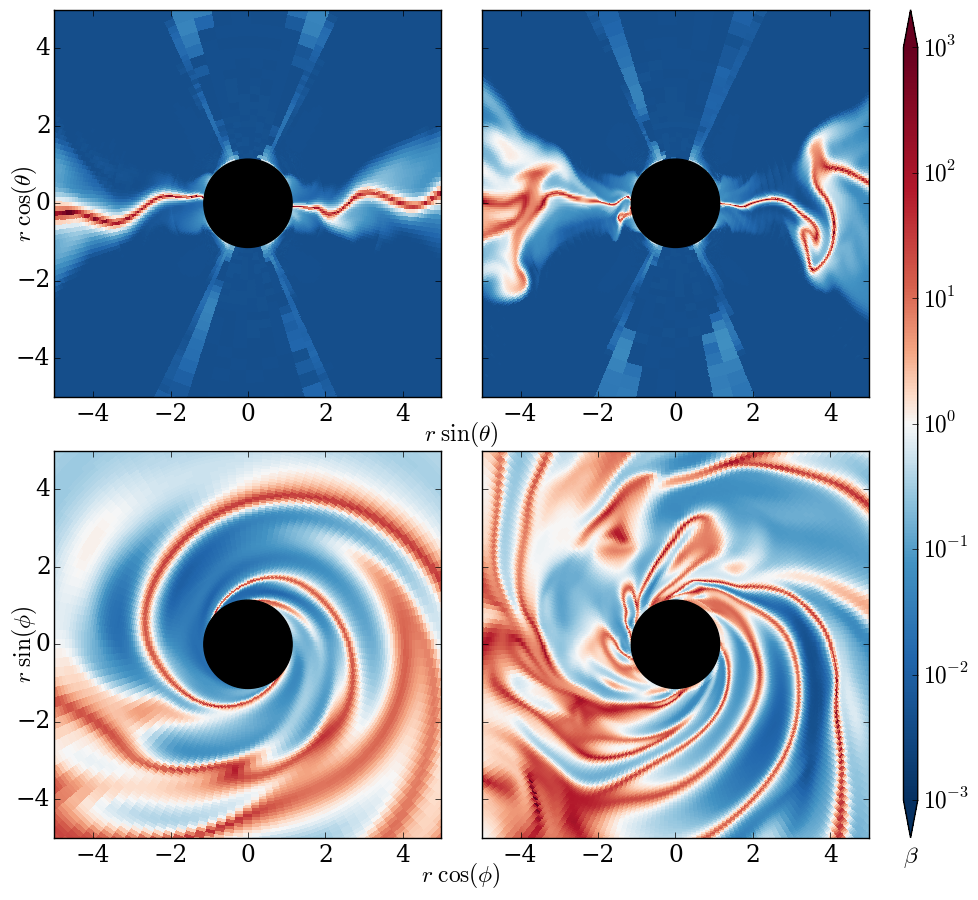}
  \caption{Zoomed in version of Figure~\ref{fig:mad_slices_high_beta}, showing poloidal (top) and equatorial (bottom) slices of plasma $\beta$ in MAD torus at time $t = 4000\ M$ with refinement levels $2$ (left) and $3$ (right). \label{fig:mad_slices_high_zoom_beta}}
\end{figure}

\begin{figure}
  \centering
  \includegraphics[width=\textwidth]{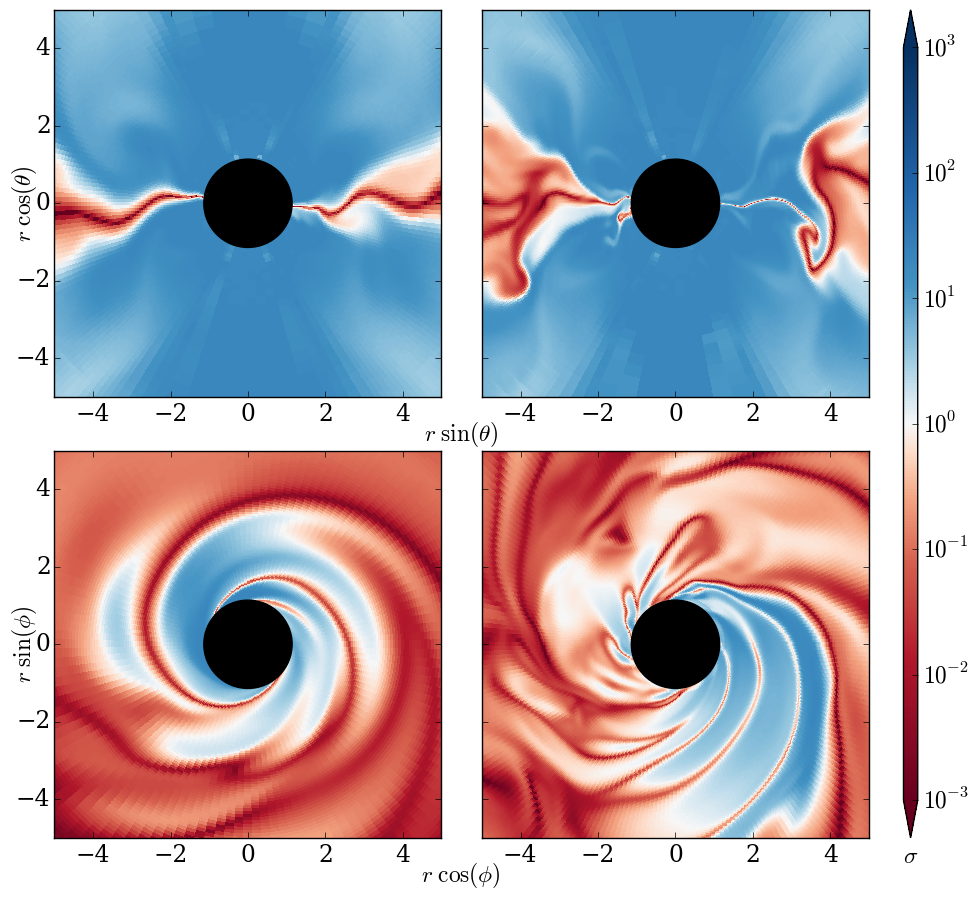}
  \caption{Zoomed in version of Figure~\ref{fig:mad_slices_high_sigma}, showing poloidal (top) and equatorial (bottom) slices of plasma $\sigma$ in MAD torus at time $t = 4000\ M$ with refinement levels $2$ (left) and $3$ (right). \label{fig:mad_slices_high_zoom_sigma}}
\end{figure}

In order to more clearly show features very close to the horizon, Figures~\ref{fig:mad_slices_high_zoom_rho}--\ref{fig:mad_slices_high_zoom_sigma} are zoomed in versions of Figures~\ref{fig:mad_slices_high_rho}--\ref{fig:mad_slices_high_sigma}. Of particular note are the disruption of the disk at high resolution in the former and the thinness of the accreting streams in the latter two.

Thus we see that at least in some aspects even the midplane flow is not entirely converged in our series of simulations. The polar funnel region, however, appears the same in all three displayed quantities between the two resolutions.

\section{Interpretation of Our Results}
\label{sec:application:interpretation}

While the lowest resolution simulation shows the large flux (Figure~\ref{fig:mad_slices_final_pmag}) and small amount of mass (Figure~\ref{fig:mad_slices_final_rho}) in the central region one expects in a MAD scenario, the picture becomes more complicated as we go to higher resolution. The flow is more turbulent, and nonaxisymmetric structures allow for mass to flow inward through the region of large flux.

This behavior is not unexpected. A magnetic field can provide pressure support, but as it has no rest mass any attempt to hold a massive fluid against gravity is likely to be Rayleigh--Taylor unstable. This is not to say that there can be no arresting of the flow in nature; MAD can still act to slow down the mass accretion.

In order to analyze the situation more quantitatively, we measure accretion in several ways as a function of time. Following \citet{Tchekhovskoy2011} we define the mass accretion rate
\begin{equation} \label{eq:m_dot}
  \dot{M} = -\int\limits_{r=r_0} \rho u^r \sqrt{-g} \, \dth\,\dph,
\end{equation}
where the quantities are evaluated in Kerr--Schild coordinates and $r_0$ is a fixed radius. Similarly we define an energy accretion rate
\begin{equation} \label{eq:e_dot}
  \dot{E} = \int\limits_{r=r_0} \tensor{T}{^r_t} \sqrt{-g} \, \dth\,\dph.
\end{equation}
These quantities measure the inflow of rest mass and energy at infinity, respectively, and in both cases a positive sign represents a flow into the black hole. We analogously define a magnetic energy flux by neglecting the density and gas pressure terms in the stress-energy tensor:
\begin{equation} \label{eq:e_dot_mag}
  \dot{E}_\mathrm{mag} = \int\limits_{r=r_0} (b_\lambda b^\lambda u^r u_t - b^r b_t) \sqrt{-g} \, \dth\,\dph.
\end{equation}
\Citeauthor{Tchekhovskoy2011}\ also define the magnetic flux threading half the horizon as
\begin{equation} \label{eq:b_flux}
  \Phi = \frac{1}{2} \int\limits_{r=r_0} \abs[\big]{\sqrt{4\pi}B^r} \sqrt{-g} \, \dth\,\dph
\end{equation}
(where we insert the factor of $\sqrt{4\pi}$ to account for the difference between our units and theirs), which they normalize to the average accretion rate over a suitablly large interval:
\begin{equation} \label{eq:b_flux_normalized}
  \phi = \frac{\Phi}{\ave{\dot{M}}^{1/2}M}.
\end{equation}

\begin{figure}
  \centering
  \includegraphics[width=\textwidth]{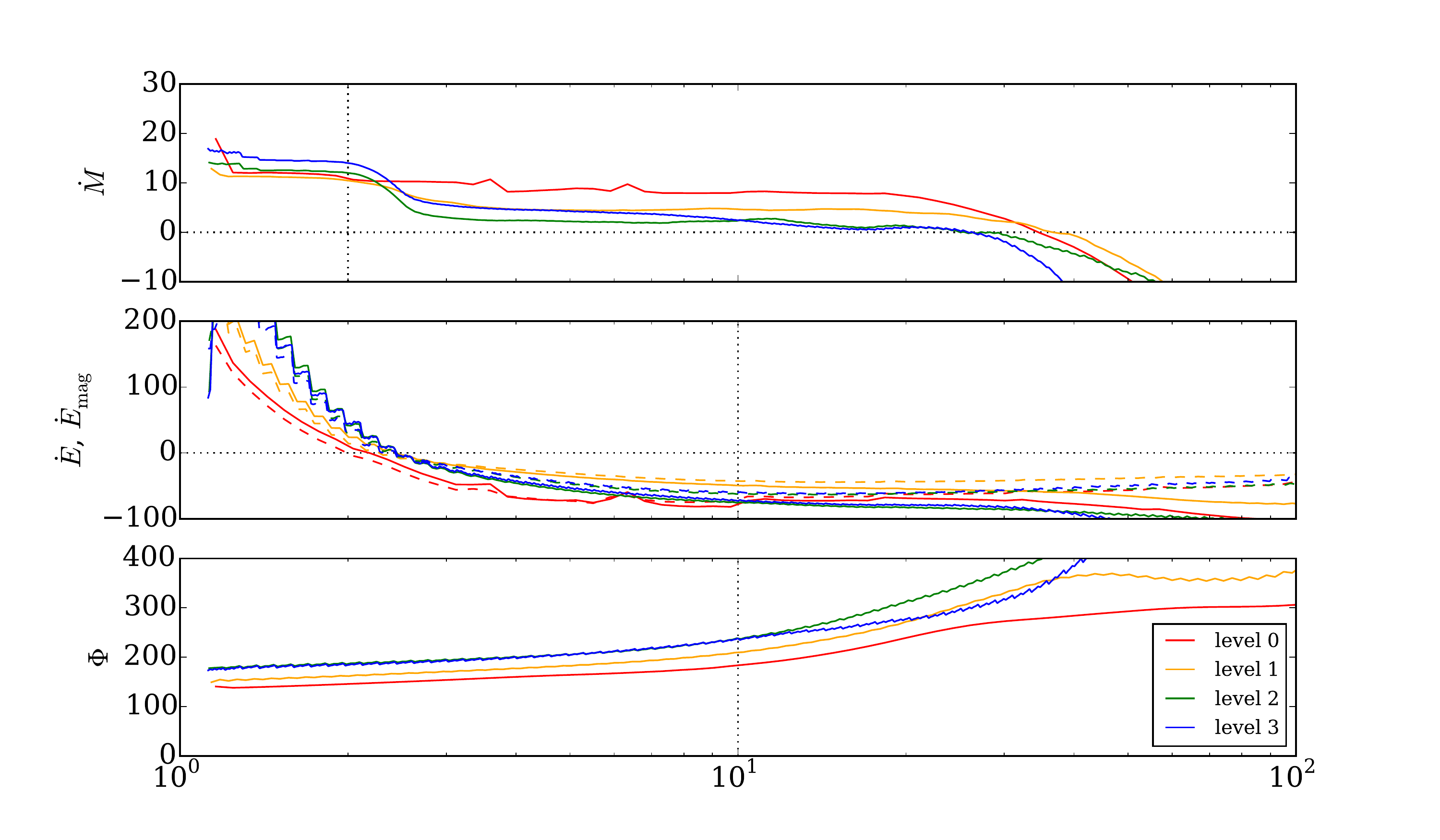}
  \caption{Mass accretion, energy accretion, and horizon-threading flux in the MAD simulations as functions of radius. Dashed lines show pure magnetic energy flux defined in \eqref{eq:e_dot_mag}. For refinement levels $0$ through $2$, values at each radius are averaged over times from $10{,}000\ M$ to $15{,}000\ M$ at intervals of $100\ M$; for level $3$ the averaging is done from $3000\ M$ to $4000\ M$, sampled every $10\ M$. Vertical dotted lines indicate selected values of $r_0$. \label{fig:mad_fluxes}}
\end{figure}

Before examining the time dependence of these quantities, we must choose $r_0$. Ideally we would choose $r_0 = \rhor$, but in practice the numerics near the horizon (especially the imposition of floors) make fluxes evaluated here suspect. In order to aid this choice, we inspect the time-averaged radial dependence of the fluxes, plotted in Figure~\ref{fig:mad_fluxes}. This includes data from refinement levels $0$ through $3$, with the highest level not run all the way to late times.

The effect of density floors is to artificially add mass the near-vacuum regions, which in our case coincide with the outflow. In order to approximately negate this effect in the analysis we zero the density of any cell within $15\%$ of the floor at that location. This has the effect of raising the values of $\dot{M}$ inside approximately $r = 2$. Note that wherever the simulation has reached a steady turbulent state we expect $\dot{M}$ to be constant. This is generally true, although the values outside approximately $r = 3$ are likely lower than they ideally would be, given that artificial mass loading closer in will appear as a massive outflow, but that at further radii this injected mass will be well above the floors and so not removed by our procedure.

By using $\tensor{T}{^r_t}$ rather than $T^{rt}$ we are essentially examining the radial flux of binding energy at infinity. This quantity should therefore also be independent of radius, at least out as far as steady state has been achieved. As can be seen in the second panel of Figure~\ref{fig:mad_fluxes}, we do see a rise in $\dot{E}$ as we move inward. This is presumably also an effect of the floors, though in a way that is less easy to correct. In particular, it is not an effect of the mass or gas pressure terms in the stress-energy tensor:\ the solid lines show $\dot{E}$ and the dashed lines show $\dot{E}_\mathrm{mag}$, and they are similar at all radii.

The magnetic flux $\Phi$ is rather constant inside $r = 10$, though it varies based on resolution. The flux is reduced at lower resolutions. Notably however the flux inside $r = 10$ is almost identical for the level $2$ and level $3$ simulations, indicating some degree of convergence.

Based on the above discussion, we choose to take $r_0 = 2$ for evaluating $\dot{M}$ in \eqref{eq:m_dot}. Meanwhile we choose $r_0 = 10$ for $\dot{E}$, $\dot{E}_\mathrm{mag}$, and $\Phi$ in \eqref{eq:e_dot}, \eqref{eq:e_dot_mag}, and \eqref{eq:b_flux}. We believe this leads to an accurate representation of the fluxes. One can use Figure~\ref{fig:mad_fluxes} to see how later results depend on this choice.

\begin{figure}
  \centering
  \includegraphics[width=\textwidth]{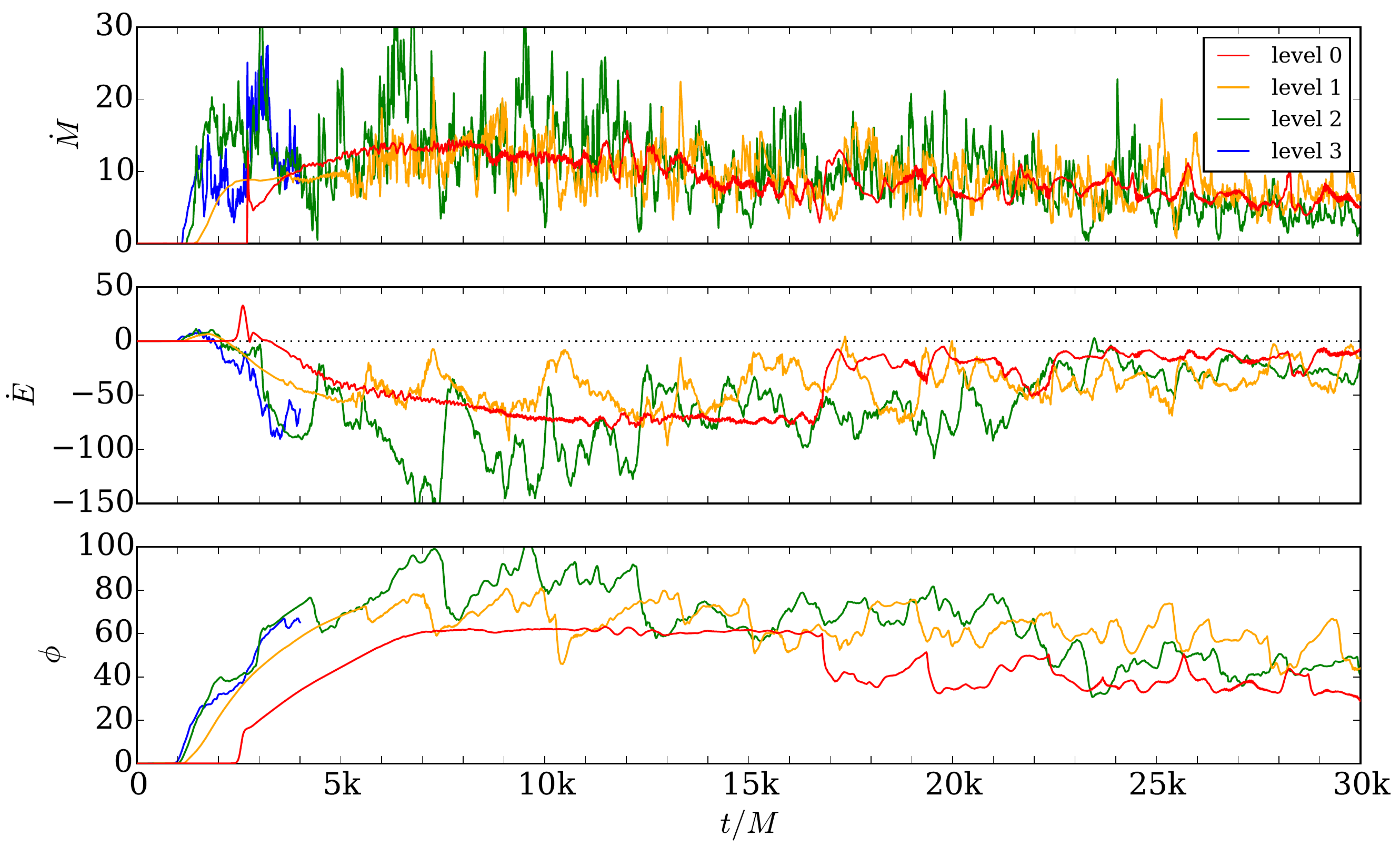}
  \caption{Mass accretion, energy accretion, and normalized horizon-threading flux in the MAD simulations as a function of time for four resolutions. All simulations are sampled at a rate of $\Delta t = 10\ M$. \label{fig:mad_accretion}}
\end{figure}

With suitable values of $r_0$ chosen, we can investigate the full time dependence of accretion. Figure~\ref{fig:mad_accretion} shows the evolution of $\dot{M}$, $\dot{E}$, and $\phi$ for the four resolutions. In order to normalize the flux $\phi$ we must choose an interval over which to average $\dot{M}$ (see \eqref{eq:b_flux_normalized}). For the lower three resolutions we choose the interval from $5000\ M$ to $30{,}000\ M$, and for the highest resolution we choose the interval from $3000\ M$ to $4000\ M$. Measures of the means and standard deviations of these quantities, as well as $\eta$ defined in \eqref{eq:eta}, over these intervals are presented in Table~\ref{tab:mad_accretion}. Though a simplistic method, analyzing standard deviations has the benefit of not being affected by the sampling rate in time at which we dump data.

\begin{table}
  \centering
  \caption{Means and standard deviations of accretion values. \label{tab:mad_accretion}}
  \begin{tabular}{ccccc}
    \toprule
    Level & $\dot{M}$ & $\dot{E}$ & $\phi$ & $\eta$ \\
    \midrule
    0 & $\phn9.2\pm2.7$ & $-40\pm26$ & $49\phd\phn\pm12\phd\phn$ & $5.4\phn\pm3.0\phn$ \\
    1 & $\phn9.4\pm3.2$ & $-40\pm17$ & $63.9\pm\phn8.5$          & $5.3\phn\pm1.9\phn$ \\
    2 & $10.0\pm5.6$    & $-62\pm34$ & $66\phd\phn\pm17\phd\phn$ & $7.2\phn\pm3.8\phn$ \\
    3 & $14.0\pm5.3$    & $-71\pm11$ & $62.8\pm\phn3.2$          & $6.10\pm0.68$       \\
    \bottomrule
  \end{tabular}
\end{table}

The relative steadiness of the level $0$ lines reflect the lack of turbulence at low resolution. We note that this resolution undergoes a qualitative jump in behavior at $t \approx 17{,}000\ M$. At this time a single large-amplitude spiral wave appears in the disk, introducing some degree of variability. With regard to $\dot{M}$ this variability is on longer timescales than seen at the other resolutions.

The level $1$ and level $2$ runs display similar accretion behavior, with $\dot{M} \approx 10$ and $\phi \approx 65$. There is around $50\%$ more energy outflow at level $2$ compared to level $1$. In all three quantities level $2$ displays approximately twice the variability (measured as the standard deviation of the values). The level $3$ data behaves very similarly to the level $2$ data over the limited time in which the former was generated.

One way in which the level $1$ and level $2$ runs differ is in the time to the onset of turbulence. The higher resolution simulation shows variability in $\dot{M}$ earlier, by $t = 1500\ M$, whereas at lower resolution there is a buildup period until $t = 5000\ M$. In some sense the first $16{,}000\ M$ time of the level $0$ resolution simulation is this same buildup phase.

Another difference is in the strength of short-term variability. The level $2$ run shows large variations between adjacent samples in $\dot{M}$. In all cases the sampling rate is fixed at $\Delta t = 10\ M$. This short-time variance is not unexpected at higher resolutions, as the finer grid allows the turbulence to cascade to smaller length scales and thus shorter characteristic times.

The fact that the level $3$ data has the same $\dot{M}$ variability as the level $2$ data (see Table~\ref{tab:mad_accretion}) may indicate a saturation in the level of turbulence. We caution however that the high-resolution simulation should be run for longer to verify this claim.

\begin{figure}
  \centering
  \includegraphics[width=\textwidth]{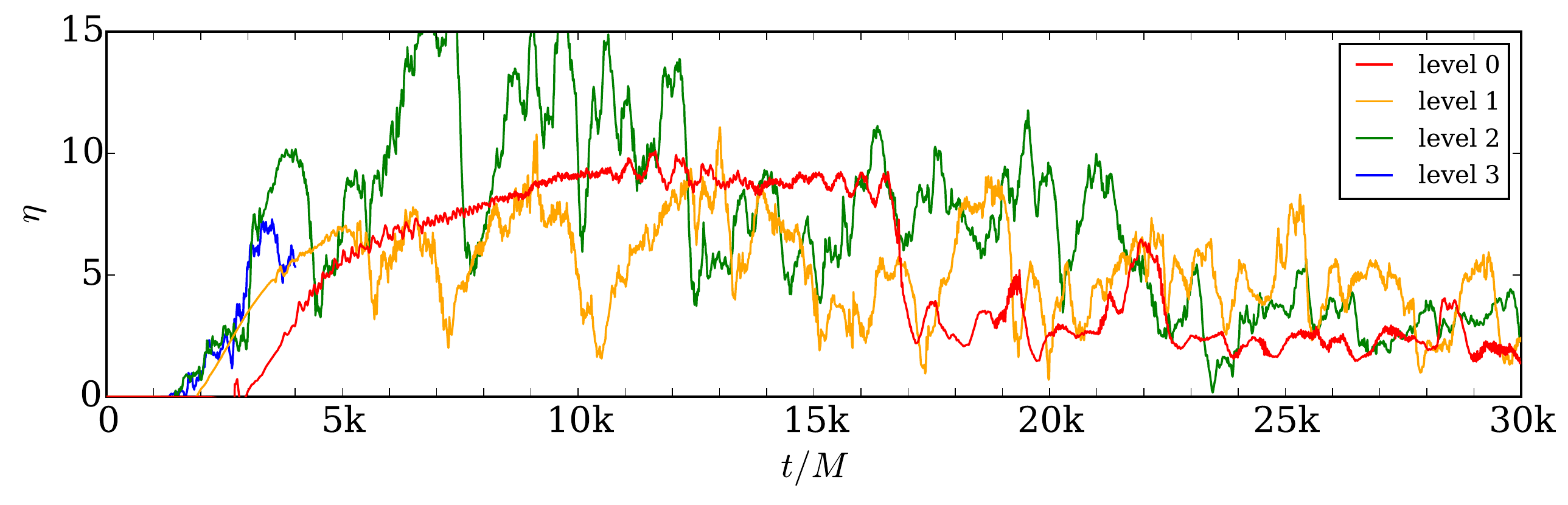}
  \caption{Efficiency parameter $\eta$ in the MAD simulations as a function of time for four resolutions. All simulations are sampled at a rate of $\Delta t = 10\ M$. \label{fig:mad_efficiency}}
\end{figure}

Finally, we return to the consideration of energy extraction, which is one of the central motivations for the study of MAD accretion onto black holes. As in \citet{Tchekhovskoy2011} we define the efficiency
\begin{equation} \label{eq:eta}
  \eta = \frac{\dot{M}-\dot{E}}{\ave{\dot{M}}}.
\end{equation}
Values greater than unity indicate more energy is being released than can be accounted for by the rest mass of the infalling material. The run of this efficiency in the various simulations is shown in Figure~\ref{fig:mad_efficiency}, with averages presented in the last column of Table~\ref{tab:mad_accretion}.

We find notably larger efficiencies than \citeauthor{Tchekhovskoy2011}, who record values of approximately $1.4$ for their run with similar parameters. To the extent that our mass accretion rates are similar to theirs, this indicates we are seeing significantly more energy outflow. Note that had we chosen a more interior $r_0$ to evaluate $\dot{E}$ (see Figure~\ref{fig:mad_fluxes}) we would have calculated lower values of $\eta$. In fact, with $r_0 = 2$ we would have found a net inflow of energy and $\eta < 0$.

There is good evidence that our accretion disks proper have converged with higher resolution -- if not in all the details then at least in important characteristics like magnetic flux and field geometry. However it is not clear that quantitative behavior of the near-vacuum outflow region is entirely free of numerical effects. This is not unexpected, given the difficulty finite-volume codes have when there is negligible matter. This is especially the case then that matter is highly magnetized, moving relativistically, and in a strongly curved region of spacetime.


\chapter{Conclusion}
\label{chap:conclusion}

Here we offer some closing remarks on the preceding material. We will consider the broader implications of our investigation into magnetically arrested disk (MAD) accretion, followed by anticipations of further applications of Athena++. We will also mention code developments slated for the future.

\section{Implications of MAD Studies}
\label{sec:conclusion:mad}

An important question about MAD accretion is how ubiquitous it is. MAD has been seen in several simulations, but it has heretofore been studied by a small number of groups using variants on the same code. We seek to understand whether the MAD scenario arises in our own simulations. As part of this we vary both resolution and numerical implementation details in order to ensure we are making statements about nature and not just about our code.

The key features of MAD as applied to black holes are the buildup of ordered poloidal magnetic field, the suppression of accretion in the inner regions, and the extraction of energy from the black hole spin. We do see this poloidal field accumulate, especially at higher resolutions (Figures~\ref{fig:mad_slices_final_sigma} and~\ref{fig:mad_slices_high_sigma}). That is, although the turbulence in the disk would seem to disorder the field, there is a net flux (by design of the initial conditions) that cannot help but accumulate. Once it is in the near-vacuum region, the field dominates the dynamics and it is no longer susceptible to being distorted by the fluid.

More quantitatively, consider the steady-state flux $\Phi$ threading spheres of various radii, as shown in Figure~\ref{fig:mad_fluxes}. The sign of the slope of the curves indicates the field lines near the black hole are not closing back at small radii. If they were, small spheres would be threaded by field loops that remained interior to large spheres, and we would see $\Phi$ increasing inward. The fact that the slope is shallow indicates that whatever field threads a sphere at large radius also threads spheres at smaller radius. That is, the magnetic field is brought all the way to the horizon. This is consistent with the MAD scenario.

MAD then proposes that the accumulated field will prevent matter from freely accreting. Indeed we do see highly magnetized regions very close to the horizon in the top panels of Figures~\ref{fig:mad_slices_high_zoom_beta} and~\ref{fig:mad_slices_high_zoom_sigma}. In the lower panels however one can clearly see some matter streaming in. On the other hand there is a clear sign of disruption at the highest resolution in the upper right panels of Figures~\ref{fig:mad_slices_high_rho} and~\ref{fig:mad_slices_high_zoom_rho}. A future investigation might focus on measuring accretion rates as a function of field geometry -- would the inflow be faster and more continuous if not for the accumulated net flux?

The energy extraction issue is more uncertain. We do see values of $\eta$ greater than $100\%$ (Figure~\ref{fig:mad_efficiency}), but these are values large enough to be suspicious. Indeed they depend highly upon the measured radius $r_0$ chosen (cf.\ Figure~\ref{fig:mad_fluxes}, especially the middle panel).

The large dynamical range of densities, magnetizations, and velocities found in simulations that include both an accretion disk and a jet presents a great challenge for a magnetohydrodynamics (MHD) code. Even to the extent that MHD is applicable across the domain, numerical issues have the potential to alter the solution. As already discussed, the need for floors on density and gas pressure, as well as the ceiling on Lorentz factors, reduces the reliability of simulations in the very jet region we may wish to study. Improvements to robustness (see \S\ref{sec:conclusion:code:improve}) can help to secure results in this area.

\section{Future Scientific Investigations}
\label{sec:conclusion:science}

Development of a new general-relativistic MHD (GRMHD) code would be a high price indeed if it were only to be applied to one particular accretion problem. Fortunately there are many problems that can be attacked with the code we have developed.

We pause to consider the comparative advantages offered by Athena++. It is fast and scalable; thus it can run models for long times covering large domains. Algorithmically it has the capability of minimizing diffusion with advanced Riemann solvers, as well as evolving the magnetic field in a natural divergence-free manner. Moreover it can accurately handle the polar singularity in spherical coordinate systems. Finally, Athena++ can also employ adaptive mesh refinement (AMR).

One potential application is the study of disks tilted relative to their black hole host spins, including the associated alignment, precession, and outflows. The vast majority of GRMHD simulations to date are done with aligned disks, to the extent that the black holes are even modeled with spin. Tilted disks challenge a number of codes by demanding resolution off the equatorial plane. This can easily be achieved with AMR without requiring large swaths of vacuum to be refined. Moreover, as momentum transport and thus alignment can be affected by standing shocks in the disk, it pays to have a code with minimal diffusion in order to get accurate results.

One of the most prominent observational features of black hole accretion is the powerful jet that can be launched. However, there is still something of a disconnect between the GRMHD simulations, focusing on accretion disk physics, and the large-scale jet propagation simulations, focusing on special relativity and emission from the jet. As a result, despite our understanding of jets themselves, we have no clear answer to the question of what accretion configurations lead to relativistic jets. A systematic study focusing on the parameter space covering plausible near-horizon conditions can yield insights into this connection. Key to such a study is the use of a coordinate system that does not contain a singularity along the very object of study, and thus our code is a prime candidate.

One last application we will mention is the relatively young field of tidal disruption events. The idea is that a star passing too close to a supermassive black hole will be tidally shredded, leading to a short-term burst of intense accretion. This scenario is believed to explain transient phenomena observed in active galactic nuclei. Again, we find an application that demands the range of scales accessible only with mesh refinement and a code that scales well computationally.

\section{Roadmap for Future Code Development}
\label{sec:conclusion:code}

The modular design of Athena++ allows it to support new components on top of the existing infrastructure. Here we briefly summarize some of the exciting prospects for the future of the code, both numerical and physical. Note that these need not be implemented by any one person, and indeed a number of them are applicable beyond the scope of GRMHD and so are likely to be included as a result of interest in other fields.

\subsection{Algorithmic Improvements and Extensions}
\label{sec:conclusion:code:improve}

A unique feature of Athena++ is its ability to run with a number of different Riemann solvers. This allows for a fair comparison between different solvers where all else is kept the same. While some comparisons were presented in \S\ref{sec:test:linear}, \S\ref{sec:test:shocks}, and \S\ref{sec:test:blast}, these are contrived examples rather than realistic turbulent flows. An easy future project then is to quantify the net benefit (or lack thereof) obtained by using advanced Riemann solvers.

Part of the difficulty in using less diffusive methods is that sharp gradients can wreak havoc with variable inversion. Sudden changes of conserved values can make the initial guess used in inversion procedures (\S\ref{sec:method:algorithm:inversion}) too far from the new solution to converge. More robust methods, such as possibly that of \citet{Newman2014}, should be explored.

All simulations discussed here are second order in space (for smooth enough flows away from coordinate singularities) and time. However the code is designed to be able to incorporate higher-order methods. These can potentially achieve a desired accuracy with less overall cost by reducing the number of cells needed. Higher-order spatial reconstruction is already implemented in some codes. There is the caveat briefly mentioned in \S\ref{sec:method:integral:discretized}, in that the source terms are only computed at second-order spatial accuracy. However even a method that is still formally second order often in practice stands to gain by having higher-order constituent parts where possible. For temporal accuracy there is no hindrance to going to higher orders in any scheme, like the van~Leer integrator, that uses partial forward-Euler substeps.

\subsection{New Physics}
\label{sec:conclusion:code:new}

The frontier of black hole accretion simulations is rapidly evolving, with ever more accurate physical models being incorporated. One area that has seen much recent development is that of fluid dynamics beyond ideal MHD. For example, \citet{Chandra2015} describes how to modify MHD in a way amenable to general relativity so as to incorporate weakly collisional effects such as anisotropic heat conduction guided by the magnetic field. Treating electrons separately from protons in the thermodynamic sense is explored in \citet{Ressler2015}. Without modifications such as these traditional GRMHD methods would be unsuitable for studying the extremely diffuse accretion in low-luminosity systems such as the supermassive black hole in our own galaxy.

At the other end of the spectrum of accretion rates, high-luminosity systems result in radiation being dynamically important. When optical depths are large across a single cell, radiation merely acts as a relativistically hot fluid and can be treated easily enough; when they are small across the entire domain the radiation streams out of the system, which can be evolved without considering the effects of the photons. The challenge is in the intermediate regime, where the equations of radiative transport must be incorporated at some level into the simulation. This is now being done in GRMHD with both closure schemes \citep{McKinney2014} and Monte Carlo methods \citep{Ryan2015}. By incorporating such a scheme, Athena++ can be used in new regions of the parameter space of accretion systems. In addition to the aforementioned methods, work is currently underway for implementing a method of short characteristics compatible with general relativity into the code.

The aforementioned physical models are examples of what can be incorporated directly into the code. At the same time, we note that simulations do not exist in a vacuum:\ work must be done to connect the results of simulations to observations. Future work with Athena++ should involve the use of post-processing methods such as the Monte Carlo radiative transport described in \citet{Dolence2009}. In this way we can make predictions more directly related to observations of electromagnetic emission from real accreting black holes.

\appendix
\renewcommand{\chaptername}{Appendix}

\chapter{Frame Transformation}
\label{chap:transformation}

Given a coordinate basis $\{\vec{\partial}_{(\mu)}\}$, we desire a new basis $\{\vec{e}_{(\hat{\mu})}\}$ satisfying the enumerated list of properties in \S\ref{sec:method:algorithm:transformation}, which we repeat here:
\begin{enumerate}
  \item \label{it:basis:orthogonal} $\vec{e}_{(\hat{\mu})}$ must be orthogonal to $\vec{e}_{(\hat{\nu})}$ for all $\mu \neq \nu$.
  \item \label{it:basis:normal} Each $\vec{e}_{(\hat{\mu})}$ must be normalized to have an inner product of $\pm1$ with itself, with $\vec{e}_{(\hat{t})}$ being timelike and $\vec{e}_{(\hat{\jmath})}$ being spacelike.
  \item \label{it:basis:time} $\vec{e}_{(\hat{t})}$ must be orthogonal to surfaces of constant $x^0$.
  \item \label{it:basis:projection} The projection of $\vec{e}_{(\hat{x})}$ onto the surface of constant $x^0$ is orthogonal to the surface of constant $x^i$ within that submanifold.
\end{enumerate}
As in the main text, we note that tensor components are indicated with unparenthesized subscripts and superscripts, with parentheses reserved for indices into generic sets. For concreteness we will work with an interface of constant $x^1$; results in the other directions follow from cyclic permutations $1 \mapsto 2 \mapsto 3$.

Property~\ref{it:basis:time} tells us
\begin{equation} \label{eq:e_t_low}
  e^{(\hat{t})}_\mu = A (1, 0, 0, 0)
\end{equation}
for some $A$, which is determined by property~\ref{it:basis:normal}:
\begin{equation}
  -1 = g^{\mu\nu} e^{(\hat{t})}_\mu e^{(\hat{t})}_\nu = A^2 g^{00}.
\end{equation}
Thus we have
\begin{equation}
  A = -(-g^{00})^{-1/2},
\end{equation}
where the sign is chosen such that the $\hat{t}$-direction is forward in time, as can be easily checked in the case $g_{\mu\nu} = \eta_{\mu\nu}$. Our first basis vector has components
\begin{equation} \label{eq:e_t}
  e_{(\hat{t})}^\mu = g^{\mu\nu} e^{(\hat{t})}_\nu = A g^{0\mu},
\end{equation}
and this is nothing more than the unit timelike normal $\hat{n}$ used in $3{+}1$ foliations of spacetime.

Next we set the $\hat{x}$ direction. It might be tempting to simply have it be parallel to $\vec{\pp}_{(1)}$, since this vector is already orthogonal to $\vec{e}_{(\hat{t})}$ by construction of the latter. However, this complicates the transformation of fluxes back to the global frame, as it results in $\hat{y}$- and $\hat{z}$-fluxes entering into the formula for $x^1$-fluxes. It is a more serious error to make $\vec{e}_{(\hat{x})}$ orthogonal to the $3$-surface of constant $x^1$. Though this would lead to a very simple formula for the flux conversion (requiring no knowledge of conserved quantities inside the wavefan), this vector can only be orthogonal to $\vec{e}_{(\hat{t})}$ if $g^{01}$ vanishes.

Instead, we enforce property~\ref{it:basis:projection}. This amounts to saying the projection of $\vec{e}_{(\hat{x})}$ onto the constant-$x^0$ subspace must have components
\begin{equation} \label{eq:e_x_3}
  \tensor*[^{(3)}]{e}{^{(\hat{x})}_i} \propto (1, 0, 0).
\end{equation}
Raising indices with the appropriate $3$-metric components $\gamma^{ij} = g^{ij} - g^{0i}g^{0j}/g^{00}$ and noting $e_{(\hat{x})}^0 = 0$ (by property~\ref{it:basis:orthogonal} and \eqref{eq:e_t_low}), we have
\begin{equation} \label{eq:e_x}
  e_{(\hat{x})}^\mu = B (g^{01}g^{0\mu} - g^{00}g^{1\mu}).
\end{equation}
The normalization is again set by property~\ref{it:basis:normal}:
\begin{equation}
  1 = g_{\mu\nu} e_{(\hat{x})}^\mu e_{(\hat{x})}^\nu = B^2 g^{00} (g^{00}g^{11} - g^{01}g^{01}).
\end{equation}
Choosing the sign such that $e_{(\hat{x})}^\mu = (0, +1, 0, 0)$ if $g_{\mu\nu} = \eta_{\mu\nu}$, we take
\begin{equation} \label{eq:factor_b}
  B = \paren[\big]{g^{00} (g^{00}g^{11} - g^{01}g^{01})}^{-1/2}.
\end{equation}

Property~\ref{it:basis:orthogonal}, together with \eqref{eq:e_t_low}, forces
\begin{equation}
  e_{(\hat{y})}^0 = e_{(\hat{z})}^0 = 0;
\end{equation}
that is, $\vec{e}_{(\hat{y})}$ and $\vec{e}_{(\hat{z})}$ must lie in the constant-$x^0$ plane. From \eqref{eq:e_x_3} and property~\ref{it:basis:orthogonal} we have
\begin{equation}
  e_{(\hat{y})}^1 = e_{(\hat{z})}^1 = 0,
\end{equation}
which conveniently implies no $\hat{y}$- or $\hat{z}$-fluxes will be needed from the special-relativistic Riemann solver.

For slight convenience we choose $e_{(\hat{z})}^2$ to be $0$. This corresponds to choosing an orientation of $\vec{e}_{(\hat{y})}$ and $\vec{e}_{(\hat{z})}$ in the plane they must span. Then we have
\begin{equation} \label{eq:e_z}
  e_{(\hat{z})}^\mu = C (0, 0, 0, 1).
\end{equation}
From property~\ref{it:basis:normal} we know
\begin{equation}
  1 = g_{\mu\nu} e_{(\hat{z})}^\mu e_{(\hat{z})}^\nu = C^2 g_{33},
\end{equation}
so
\begin{equation}
  C = (g_{33})^{-1/2},
\end{equation}
where again we check that we have a sensible definition when $g_{\mu\nu} = \eta_{\mu\nu}$:\ $e_{(\hat{z})}^\mu = (0, 0, 0, +1)$.

Finally, we solve for the two nonzero components of $\vec{e}_{(\hat{y})}$. Orthogonality (property~\ref{it:basis:orthogonal}) with $\vec{e}_{(\hat{z})}$, together with normalization (property~\ref{it:basis:normal}), constrain the solution to be
\begin{equation} \label{eq:e_y}
  e_{(\hat{y})}^\mu = D (0, 0, g_{33}, -g_{23}),
\end{equation}
where
\begin{equation}
  D = \paren[\big]{g_{33} (g_{22}g_{33} - g_{23}g_{23})}^{-1/2}.
\end{equation}
Once again we check the sign choice:\ $e_{(\hat{y})}^\mu = (0, 0, +1, 0)$ if $g_{\mu\nu} = \eta_{\mu\nu}$.

Given the formulas \eqref{eq:e_t}, \eqref{eq:e_x}, \eqref{eq:e_z}, \eqref{eq:e_y}, it is a simple matter to assemble the transformation matrix \eqref{eq:to_global}:
\begin{equation}
  \tensor{M}{^\mu_{\hat{\nu}}} = e_{(\hat{\nu})}^\mu =
  \begin{pmatrix}
    A g^{00} & 0                               & 0                   & 0 \\
    A g^{01} & B (g^{01}g^{01} - g^{00}g^{11}) & 0                   & 0 \\
    A g^{02} & B (g^{01}g^{02} - g^{00}g^{12}) & \phantom{-}D g_{33} & 0 \\
    A g^{03} & B (g^{01}g^{03} - g^{00}g^{13}) & -D g_{23}           & C
  \end{pmatrix},
\end{equation}
with
\begin{subequations} \begin{align}
  A & = -(-g^{00})^{-1/2}, \\
  B & = \paren[\big]{g^{00} (g^{00}g^{11} - g^{01}g^{01})}^{-1/2}, \\
  C & = (g_{33})^{-1/2}, \\
  D & = \paren[\big]{g_{33} (g_{22}g_{33} - g_{23}g_{23})}^{-1/2}.
\end{align} \end{subequations}
The inverse transformation $\eqref{eq:to_local}$ is the inverse matrix
\begin{equation}
  \tensor{M}{^{\hat{\mu}}_\nu} =
  \begin{pmatrix}
    -A                     & 0                      & 0                     & 0   \\
    B g^{01}               & -B g^{00}              & 0                     & 0   \\
    B^2E g^{00} / D g_{33} & B^2F g^{00} / D g_{33} & 1 / D g_{33}          & 0   \\
    I                      & J                      & (1/C) g_{23} / g_{33} & 1/C
  \end{pmatrix},
\end{equation}
with
\begin{subequations} \begin{align}
  E & = g^{01} g^{12} - g^{11} g^{02}, \\
  F & = g^{01} g^{02} - g^{00} g^{12}, \\
  G & = g^{01} g^{13} - g^{11} g^{03}, \\
  H & = g^{01} g^{03} - g^{00} g^{13}, \\
  I & = \frac{B^2g^{00}}{C} \paren[\bigg]{G + \frac{Eg_{23}}{g_{33}}}, \\
  J & = \frac{B^2g^{00}}{C} \paren[\bigg]{H + \frac{Fg_{23}}{g_{33}}},
\end{align} \end{subequations}
Both matrices are lower-diagonal, mostly as a result of our enumerated properties but also thanks to our choice of rotation $e_{(\hat{z})}^2 = 0$.


\chapter[Wavespeeds in SRMHD]{Extremal Wavespeeds in Special-Relativistic Magnetohydrodynamics}
\label{chap:wavespeeds}

Our Riemann solvers all require knowing extremal leftgoing and rightgoing signal speeds. If information propagates faster than the values we use, numerical instability ensues. On the other hand, if we overestimate these speeds (for example by always using the speed of light) we introduce unnecessary numerical dissipation.

In the case of magnetohydrodynamics, the speeds to use are the fast magnetosonic wavespeeds. As these are nontrivial to find, we detail here the method in the case of special relativity. Note that when using frame transformations in general relativity we will be computing wavespeeds in the transformed system, so the following procedure still applies.

Our procedure is based on the equations given in \citet{Mignone2006} for the magnetosonic speeds. We consider only a constant-$\Gamma$ equation of state. Let $x$ denote the direction in which we are interested in calculating propagation speeds (e.g.\ for interfaces of constant $x$). We assume the density $\rho$, gas pressure $\pgas$, $3$-velocity components $v^i$, magnetic field components $B^i$, contravariant magnetic field components $b^t$ and $b^x$, and magnetic pressure $\pmag$ are known. We can easily calculate the gas and total enthalpies,
\begin{subequations} \begin{align}
  \wgas & = \rho + \frac{\Gamma}{\Gamma-1} \pgas, \\
  \wtot & = \wgas + 2\pmag,
\end{align} \end{subequations}
as well as the square of the sound speed,
\begin{equation}
  \cs^2 = \frac{\Gamma\pgas}{\wgas}
\end{equation}
\citep{Mignone2005}, and the square of the Lorentz factor,
\begin{equation}
  \gamma^2 = \frac{1}{1-v_iv^i}.
\end{equation}

Before solving the general case, we consider two degenerate limits. In all that follows, explicit square roots are understood to return $0$ if their arguments are negative.

\section{Vanishing Velocity}
\label{sec:wavespeeds:velocity}

If there is no velocity, then the magnetosonic speeds will be symmetrically distributed around $0$. To be precise, if $v_i v^i < 10^{-12}$, we apply \extref{57} from \citet{Mignone2006}. Rewritten, this becomes
\begin{equation}
  \lambda^4 + a_2 \lambda^2 + a_0 = 0,
\end{equation}
where we define
\begin{subequations} \begin{align}
  a_2 & = -\frac{1}{\wtot} \paren[\big]{2\pmag + \cs^2 (\wgas + B_x^2)}, \\
  a_0 & = \frac{\cs^2B_x^2}{\wtot}.
\end{align} \end{subequations}
We can then calculate the wavespeeds of interest according to
\begin{align}
  s^2 & = a_2^2 - 4a_0, \\
  s & = \sqrt{s^2}, \\
  \lambda_\pm & = \pm\sqrt{\frac{s-a_2}{2}}.
\end{align}

\section{Vanishing Longitudinal Magnetic Field}
\label{sec:wavespeeds:field}

If the fluid velocity does not vanish but the longitudinal magnetic field effectively does ($\abs{B^x} < 10^{-7}$ in our code), then \extref{58} from \citeauthor{Mignone2006}\ applies. Like them, we define
\begin{equation}
  Q = 2\pmag - \cs^2 (v^y B^y + v^z B^z)^2.
\end{equation}
Then the fast magnetosonic speeds satisfy
\begin{equation}
  \lambda^2 + a_1 \lambda + a_0 = 0,
\end{equation}
where we have
\begin{subequations} \begin{align}
  a_1 & = -\frac{2\wgas\gamma^2v^x(1-\cs^2)}{\wgas(\cs^2+\gamma^2(1-\cs^2))+Q}, \\
  a_0 & = \frac{\wgas(\gamma^2v_x^2(1-\cs^2)-\cs^2)-Q}{\wgas(\cs^2+\gamma^2(1-\cs^2))+Q}.
\end{align} \end{subequations}
Selectively applying the alternate form of the quadratic formula to avoid cancellation errors \citep{NR}, we find
\begin{align}
  s^2 & = a_1^2 - 4a_0; \\
  s & = \sqrt{s^2}; \\
  \lambda_\pm & =
  \begin{cases}
    \displaystyle -\frac{2a_0}{a_1\pm s}, \\
    \displaystyle -\frac{1}{2} (a_1 \mp s);
  \end{cases}
\end{align}
where the first case is taken if $s^2 >= 0$ and either $a_1 >= 0$ ($+$ case) or $a_1 < 0$ ($-$ case).

\section{General Case}
\label{sec:wavespeeds:general}

If neither of the former two cases apply, we are forced to solve the quartic equation \extref{56} from \citet{Mignone2006}. Here we detail the method used to solve this equation, which has the advantage of using only real arithmetic. This particular procedure is justified on the physical grounds that the roots of our equation are magnetosonic speeds and so must be real.

Begin by rewriting the equation as
\begin{equation}
  \lambda^4 + a_3 \lambda^3 + a_2 \lambda^2 + a_1 \lambda + a_0 = 0,
\end{equation}
with
\begin{subequations} \begin{align}
  a_3 & = -\frac{1}{C} \paren[\big]{(4A + 2B) v^x - 2\cs^2 b^t b^x}, \\
  a_2 & = \frac{1}{C} \paren[\big]{6A v_x^2 + B (v_x^2 - 1) + \cs^2 (b_t^2 - b_x^2)}, \\
  a_1 & = -\frac{1}{C} (4A v_x^3 - 2B v^x + 2\cs^2 b^t b^x), \\
  a_0 & = \frac{1}{C} (A v_x^4 - B v_x^2 + \cs^2 b_x^2).
\end{align} \end{subequations}
Here we define
\begin{subequations} \begin{align}
  A & = \gamma^4 \wgas (1 - \cs^2), \\
  B & = \gamma^2 (2\pmag + \wgas \cs^2), \\
  C & = A + B - \cs^2 b_t^2.
\end{align} \end{subequations}

Next transform to the reduced quartic equation in $\lambda' = \lambda + a_3/4$:
\begin{equation} \label{eq:reduced_quartic}
  (\lambda')^4 + b_2 (\lambda')^2 + b_1 \lambda' + b_0 = 0,
\end{equation}
with
\begin{subequations} \begin{align}
  b_2 & = a_2 - \frac{3}{8} a_3^2, \\
  b_1 & = a_1 - \frac{1}{2} a_2 a_3 + \frac{1}{8} a_3^3, \\
  b_0 & = a_0 - \frac{1}{4} a_1 a_3 + \frac{1}{16} a_2 a_3^2 - \frac{3}{256} a_3^4.
\end{align} \end{subequations}

In order to write \eqref{eq:reduced_quartic} in terms of quadratic equations, we must find a real root $x$ to the resolvent cubic equation
\begin{equation}
  x^3 + c_2 x^2 + c_1 x + c_0 = 0,
\end{equation}
with
\begin{subequations} \begin{align}
  c_2 & = -b_2, \\
  c_1 & = -4b_0, \\
  c_0 & = 4b_0 b_2 - b_1^2.
\end{align} \end{subequations}
\Citet{NR} provide just such a method. Define the quantities
\begin{subequations} \begin{align}
  D & = \frac{1}{9} (c_2^2 - 3c_1), \\
  E & = \frac{1}{54} (2c_2^3 - 9c_1 c_2 + 27c_0), \\
  s^2 & = E^2 - D^3.
\end{align} \end{subequations}
If $s^2 < 0$, then our root $x_0$ is given by
\begin{align}
  \theta & = \cos^{-1}\paren[\bigg]{\frac{E}{\sqrt{D^3}}}, \\
  x_0 & = -2 \sqrt{D} \cos\paren[\bigg]{\frac{\theta}{3}} - \frac{1}{3} c_2.
\end{align}
Otherwise, we have
\begin{align}
  s & = \sqrt{s^2}, \\
  F & = -\sgn(E) \sqrt[3]{\abs{E}+s}, \\
  G & = \frac{D}{F}, \\
  x_0 & = F + G - \frac{1}{3} c_2.
\end{align}
where $G$ is taken to be $0$ if $F$ vanishes. These are essentially the same formula with different branch cut considerations.

Using the real cubic root $x_0$, we define quadratic coefficients
\begin{subequations} \begin{align}
  d_1 & = +\sqrt{x_0-b_2}, & e_1 & = -\sqrt{x_0-b_2}, \\
  d_0 & = \frac{1}{2} x_0 - \sgn(b_1) t, & e_0 & = \frac{1}{2} x_0 + \sgn(b_1) t,
\end{align} \end{subequations}
where
\begin{equation}
  t = \sqrt{\frac{1}{4}x_0^2-b_0}.
\end{equation}
The roots of \eqref{eq:reduced_quartic} are then given by
\begin{align}
  u^2 & = d_1^2 - 4d_0; & v^2 & = e_1^2 - 4e_0; \\
  u & = \sqrt{u^2}; & v & = \sqrt{v^2}; \\
  \lambda'_1 & =
  \begin{cases}
    \displaystyle -\frac{2d_0}{d_1-u}, \\
    \displaystyle -\frac{1}{2} (d_1 + u);
  \end{cases}
  & \lambda'_3 & =
  \begin{cases}
    \displaystyle -\frac{2e_0}{e_1-v}, \\
    \displaystyle -\frac{1}{2} (e_1 + v);
  \end{cases} \\
  \lambda'_2 & =
  \begin{cases}
    \displaystyle -\frac{2d_0}{d_1+u}, \\
    \displaystyle -\frac{1}{2} (d_1 - u);
  \end{cases}
  & \lambda'_4 & =
  \begin{cases}
    \displaystyle -\frac{2e_0}{e_1+v}, \\
    \displaystyle -\frac{1}{2} (e_1 - v).
  \end{cases}
\end{align}
In the above equations the first case is taken if the linear coefficient ($d_1$ or $e_1$ as appropriate) is nonnegative.

Finally, we take the fast magnetosonic speeds of interest to be
\begin{subequations} \begin{align}
  \lambda_- & = \max\set[\bigg]{\min\{\lambda'_1, \lambda'_3\} - \frac{1}{4} a_3, -1}, \\
  \lambda_+ & = \min\set[\bigg]{\max\{\lambda'_2, \lambda'_4\} - \frac{1}{4} a_3, +1}.
\end{align} \end{subequations}

We note that a very similar procedure can be used for variable inversion in the case of special-relativistic hydrodynamics, as this is also given by a quartic equation, \eqref{eq:quartic_inversion}.


\chapter{Coordinate System Examples}
\label{chap:examples}

Here we explicitly provide the core formulas for several metrics. For each metric we are interested in three sets of quantities:
\begin{itemize}
  \item There are the fundamental, pointwise-defined geometric quantities $g_{\mu\nu}$, $g^{\mu\nu}$, $\sqrt{-g}$, and $\Gamma^\sigma_{\mu\nu}$, recorded here in their entirety.
  \item Next there are the finite-volume quantities $\Delta V$, $\Delta A$, $\Delta L$, and $\Delta x^i$ defined in \eqref{eq:volume}, \eqref{eq:area}, \eqref{eq:length}, and \eqref{eq:width}, as well as the volume-averaged midpoints $\bar{x}^i$ at which various quantities are defined.
  \item Finally, there are the transformation matrices with components $\tensor{M}{^{\hat{\mu}}_\nu}$ and $\tensor{M}{^\mu_{\hat{\nu}}}$. There is a pair of such matrices for each of the three types of interfaces, with the formulas for the $x^1$-interface versions given in \eqref{eq:to_local} and \eqref{eq:to_global}.
\end{itemize}

The $i,j,k$ index of cells is understood if not specified. Interfaces and edges will be indexed with $i_+$ between $i$ and $i+1$ and $i_-$ between $i-1$ and $i$ as appropriate, and likewise for $j$ and $k$. Coordinates will be subscripted with $-$ or $+$ when they refer to interface values.

\section{Minkowski Spacetime}
\label{sec:examples:minkowski}

The simplest spacetime we can have is the flat Minkowski spacetime. Several coordinate systems are given.

\subsection{Minkowski Coordinates}
\label{sec:examples:minkowski:minkowski}

First we consider the Minkowski (Cartesian) coordinates of special relativity:\ $x^\mu = (t,x,y,z)$. The metric and its inverse are
\begin{equation}
  g_{\mu\nu} = g^{\mu\nu} =
  \begin{pmatrix}
    -1 & 0 & 0 & 0 \\
    0  & 1 & 0 & 0 \\
    0  & 0 & 1 & 0 \\
    0  & 0 & 0 & 1
  \end{pmatrix},
\end{equation}
and we have
\begin{equation}
  \sqrt{-g} = 1.
\end{equation}
The connection coefficients all vanish:
\begin{equation}
  \Gamma^\sigma_{\mu\nu} = 0.
\end{equation}

The volumes are given by
\begin{equation}
  \Delta V = (x_+ - x_-) (y_+ - y_-) (z_+ - z_-),
\end{equation}
the areas by
\begin{subequations} \begin{align}
  \Delta A_{i_-,j,k} & = (y_+ - y_-) (z_+ - z_-), \\
  \Delta A_{i,j_-,k} & = (x_+ - x_-) (z_+ - z_-), \\
  \Delta A_{i,j,k_-} & = (x_+ - x_-) (y_+ - y_-),
\end{align} \end{subequations}
the lengths by
\begin{subequations} \begin{align}
  \Delta L_{i,j_-,k_-} & = x_+ - x_i, \\
  \Delta L_{i_-,j,k_-} & = y_+ - y_i, \\
  \Delta L_{i_-,j_-,k} & = z_+ - z_i,
\end{align} \end{subequations}
and the widths by
\begin{subequations} \begin{align}
  \Delta W_x & = x_+ - x_-, \\
  \Delta W_y & = y_+ - y_-, \\
  \Delta W_z & = z_+ - z_-.
\end{align} \end{subequations}
The volume-averaged midpoints are
\begin{subequations} \begin{align}
  \bar{x} & = \frac{1}{2} (x_- + x_+), \\
  \bar{y} & = \frac{1}{2} (y_- + y_+), \\
  \bar{z} & = \frac{1}{2} (z_- + z_+).
\end{align} \end{subequations}

The transformation matrices are very simple:
\begin{subequations} \begin{align}
  \paren[\big]{\tensor{M}{^{\hat{\mu}}_\nu}}_{i_-,j,k} & =
  \begin{pmatrix}
    1 & 0 & 0 & 0 \\
    0 & 1 & 0 & 0 \\
    0 & 0 & 1 & 0 \\
    0 & 0 & 0 & 1
  \end{pmatrix},
  & \paren[\big]{\tensor{M}{^\mu_{\hat{\nu}}}}_{i_-,j,k} & =
  \begin{pmatrix}
    1 & 0 & 0 & 0 \\
    0 & 1 & 0 & 0 \\
    0 & 0 & 1 & 0 \\
    0 & 0 & 0 & 1
  \end{pmatrix}, \\
  \paren[\big]{\tensor{M}{^{\hat{\mu}}_\nu}}_{i,j_-,k} & =
  \begin{pmatrix}
    1 & 0 & 0 & 0 \\
    0 & 0 & 1 & 0 \\
    0 & 0 & 0 & 1 \\
    0 & 1 & 0 & 0
  \end{pmatrix},
  & \paren[\big]{\tensor{M}{^\mu_{\hat{\nu}}}}_{i,j_-,k} & =
  \begin{pmatrix}
    1 & 0 & 0 & 0 \\
    0 & 0 & 0 & 1 \\
    0 & 1 & 0 & 0 \\
    0 & 0 & 1 & 0
  \end{pmatrix}, \\
  \paren[\big]{\tensor{M}{^{\hat{\mu}}_\nu}}_{i,j,k_-} & =
  \begin{pmatrix}
    1 & 0 & 0 & 0 \\
    0 & 0 & 0 & 1 \\
    0 & 1 & 0 & 0 \\
    0 & 0 & 1 & 0
  \end{pmatrix},
  & \paren[\big]{\tensor{M}{^\mu_{\hat{\nu}}}}_{i,j,k_-} & =
  \begin{pmatrix}
    1 & 0 & 0 & 0 \\
    0 & 0 & 1 & 0 \\
    0 & 0 & 0 & 1 \\
    0 & 1 & 0 & 0
  \end{pmatrix}.
\end{align} \end{subequations}

\subsection{Spherical Coordinates}
\label{sec:examples:minkowski:spherical}

Next we consider flat spacetime but recast in terms of spherical spatial variables $r$, $\theta$, and $\phi$. The metric and its inverse are diagonal:
\begin{align}
  g_{\mu\nu} & =
  \begin{pmatrix}
    -1 & 0 & 0   & 0                  \\
    0  & 1 & 0   & 0                  \\
    0  & 0 & r^2 & 0                  \\
    0  & 0 & 0   & r^2 \sin^2\!\theta
  \end{pmatrix},
  & g^{\mu\nu} & =
  \begin{pmatrix}
    -1 & 0 & 0     & 0                  \\
    0  & 1 & 0     & 0                  \\
    0  & 0 & 1/r^2 & 0                  \\
    0  & 0 & 0     & \csc^2\!\theta/r^2
  \end{pmatrix}.
\end{align}
We have the factor
\begin{equation}
  \sqrt{-g} = r^2 \abs{\sin\theta}.
\end{equation}
The nonzero connection coefficients are
\begin{subequations} \begin{align}
  \Gamma^1_{22} & = -r, \\
  \Gamma^1_{33} & = -r \sin^2\!\theta, \\
  \Gamma^2_{12} = \Gamma^2_{21} & = \frac{1}{r}, \\
  \Gamma^2_{33} & = -\sin\theta \cos\theta, \\
  \Gamma^3_{13} = \Gamma^3_{31} & = \frac{1}{r}, \\
  \Gamma^3_{23} = \Gamma^3_{32} & = \cot\theta.
\end{align} \end{subequations}

The volumes are given by
\begin{equation}
  \Delta V = \frac{1}{3} (r_+^3 - r_-^3) \abs{\cos\theta_--\cos\theta_+} (\phi_+ - \phi_-),
\end{equation}
the areas by
\begin{subequations} \begin{align}
  \Delta A_{i_-,j,k} & = r_-^2 \abs{\cos\theta_--\cos\theta_+} (\phi_+ - \phi_-), \\
  \Delta A_{i,j_-,k} & = \frac{1}{3} (r_+^3 - r_-^3) \abs{\sin\theta_-} (\phi_+ - \phi_-), \\
  \Delta A_{i,j,k_-} & = \frac{1}{3} (r_+^3 - r_-^3) \abs{\cos\theta_--\cos\theta_+},
\end{align} \end{subequations}
the lengths by
\begin{subequations} \begin{align}
  \Delta L_{i,j_-,k_-} & = \frac{1}{3} (r_+^3 - r_-^3) \abs{\sin\theta_-}, \\
  \Delta L_{i_-,j,k_-} & = r_-^2 \abs{\cos\theta_--\cos\theta_+}, \\
  \Delta L_{i_-,j_-,k} & = r_-^2 \abs{\sin\theta_-} (\phi_+ - \phi_-),
\end{align} \end{subequations}
and the widths by
\begin{subequations} \begin{align}
  \Delta W_r & = r_+ - r_-, \\
  \Delta W_\theta & = \bar{r} (\theta_+ - \theta_-), \\
  \Delta W_\phi & = \bar{r} \abs{\sin\bar{\theta}} (\phi_+ - \phi_-).
\end{align} \end{subequations}
The volume-averaged midpoints are
\begin{subequations} \begin{align}
  \bar{r} & = \paren[\bigg]{\frac{r_-^3+r_+^3}{2}}^{\!1/3}, \\
  \bar{\theta} & = \cos^{-1}\paren[\bigg]{\frac{\cos\theta_-+\cos\theta_+}{2}}, \\
  \bar{\phi} & = \frac{1}{2} (\phi_- + \phi_+).
\end{align} \end{subequations}

The transformation matrices are as follows:
\begin{subequations} \begin{align}
  \paren[\big]{\tensor{M}{^{\hat{\mu}}_\nu}}_{i_-,j,k} & =
  \begin{pmatrix}
    1 & 0 & 0 & 0                  \\
    0 & 1 & 0 & 0                  \\
    0 & 0 & r & 0                  \\
    0 & 0 & 0 & r \abs{\sin\theta}
  \end{pmatrix},
  & \paren[\big]{\tensor{M}{^\mu_{\hat{\nu}}}}_{i_-,j,k} & =
  \begin{pmatrix}
    1 & 0 & 0   & 0                  \\
    0 & 1 & 0   & 0                  \\
    0 & 0 & 1/r & 0                  \\
    0 & 0 & 0   & \abs{\csc\theta}/r
  \end{pmatrix}, \\
  \paren[\big]{\tensor{M}{^{\hat{\mu}}_\nu}}_{i,j_-,k} & =
  \begin{pmatrix}
    1 & 0 & 0 & 0                  \\
    0 & 0 & r & 0                  \\
    0 & 0 & 0 & r \abs{\sin\theta} \\
    0 & 1 & 0 & 0
  \end{pmatrix},
  & \paren[\big]{\tensor{M}{^\mu_{\hat{\nu}}}}_{i,j_-,k} & =
  \begin{pmatrix}
    1 & 0   & 0                  & 0 \\
    0 & 0   & 0                  & 1 \\
    0 & 1/r & 0                  & 0 \\
    0 & 0   & \abs{\csc\theta}/r & 0
  \end{pmatrix}, \\
  \paren[\big]{\tensor{M}{^{\hat{\mu}}_\nu}}_{i,j,k_-} & =
  \begin{pmatrix}
    1 & 0 & 0 & 0                  \\
    0 & 0 & 0 & r \abs{\sin\theta} \\
    0 & 1 & 0 & 0                  \\
    0 & 0 & r & 0
  \end{pmatrix},
  & \paren[\big]{\tensor{M}{^\mu_{\hat{\nu}}}}_{i,j,k_-} & =
  \begin{pmatrix}
    1 & 0                  & 0 & 0   \\
    0 & 0                  & 1 & 0   \\
    0 & 0                  & 0 & 1/r \\
    0 & \abs{\csc\theta}/r & 0 & 0
  \end{pmatrix}.
\end{align} \end{subequations}

\subsection{Sinusoidal Coordinates}
\label{sec:examples:minkowski:sinusoidal}

A useful coordinate system for testing the correctness of the code is obtained from Minkowski coordinates $(t,x,y,z)$ according to
\begin{subequations} \begin{align}
  T & = t, \\
  X & = x, \\
  Y & = y + a \sin(kx), \\
  Z & = z,
\end{align} \end{subequations}
where $a$ and $k$ are parameters for the transformation. These ``snake'' coordinates $(T,X,Y,Z)$ have lines of constant $Y$ that are sinusoidal in the $xy$-plane. For these coordinates define
\begin{subequations} \begin{align}
  \alpha & = \sqrt{1+a^2k^2\cos^2(kX)}, \\
  \beta & = ak \cos(kX),
\end{align} \end{subequations}
where bars and subscripts applied to these variables are inherited by the $X$'s in the definition.

The metric and its inverse are given by
\begin{align}
  g_{\mu\nu} & =
  \begin{pmatrix}
    -1 & 0        & 0      & 0 \\
    0  & \alpha^2 & -\beta & 0 \\
    0  & -\beta   & 1      & 0 \\
    0  & 0        & 0      & 1
  \end{pmatrix},
  & g^{\mu\nu} & =
  \begin{pmatrix}
    -1 & 0     & 0        & 0 \\
    0  & 1     & \beta    & 0 \\
    0  & \beta & \alpha^2 & 0 \\
    0  & 0     & 0        & 1
  \end{pmatrix},
\end{align}
and we have
\begin{equation}
  \sqrt{-g} = 1.
\end{equation}
The only nonzero connection coefficient is
\begin{equation}
  \Gamma^2_{11} = ak^2 \sin(kX).
\end{equation}

The volumes are given by
\begin{equation}
  \Delta V = (X_+ - X_-) (Y_+ - Y_-) (Z_+ - Z_-),
\end{equation}
the areas by
\begin{subequations} \begin{align}
  \Delta A_{i_-,j,k} & = (Y_+ - Y_-) (Z_+ - Z_-), \\
  \Delta A_{i,j_-,k} & = (X_+ - X_-) (Z_+ - Z_-), \\
  \Delta A_{i,j,k_-} & = (X_+ - X_-) (Y_+ - Y_-),
\end{align} \end{subequations}
the lengths by
\begin{subequations} \begin{align}
  \Delta L_{i,j_-,k_-} & = X_+ - X_i, \\
  \Delta L_{i_-,j,k_-} & = Y_+ - Y_i, \\
  \Delta L_{i_-,j_-,k} & = Z_+ - Z_i,
\end{align} \end{subequations}
and the widths by
\begin{subequations} \begin{align}
  \Delta W_X & = \int_{X_-}^{X_+} \alpha \, \dd X \geq \int_{X_-}^{X_+} \dd X = X_+ - X_-, \\
  \Delta W_Y & = Y_+ - Y_-, \\
  \Delta W_Z & = Z_+ - Z_-.
\end{align} \end{subequations}
The bound on the elliptic integral can be used to conservatively limit the step sizes ($\Delta X$ being used for no other purpose). The volume-averaged midpoints are
\begin{subequations} \begin{align}
  \bar{X} & = \frac{1}{2} (X_- + X_+), \\
  \bar{Y} & = \frac{1}{2} (Y_- + Y_+), \\
  \bar{Z} & = \frac{1}{2} (Z_- + Z_+).
\end{align} \end{subequations}

The transformation matrices are given by
\begin{subequations} \begin{align}
  \paren[\big]{\tensor{M}{^{\hat{\mu}}_\nu}}_{i_-,j,k} & =
  \begin{pmatrix}
    1 & 0      & 0 & 0 \\
    0 & 1      & 0 & 0 \\
    0 & -\beta & 1 & 0 \\
    0 & 0      & 0 & 1
  \end{pmatrix},
  & \paren[\big]{\tensor{M}{^\mu_{\hat{\nu}}}}_{i_-,j,k} & =
  \begin{pmatrix}
    1 & 0     & 0 & 0 \\
    0 & 1     & 0 & 0 \\
    0 & \beta & 1 & 0 \\
    0 & 0     & 0 & 1
  \end{pmatrix}, \\
  \paren[\big]{\tensor{M}{^{\hat{\mu}}_\nu}}_{i,j_-,k} & =
  \begin{pmatrix}
    1 & 0      & 0             & 0 \\
    0 & 0      & 1/\alpha      & 0 \\
    0 & 0      & 0             & 1 \\
    0 & \alpha & -\beta/\alpha & 0
  \end{pmatrix},
  & \paren[\big]{\tensor{M}{^\mu_{\hat{\nu}}}}_{i,j_-,k} & =
  \begin{pmatrix}
    1 & 0            & 0 & 0        \\
    0 & \beta/\alpha & 0 & 1/\alpha \\
    0 & \alpha       & 0 & 0        \\
    0 & 0            & 1 & 0
  \end{pmatrix}, \\
  \paren[\big]{\tensor{M}{^{\hat{\mu}}_\nu}}_{i,j,k_-} & =
  \begin{pmatrix}
    1 & 0      & 0 & 0 \\
    0 & 0      & 0 & 1 \\
    0 & 1      & 0 & 0 \\
    0 & -\beta & 1 & 0
  \end{pmatrix},
  & \paren[\big]{\tensor{M}{^\mu_{\hat{\nu}}}}_{i,j,k_-} & =
  \begin{pmatrix}
    1 & 0 & 0     & 0 \\
    0 & 0 & 1     & 0 \\
    0 & 0 & \beta & 1 \\
    0 & 1 & 0     & 0
  \end{pmatrix}.
\end{align} \end{subequations}

\subsection{Tilted Coordinates}
\label{sec:examples:minkowski:tilted}

Just as sinusoidal coordinates can be used to test space--space terms in the metric, we use tilted coordinates to test time--space terms. For a shift parameterized by $a$, $\abs{a} < 1$, define
\begin{subequations} \begin{align}
  \alpha & = \sqrt{1+a^2}, \\
  \beta & = \sqrt{1-a^2}.
\end{align} \end{subequations}
Then we have the transformation from Minkowski coordinates $(t,x,y,z)$ according to
\begin{subequations} \begin{align}
  T & = \frac{1}{\alpha} (t + ax), \\
  X & = \frac{1}{\alpha} (x - at), \\
  Y & = y, \\
  Z & = z.
\end{align} \end{subequations}
Observers moving with $x$-velocity $a$ have worldlines in the $T$-direction, while surfaces of constant $T$ correspond to surfaces of constant time for observers with $x$-velocity $-a$.

The metric is its own inverse,
\begin{equation}
  g_{\mu\nu} = g^{\mu\nu} =
  \begin{pmatrix}
    -\beta^2/\alpha^2 & 2a/\alpha^2      & 0 & 0 \\
    2a/\alpha^2       & \beta^2/\alpha^2 & 0 & 0 \\
    0                 & 0                & 1 & 0 \\
    0                 & 0                & 0 & 1
  \end{pmatrix},
\end{equation}
and we have
\begin{equation}
  \sqrt{-g} = 1.
\end{equation}
The connection coefficients all vanish:
\begin{equation}
  \Gamma^\sigma_{\mu\nu} = 0.
\end{equation}

The volumes are given by
\begin{equation}
  \Delta V = (X_+ - X_-) (Y_+ - Y_-) (Z_+ - Z_-),
\end{equation}
the areas by
\begin{subequations} \begin{align}
  \Delta A_{i_-,j,k} & = (Y_+ - Y_-) (Z_+ - Z_-), \\
  \Delta A_{i,j_-,k} & = (X_+ - X_-) (Z_+ - Z_-), \\
  \Delta A_{i,j,k_-} & = (X_+ - X_-) (Y_+ - Y_-),
\end{align} \end{subequations}
the lengths by
\begin{subequations} \begin{align}
  \Delta L_{i,j_-,k_-} & = X_+ - X_i, \\
  \Delta L_{i_-,j,k_-} & = Y_+ - Y_i, \\
  \Delta L_{i_-,j_-,k} & = Z_+ - Z_i,
\end{align} \end{subequations}
and the widths by
\begin{subequations} \begin{align}
  \Delta W_X & = \frac{\beta}{\alpha} (X_+ - X_-), \\
  \Delta W_Y & = Y_+ - Y_-, \\
  \Delta W_Z & = Z_+ - Z_-.
\end{align} \end{subequations}
The volume-averaged midpoints are
\begin{subequations} \begin{align}
  \bar{X} & = \frac{1}{2} (X_- + X_+), \\
  \bar{Y} & = \frac{1}{2} (Y_- + Y_+), \\
  \bar{Z} & = \frac{1}{2} (Z_- + Z_+).
\end{align} \end{subequations}

The transformation matrices are given by
\begin{subequations} \begin{align}
  \paren[\big]{\tensor{M}{^{\hat{\mu}}_\nu}}_{i_-,j,k} & =
  \begin{pmatrix}
    \alpha/\beta   & 0            & 0 & 0 \\
    2a/\alpha\beta & \beta/\alpha & 0 & 0 \\
    0              & 0            & 1 & 0 \\
    0              & 0            & 0 & 1
  \end{pmatrix},
  & \paren[\big]{\tensor{M}{^\mu_{\hat{\nu}}}}_{i_-,j,k} & =
  \begin{pmatrix}
    \beta/\alpha    & 0            & 0 & 0 \\
    -2a/\alpha\beta & \alpha/\beta & 0 & 0 \\
    0               & 0            & 1 & 0 \\
    0               & 0            & 0 & 1
  \end{pmatrix}, \\
  \paren[\big]{\tensor{M}{^{\hat{\mu}}_\nu}}_{i,j_-,k} & =
  \begin{pmatrix}
    \alpha/\beta   & 0            & 0 & 0 \\
    0              & 0            & 1 & 0 \\
    0              & 0            & 0 & 1 \\
    2a/\alpha\beta & \beta/\alpha & 0 & 0
  \end{pmatrix},
  & \paren[\big]{\tensor{M}{^\mu_{\hat{\nu}}}}_{i,j_-,k} & =
  \begin{pmatrix}
    \beta/\alpha    & 0 & 0 & 0            \\
    -2a/\alpha\beta & 0 & 0 & \alpha/\beta \\
    0               & 1 & 0 & 0            \\
    0               & 0 & 1 & 0
  \end{pmatrix}, \\
  \paren[\big]{\tensor{M}{^{\hat{\mu}}_\nu}}_{i,j,k_-} & =
  \begin{pmatrix}
    \alpha/\beta   & 0            & 0 & 0 \\
    0              & 0            & 0 & 1 \\
    2a/\alpha\beta & \beta/\alpha & 0 & 0 \\
    0              & 0            & 1 & 0
  \end{pmatrix},
  & \paren[\big]{\tensor{M}{^\mu_{\hat{\nu}}}}_{i,j,k_-} & =
  \begin{pmatrix}
    \beta/\alpha    & 0            & 0 & 0 \\
    -2a/\alpha\beta & 0 & \alpha/\beta & 0 \\
    0               & 0            & 0 & 1 \\
    0               & 1            & 0 & 0
  \end{pmatrix}.
\end{align} \end{subequations}

\section{Schwarzschild Spacetime}
\label{sec:examples:schwarzschild}

Perhaps the simplest nontrivial metric used in practical applications is the Schwarz\-schild metric. The only parameter is the mass $M$, and we will use coordinates $x^\mu = (t,r,\theta,\phi)$. For convenience we introduce the lapse function
\begin{equation}
  \alpha = \sqrt{1-\frac{2M}{r}}.
\end{equation}

The metric and its inverse are
\begin{align}
  g_{\mu\nu} & =
  \begin{pmatrix}
    -\alpha^2 & 0          & 0   & 0                  \\
    0         & 1/\alpha^2 & 0   & 0                  \\
    0         & 0          & r^2 & 0                  \\
    0         & 0          & 0   & r^2 \sin^2\!\theta
  \end{pmatrix},
  & g^{\mu\nu} & =
  \begin{pmatrix}
    -1/\alpha^2 & 0        & 0     & 0                  \\
    0           & \alpha^2 & 0     & 0                  \\
    0           & 0        & 1/r^2 & 0                  \\
    0           & 0        & 0     & \csc^2\!\theta/r^2
  \end{pmatrix},
\end{align}
and we have
\begin{equation}
  \sqrt{-g} = r^2 \abs{\sin\theta}.
\end{equation}
The nonzero connection coefficients are
\begin{subequations} \begin{align}
  \Gamma^0_{01} = \Gamma^0_{10} & = \frac{M}{r^2\alpha^2}, \\
  \Gamma^1_{00} & = \frac{M\alpha^2}{r^2}, \\
  \Gamma^1_{11} & = -\frac{M}{r^2\alpha^2}, \\
  \Gamma^1_{22} & = -r\alpha^2, \\
  \Gamma^1_{33} & = -r\alpha^2 \sin^2\!\theta, \\
  \Gamma^2_{12} = \Gamma^2_{21} & = \frac{1}{r}, \\
  \Gamma^2_{33} & = -\sin\theta \cos\theta, \\
  \Gamma^3_{13} = \Gamma^3_{31} & = \frac{1}{r}, \\
  \Gamma^3_{23} = \Gamma^3_{32} & = \cot\theta.
\end{align} \end{subequations}

So long as a cell does not include the polar axis in its interior, the volumes are given by
\begin{equation}
  \Delta V = \frac{1}{3} (r_+^3 - r_-^3) \abs{\cos\theta_--\cos\theta_+} (\phi_+ - \phi_-),
\end{equation}
the areas by
\begin{subequations} \begin{align}
  \Delta A_{i_-,j,k} & = r_-^2 \abs{\cos\theta_--\cos\theta_+} (\phi_+ - \phi_-), \\
  \Delta A_{i,j_-,k} & = \frac{1}{3} (r_+^3 - r_-^3) \abs{\sin\theta_-} (\phi_+ - \phi_-), \\
  \Delta A_{i,j,k_-} & = \frac{1}{3} (r_+^3 - r_-^3) \abs{\cos\theta_--\cos\theta_+},
\end{align} \end{subequations}
and the lengths by
\begin{subequations} \begin{align}
  \Delta L_{i,j_-,k_-} & = \frac{1}{3} (r_+^3 - r_-^3) \abs{\sin\theta_-}, \\
  \Delta L_{i_-,j,k_-} & = r_-^2 \abs{\cos\theta_--\cos\theta_+}, \\
  \Delta L_{i_-,j_-,k} & = r_-^2 \abs{\sin\theta_-} (\phi_+ - \phi_-),
\end{align} \end{subequations}
just as in spherical coordinates in Minkowski spacetime. The widths are given by
\begin{subequations} \begin{align}
  \Delta W_r & = r_+\alpha_+ - r_-\alpha_- + M \log\paren[\bigg]{\frac{r_+(1+\alpha_+)-M}{r_-(1+\alpha_-)-M}}, \\
  \Delta W_\theta & = \bar{r} (\theta_+ - \theta_-), \\
  \Delta W_\phi & = \bar{r} \abs{\sin\bar{\theta}} (\phi_+ - \phi_-).
\end{align} \end{subequations}
The volume-averaged midpoints are
\begin{subequations} \begin{align}
  \bar{r} & = \paren[\bigg]{\frac{r_-^3+r_+^3}{2}}^{\!1/3}, \\
  \bar{\theta} & = \cos^{-1}\paren[\bigg]{\frac{\cos\theta_-+\cos\theta_+}{2}}, \\
  \bar{\phi} & = \frac{1}{2} (\phi_- + \phi_+),
\end{align} \end{subequations}
again just as in spherical coordinates.

The transformation matrices are very simple:
\begin{subequations} \begin{align}
  \paren[\big]{\tensor{M}{^{\hat{\mu}}_\nu}}_{i_-,j,k} & =
  \begin{pmatrix}
    \alpha & 0        & 0 & 0                  \\
    0      & 1/\alpha & 0 & 0                  \\
    0      & 0        & r & 0                  \\
    0      & 0        & 0 & r \abs{\sin\theta}
  \end{pmatrix},
  & \paren[\big]{\tensor{M}{^\mu_{\hat{\nu}}}}_{i_-,j,k} & =
  \begin{pmatrix}
    1/\alpha & 0      & 0   & 0                  \\
    0        & \alpha & 0   & 0                  \\
    0        & 0      & 1/r & 0                  \\
    0        & 0      & 0   & \abs{\csc\theta}/r
  \end{pmatrix}, \\
  \paren[\big]{\tensor{M}{^{\hat{\mu}}_\nu}}_{i,j_-,k} & =
  \begin{pmatrix}
    \alpha & 0        & 0 & 0                  \\
    0      & 0        & r & 0                  \\
    0      & 0        & 0 & r \abs{\sin\theta} \\
    0      & 1/\alpha & 0 & 0
  \end{pmatrix},
  & \paren[\big]{\tensor{M}{^\mu_{\hat{\nu}}}}_{i,j_-,k} & =
  \begin{pmatrix}
    1/\alpha & 0   & 0                  & 0      \\
    0        & 0   & 0                  & \alpha \\
    0        & 1/r & 0                  & 0      \\
    0        & 0   & \abs{\csc\theta}/r & 0
  \end{pmatrix}, \\
  \paren[\big]{\tensor{M}{^{\hat{\mu}}_\nu}}_{i,j,k_-} & =
  \begin{pmatrix}
    \alpha & 0        & 0 & 0                  \\
    0      & 0        & 0 & r \abs{\sin\theta} \\
    0      & 1/\alpha & 0 & 0                  \\
    0      & 0        & r & 0
  \end{pmatrix},
  & \paren[\big]{\tensor{M}{^\mu_{\hat{\nu}}}}_{i,j,k_-} & =
  \begin{pmatrix}
    1/\alpha & 0                  & 0      & 0   \\
    0        & 0                  & \alpha & 0   \\
    0        & 0                  & 0      & 1/r \\
    0        & \abs{\csc\theta}/r & 0      & 0
  \end{pmatrix}.
\end{align} \end{subequations}

\section{Kerr Spacetime}
\label{sec:examples:kerr}

Kerr spacetime applies to spinning black holes, and more generally Kerr--Newman spacetime allows black holes to carry charge as well. Because of the ubiquity and importance of this spacetime, we present several different forms for the metric, some of which may be better than others in numerical applications.

In all cases the black hole is parameterized by mass $M$, spin $a = J/M$, and charge $Q$. We employ the conventional shorthand symbols
\begin{align}
  \Delta & = r^2 - 2Mr + a^2 + Q^2, \\
  \Sigma & = r^2 + a^2 \cos^2\!\theta,
\end{align}
where $r$ and $\theta$ are the same radial and polar coordinates in all of the following examples. The timelike and azimuthal coordinates, however, may vary between coordinate systems.

In what follows we will always take $Q = 0$.

\subsection{Boyer--Lindquist Coordinates}
\label{sec:examples:kerr:boyer-lindquist}

In traditional Boyer--Lindquist coordinates $x^\mu = (t,r,\theta,\phi)$ the metric and its inverse are written
\begin{subequations} \begin{align}
  g_{\mu\nu} & =
  \begin{pmatrix}
    -(1-2Mr/\Sigma)               & 0             & 0      & -(2Mar/\Sigma) \sin^2\!\theta                          \\
    0                             & \Sigma/\Delta & 0      & 0                                                      \\
    0                             & 0             & \Sigma & 0                                                      \\
    -(2Mar/\Sigma) \sin^2\!\theta & 0             & 0      & (r^2+a^2+(2Ma^2r/\Sigma)\sin^2\!\theta) \sin^2\!\theta
  \end{pmatrix}, \\
  g^{\mu\nu} & =
  \begin{pmatrix}
    -1 - 2Mr(r^2+a^2)/\Delta\Sigma & 0             & 0        & -2Mar/\Delta\Sigma                      \\
    0                              & \Delta/\Sigma & 0        & 0                                       \\
    0                              & 0             & 1/\Sigma & 0                                       \\
    -2Mar/\Delta\Sigma             & 0             & 0        & (\Sigma-2Mr)/\Delta\Sigma\sin^2\!\theta
  \end{pmatrix}
\end{align} \end{subequations}
\citep[33.2]{MTW}. We have the factor
\begin{equation}
  \sqrt{-g} = \Sigma \abs{\sin\theta}.
\end{equation}

There are a large number of nonzero connection coefficients; we have the following formulas for them:
\begin{subequations} \begin{align}
  \Gamma^0_{01} = \Gamma^0_{10} & = \frac{1}{\Delta\rho^4} M (r^2 + a^2) (r^2 - a^2 \cos^2\!\theta), \\
  \Gamma^0_{02} = \Gamma^0_{20} & = -\frac{Ma^2r}{\rho^4} \sin(2\theta), \\
  \Gamma^0_{13} = \Gamma^0_{31} & = -\frac{Ma}{\Delta\rho^4} \paren[\big]{3r^4+a^2r^2(1+\cos^2\!\theta)-a^4\cos^2\!\theta} \sin^2\!\theta, \\
  \Gamma^0_{23} = \Gamma^0_{32} & = \frac{Ma^3r}{\rho^4} \sin(2\theta) \sin^2\!\theta, \\
  \Gamma^1_{00} & = \frac{M\Delta}{\rho^6} (r^2 - a^2 \cos^2\!\theta), \\
  \Gamma^1_{03} = \Gamma^1_{30} & = -\frac{Ma\Delta}{\rho^6} (r^2 - a^2 \cos^2\!\theta) \sin^2\!\theta, \\
  \Gamma^1_{11} & = \frac{r}{\Sigma} - \frac{r-M}{\Delta}, \\
  \Gamma^1_{12} = \Gamma^1_{21} & = -\frac{a^2}{2\Sigma} \sin(2\theta), \\
  \Gamma^1_{22} & = -\frac{\Delta r}{\Sigma}, \\
  \Gamma^1_{33} & = \frac{\Delta}{\Sigma} \paren[\bigg]{\frac{Ma^2}{\rho^4} (r^2 - a^2 \cos^2\!\theta) \sin^2\!\theta - r} \sin^2(\theta), \\
  \Gamma^2_{00} & = -\frac{Ma^2r}{\rho^6} \sin(2\theta), \\
  \Gamma^2_{03} = \Gamma^2_{30} & = \frac{Mar}{\rho^6} (r^2 + a^2) \sin(2\theta), \\
  \Gamma^2_{11} & = \frac{a^2}{2\Delta\Sigma} \sin(2\theta), \\
  \Gamma^2_{12} = \Gamma^2_{21} & = \frac{r}{\Sigma}, \\
  \Gamma^2_{22} & = -\frac{a^2}{2\Sigma} \sin(2\theta), \\
  \Gamma^2_{33} & = -\frac{1}{2\Sigma} \paren[\bigg]{\Delta + \frac{2Mr}{\rho^4} (r^2 + a^2)^2} \sin(2\theta), \\
  \Gamma^3_{01} = \Gamma^3_{10} & = \frac{Ma}{\Delta\rho^4} (r^2 - a^2 \cos^2\!\theta), \\
  \Gamma^3_{02} = \Gamma^3_{20} & = -\frac{2Mar}{\rho^4} \cot\theta, \\
  \Gamma^3_{13} = \Gamma^3_{31} & = \frac{1}{\Delta\Sigma} \paren[\bigg]{r (\Sigma - 2Mr) - \frac{Ma^2}{\Sigma} (r^2 - a^2 \cos^2\!\theta) \sin^2\!\theta}, \\
  \Gamma^3_{23} = \Gamma^3_{32} & = \frac{Ma^2r}{\rho^4} \sin(2\theta) + \cot\theta.
\end{align} \end{subequations}

The volumes are given by
\begin{equation} \begin{split}
  \Delta V & = \frac{1}{3} (r_+^3 - r_-^3) \abs{\cos\theta_--\cos\theta_+} (\phi_+ - \phi_-) \\
  & \qquad \times \paren[\big]{r_-^2 + r_- r_+ + r_+^2 + a^2 (\cos^2\!\theta_- + \cos\theta_- \cos\theta_+ + \cos^2\!\theta_+)},
\end{split} \end{equation}
the areas by
\begin{subequations} \begin{align}
  \Delta A_{i_-,j,k} & = \frac{1}{3} (\phi_+ - \phi_-) \abs{\cos\theta_--\cos\theta_+} \paren[\big]{3r_-^2+a^2(\cos^2\!\theta_-+\cos\theta_-\cos\theta_++\cos^2\!\theta_+)}, \\
  \Delta A_{i,j_-,k} & = \frac{1}{3} (r_+ - r_-) \abs{\sin\theta_-} (\phi_+ - \phi_-) (r_-^2 + r_- r_+ + r_+^2 + 3 a^2 \cos^2\!\theta_-), \\
  \Delta A_{i,j,k-} & = \frac{1}{3} (r_+ - r_-) \abs{\cos\!\theta_--\cos\!\theta_+} \notag \\
  & \qquad \times \paren[\big]{r_-^2 + r_- r_+ + r_+^2 + a^2 (\cos^2\!\theta_- + \cos\theta_- \cos\theta_+ + \cos^2\!\theta_+)},
\end{align} \end{subequations}
and the lengths by
\begin{subequations} \begin{align}
  \Delta L_{i,j_-,k_-} & = \frac{1}{3} (r_+ - r_-) \abs{\sin\theta_-} (r_-^2 + r_- r_+ + r_+^2 + 3 a^2 \cos^2\!\theta_-), \\
  \Delta L_{i_-,j,k_-} & = r_-^2 \abs{\cos\theta_--\cos\theta_+} + \frac{1}{3} a^2 (\cos^3\!\theta_- - \cos^3\!\theta_+), \\
  \Delta L_{i_-,j_-,k} & = \Sigma_- \abs{\sin\theta_-} (\phi_+ - \phi_-).
\end{align} \end{subequations}
Note that while these cannot be factored into the schematic form $f(r) g(\theta) h(\phi)$, they can still be separated into a sum of such terms, and therefore only require 1D storage. As was the case with sinusoidal coordinates, we find it convenient to bound some of the cell widths. By replacing the integrands $\sqrt{g_{ii}}$ with more tractable forms, we find
\begin{subequations} \begin{align}
  \Delta W_r & \geq \int_{r_-}^{r_+} \frac{r}{\sqrt{\Delta}} \, \dr = \sqrt{\Delta_+} - \sqrt{\Delta_-} + M \log\paren[\bigg]{\frac{r_++\sqrt{\Delta_+}-M}{r_-+\sqrt{\Delta_-}-M}}, \\
  \Delta W_\theta & \geq \int_{\theta_-}^{\theta_+} r \, \dth = \bar{r} (\theta_+ - \theta_-), \\
  \Delta W_\phi & = \abs{\sin\bar{\theta}} (\phi_+ - \phi_-) \sqrt{\bar{r}^2+a^2+\frac{2Ma^2\bar{r}}{\bar{\Sigma}}\sin^2\!\bar{\theta}}.
\end{align} \end{subequations}
The midpoints somewhat more complicated. While the azimuthal coordinate is trivial, the other two obey cubic equations:
\begin{subequations} \begin{gather}
  \bar{r}^3 + \beta \bar{r} - \frac{1}{2} \paren[\big]{r_-^3+r_+^3+\beta(r_-+r_+)} = 0, \\
  \cos^3(\theta_j) + \frac{4\alpha}{3a^2} \cos\bar{\theta} - \frac{2}{3} (\cos^3\!\theta_- + \cos^3\!\theta_+) - \frac{2\alpha}{3a^2} (\cos\theta_- + \cos\theta_+) = 0, \\
  \bar{\phi} = \frac{1}{2} (\phi_- + \phi_+),
\end{gather} \end{subequations}
with
\begin{align}
  \alpha & = r_-^2 + r_- r_+ + r_+^2, \\
  \beta & = a^2 \paren{\cos^2\!\theta_-+\cos\theta_-\cos\theta_++\cos^2\!\theta_+}.
\end{align}
While solving these equations numerically is not difficult, the problem is in the coupling between $r$ and $\theta$; the exact midpoint formulas are not separable. For simplicity of code design, then, the averages
\begin{subequations} \begin{align}
  \bar{r} & \approx \frac{1}{2} (r_- + r_+), \\
  \bar{\theta} & \approx \frac{1}{2} (\theta_- + \theta_+), \\
  \bar{\phi} & = \frac{1}{2} (\phi_- + \phi_+)
\end{align} \end{subequations}
are used instead. These are still second-order accurate in space.

The transformations \eqref{eq:to_local} are
\begin{subequations} \begin{align}
  \paren[\big]{\tensor{M}{^{\hat{\mu}}_\nu}}_{i_-,j,k} & =
  \begin{pmatrix}
    \tensor{M}{^{\hat{t}}_0} & 0                    & 0             & 0             \\
    0                        & \sqrt{\Sigma/\Delta} & 0             & 0             \\
    0                        & 0                    & \sqrt{\Sigma} & 0             \\
    \tensor{M}{^{\hat{z}}_0} & 0                    & 0             & \sqrt{g_{33}}
  \end{pmatrix}, \\
  \paren[\big]{\tensor{M}{^{\hat{\mu}}_\nu}}_{i,j_-,k} & =
  \begin{pmatrix}
    \tensor{M}{^{\hat{t}}_0} & 0                    & 0             & 0             \\
    0                        & 0                    & \sqrt{\Sigma} & 0             \\
    \tensor{M}{^{\hat{y}}_0} & 0                    & 0             & \sqrt{g_{33}} \\
    0                        & \sqrt{\Sigma/\Delta} & 0             & 0
  \end{pmatrix}, \\
  \paren[\big]{\tensor{M}{^{\hat{\mu}}_\nu}}_{i,j,k_-} & =
  \begin{pmatrix}
    \tensor{M}{^{\hat{t}}_0} & 0                    & 0             & 0                                   \\
    \tensor{M}{^{\hat{x}}_0} & 0                    & 0             & \sin\theta/\tensor{M}{^{\hat{t}}_0} \\
    0                        & \sqrt{\Sigma/\Delta} & 0             & 0                                   \\
    0                        & 0                    & \sqrt{\Sigma} & 0
  \end{pmatrix},
\end{align} \end{subequations}
where we have
\begin{subequations} \begin{align}
  \tensor{M}{^{\hat{t}}_0} & = \paren[\bigg]{\frac{\Delta\Sigma}{\Delta\Sigma+2Mr(r^2+a^2)}}^{1/2}, \\
  \tensor{M}{^{\hat{z}}_0} & = -\frac{2Mar\sqrt{g_{33}}}{r^4+(r^2+2Mr+\Delta\cos^2\!\theta)a^2}, \\
  \tensor{M}{^{\hat{y}}_0} & = -\frac{2Mar\sin\theta}{r^4+(r^2+2Mr+\Delta\cos^2\!\theta)a^2} \paren[\bigg]{r^2+a^2+\frac{2Ma^2r}{\Sigma}\sin^2\!\theta}^{1/2}, \\
  \tensor{M}{^{\hat{x}}_0} & = -\frac{2Mar\sin\theta}{\sqrt{\Delta\Sigma^2+2Mr\Sigma(r^2+a^2)}}.
\end{align} \end{subequations}
The inverse transformations \eqref{eq:to_global} are
\begin{subequations} \begin{align}
  \paren[\big]{\tensor{M}{^\mu_{\hat{\nu}}}}_{i_-,j,k} & =
  \begin{pmatrix}
    \tensor{M}{^0_{\hat{t}}} & 0                    & 0               & 0                        \\
    0                        & \sqrt{\Delta/\Sigma} & 0               & 0                        \\
    0                        & 0                    & 1/\sqrt{\Sigma} & 0                        \\
    \tensor{M}{^3_{\hat{t}}} & 0                    & 0               & \tensor{M}{^3_{\hat{z}}}
  \end{pmatrix}, \\
  \paren[\big]{\tensor{M}{^\mu_{\hat{\nu}}}}_{i,j_-,k} & =
  \begin{pmatrix}
    \tensor{M}{^0_{\hat{t}}} & 0               & 0                        & 0                    \\
    0                        & 0               & 0                        & \sqrt{\Delta/\Sigma} \\
    0                        & 1/\sqrt{\Sigma} & 0                        & 0                    \\
    \tensor{M}{^3_{\hat{t}}} & 0               & \tensor{M}{^3_{\hat{y}}} & 0
  \end{pmatrix}, \\
  \paren[\big]{\tensor{M}{^\mu_{\hat{\nu}}}}_{i,j,k_-} & =
  \begin{pmatrix}
    \tensor{M}{^0_{\hat{t}}} & 0                        & 0                    & 0               \\
    0                        & 0                        & \sqrt{\Delta/\Sigma} & 0               \\
    0                        & 0                        & 0                    & 1/\sqrt{\Sigma} \\
    \tensor{M}{^3_{\hat{t}}} & \tensor{M}{^3_{\hat{x}}} & 0                    & 0
  \end{pmatrix},
\end{align} \end{subequations}
where we have
\begin{subequations} \begin{align}
  \tensor{M}{^0_{\hat{t}}} & = \paren[\bigg]{\frac{\Delta\Sigma+2Mr(r^2+a^2)}{\Delta\Sigma}}^{1/2}, \\
  \tensor{M}{^3_{\hat{t}}} & = \frac{2Mar}{\Delta\Sigma} \paren[\bigg]{\frac{\Delta\Sigma}{\Delta\Sigma+2Mr(r^2+a^2)}}^{1/2}, \\
  \tensor{M}{^3_{\hat{z}}} = \tensor{M}{^3_{\hat{y}}} = \tensor{M}{^3_{\hat{x}}} & = \csc\theta \paren[\bigg]{\frac{\Sigma}{r^2+a^2+2Ma^2r\sin^2\!\theta}}^{1/2}.
\end{align} \end{subequations}

\subsection{Kerr--Schild Coordinates}
\label{sec:examples:kerr:kerr-schild}

From Boyer--Lindquist coordinates, define ingoing Kerr coordinates $(\tilde{V},r,\theta,\tilde{\phi})$ according to
\begin{subequations} \begin{align}
  \dd\tilde{V} & = \dt + \frac{r^2+a^2}{\Delta} \dr, \\
  \dd\tilde{\phi} & = \dph + \frac{a}{\Delta} \dr
\end{align} \end{subequations}
\citep[Box~33.2.4]{MTW}.
We then define the so-called Kerr--Schild coordinates $(\tilde{t},r,\theta,\tilde{\phi})$ with
\begin{equation}
  \dd\tilde{t} = \dd\tilde{V} - \dr
\end{equation}
\citep[cf.][\extref{35}]{Gammie2003}. Note that there are multiple coordinate systems in the literature named ``Kerr--Schild.''

Hereafter in this section we will omit tildes distinguishing Kerr--Schild from Boyer--Lindquist quantities. For convenience we also introduce the shorthand
\begin{equation}
  \Xi = r^2 - a^2 \cos^2\!\theta.
\end{equation}

The metric and its inverse are
\begin{subequations} \begin{align}
  g_{\mu\nu} & =
  \begin{pmatrix}
    -(1-2Mr/\Sigma)               & 2Mr/\Sigma                      & 0      & -(2Mar/\Sigma) \sin^2\!\theta   \\
    2Mr/\Sigma                    & 1+2Mr/\Sigma                    & 0      & -(1+2Mr/\Sigma) a\sin^2\!\theta \\
    0                             & 0                               & \Sigma & 0                               \\
    -(2Mar/\Sigma) \sin^2\!\theta & -(1+2Mr/\Sigma) a\sin^2\!\theta & 0      & g_{33}
  \end{pmatrix}, \\
  g^{\mu\nu} & =
  \begin{pmatrix}
    -(1+2Mr/\Sigma) & 2Mr/\Sigma    & 0        & 0                      \\
    2Mr/\Sigma      & \Delta/\Sigma & 0        & a/\Sigma               \\
    0               & 0             & 1/\Sigma & 0                      \\
    0               & a/\Sigma      & 0        & 1/\Sigma\sin^2\!\theta
  \end{pmatrix},
\end{align} \end{subequations}
with
\begin{equation}
  g_{33} = \paren[\Bigg]{r^2 + a^2 + \paren[\bigg]{\frac{2Ma^2r}{\Sigma}} \sin^2\!\theta} \sin^2\!\theta,
\end{equation}
and we still have
\begin{equation}
  \sqrt{-g} = \Sigma \abs{\sin\theta}.
\end{equation}

None of the connection coefficients vanish. The full set of them is
\begin{subequations} \begin{align}
  \Gamma^0_{00} & = \frac{2M^2\Xi r}{\Sigma^3}, \\
  \Gamma^0_{01} = \Gamma^0_{10} & = \frac{M\Xi}{\Sigma^2} \paren[\bigg]{1+\frac{2Mr}{\Sigma}}, \\
  \Gamma^0_{02} = \Gamma^0_{20} & = -\frac{2Ma^2r}{\Sigma^2} \sin\theta \cos\theta, \\
  \Gamma^0_{03} = \Gamma^0_{30} & = -\frac{2M^2ar\Xi}{\Sigma^3} \sin^2\!\theta, \\
  \Gamma^0_{11} & = \frac{2M\Xi}{\Sigma^2} \paren[\bigg]{1+\frac{Mr}{\Sigma}}, \\
  \Gamma^0_{12} = \Gamma^0_{21} & = -\frac{2Ma^2r}{\Sigma^2} \sin\theta \cos\theta, \\
  \Gamma^0_{13} = \Gamma^0_{31} & = -\frac{Ma\Xi}{\Sigma^2} \paren[\bigg]{1+\frac{2Mr}{\Sigma}} \sin^2\!\theta, \\
  \Gamma^0_{22} & = -\frac{2Mr^2}{\Sigma}, \\
  \Gamma^0_{23} = \Gamma^0_{32} & = \frac{2Ma^3r}{\Sigma^2} \sin^3\!\theta \cos\theta, \\
  \Gamma^0_{33} & = -\frac{2Mr}{\Sigma} \paren[\bigg]{r-\frac{Ma^2\Xi}{\Sigma^2}\sin^2\!\theta} \sin^2\!\theta;
\end{align} \end{subequations}
\begin{subequations} \begin{align}
  \Gamma^1_{00} & = \frac{M\Delta\Xi}{\Sigma^3}, \\
  \Gamma^1_{01} = \Gamma^1_{10} & = \frac{M\Xi(\Delta-\Sigma)}{\Sigma^3}, \\
  \Gamma^1_{02} = \Gamma^1_{20} & = 0, \\
  \Gamma^1_{03} = \Gamma^1_{30} & = -\frac{Ma\Delta\Xi}{\Sigma^3} \sin^2\!\theta, \\
  \Gamma^1_{11} & = \frac{M\Xi(\Delta-2\Sigma)}{\Sigma^3}, \\
  \Gamma^1_{12} = \Gamma^1_{21} & = -\frac{a^2}{\Sigma} \sin\theta \cos\theta, \\
  \Gamma^1_{13} = \Gamma^1_{31} & = \frac{a}{\Sigma^3} \parenopen[\bigg]{r^3 (r^2 + 2M^2)} \notag \\
  & \qquad \parenclose[\bigg]{+ a^2 \paren[\Big]{ra^2\cos^4\!\theta + \paren[\big]{2r^3 + M (\Delta - \Sigma)} \cos^2\!\theta - Mr^2 \sin^2\!\theta}}, \\
  \Gamma^1_{22} & = -\frac{r\Delta}{\Sigma}, \\
  \Gamma^1_{23} = \Gamma^1_{32} & = 0, \\
  \Gamma^1_{33} & = -\frac{\Delta}{\Sigma} \paren[\bigg]{r-\frac{Ma^2\Xi}{\Sigma^2}\sin^2\!\theta} \sin^2\!\theta;
\end{align} \end{subequations}
\begin{subequations} \begin{align}
  \Gamma^2_{00} & = -\frac{2Ma^2r}{\Sigma^3} \sin\theta \cos\theta, \\
  \Gamma^2_{01} = \Gamma^2_{10} & = -\frac{2Ma^2r}{\Sigma^3} \sin\theta \cos\theta, \\
  \Gamma^2_{02} = \Gamma^2_{20} & = 0, \\
  \Gamma^2_{03} = \Gamma^2_{30} & = \frac{2Mar}{\Sigma^3} (r^2+a^2) \sin\theta \cos\theta, \\
  \Gamma^2_{11} & = -\frac{2Ma^2r}{\Sigma^3} \sin\theta \cos\theta, \\
  \Gamma^2_{12} = \Gamma^2_{21} & = \frac{r}{\Sigma}, \\
  \Gamma^2_{13} = \Gamma^2_{31} & = \frac{a}{\Sigma} \paren[\bigg]{1+\frac{2Mr}{\Sigma^2}(r^2+a^2)} \sin\theta \cos\theta, \\
  \Gamma^2_{22} & = -\frac{a^2}{\Sigma} \sin\theta \cos\theta, \\
  \Gamma^2_{23} = \Gamma^2_{32} & = 0, \\
  \Gamma^2_{33} & = -\frac{1}{\Sigma} \paren[\bigg]{\Delta+\frac{2Mr}{\Sigma^2}(r^2+a^2)^2} \sin\theta \cos\theta;
\end{align} \end{subequations}
and
\begin{subequations} \begin{align}
  \Gamma^3_{00} & = \frac{Ma\Xi}{\Sigma^3}, \\
  \Gamma^3_{01} = \Gamma^3_{10} & = \frac{Ma\Xi}{\Sigma^3}, \\
  \Gamma^3_{02} = \Gamma^3_{20} & = -\frac{2Mar}{\Sigma^2} \cot\theta, \\
  \Gamma^3_{03} = \Gamma^3_{30} & = -\frac{Ma^2\Xi}{\Sigma^3} \sin^2\!\theta, \\
  \Gamma^3_{11} & = \frac{Ma\Xi}{\Sigma^3}, \\
  \Gamma^3_{12} = \Gamma^3_{21} & = -\frac{a}{\Sigma^2} (2Mr+\Sigma) \cot\theta, \\
  \Gamma^3_{13} = \Gamma^3_{31} & = \frac{1}{\Sigma^3} (r\Sigma^2 - Ma^2\Xi\sin^2\!\theta), \\
  \Gamma^3_{22} & = -\frac{ar}{\Sigma}, \\
  \Gamma^3_{23} = \Gamma^3_{32} & = \frac{2Ma^2r}{\Sigma^2} \sin\theta \cos\theta + \cot\theta, \\
  \Gamma^3_{33} & = \frac{a}{\Sigma^3} (Ma^2\Xi\sin^2\!\theta - r\Sigma^2) \sin^2\!\theta.
\end{align} \end{subequations}

The volumes, areas, and lengths have the same formulas as in Boyer--Lindquist coordinates. For completeness we record them here. The volumes are given by
\begin{equation} \begin{split}
  \Delta V & = \frac{1}{3} (r_+ - r_-) \abs{\cos\theta_--\cos\theta_+} (\phi_+ - \phi_-) \\
  & \qquad \times \paren[\big]{r_-^2 + r_- r_+ + r_+^2 + a^2 (\cos^2\!\theta_- + \cos\theta_- \cos\theta_+ + \cos^2\!\theta_+)},
\end{split} \end{equation}
the areas by
\begin{subequations} \begin{align}
  \Delta A_{i_-,j,k} & = \frac{1}{3} \abs{\cos\theta_--\cos\theta_+} (\phi_+ - \phi_-) \paren[\big]{3r_-^2+a^2(\cos^2\!\theta_-+\cos\theta_-\cos\theta_++\cos^2\!\theta_+)}, \\
  \Delta A_{i,j_-,k} & = \frac{1}{3} (r_+ - r_-) \abs{\sin\theta_-} (\phi_+ - \phi_-) (r_-^2 + r_- r_+ + r_+^2 + 3 a^2 \cos^2\!\theta_-), \\
  \Delta A_{i,j,k-} & = \frac{1}{3} (r_+ - r_-) \abs{\cos\!\theta_--\cos\!\theta_+} \notag \\
  & \qquad \times \paren[\big]{r_-^2 + r_- r_+ + r_+^2 + a^2 (\cos^2\!\theta_- + \cos\theta_- \cos\theta_+ + \cos^2\!\theta_+)},
\end{align} \end{subequations}
and the lengths by
\begin{subequations} \begin{align}
  \Delta L_{i,j_-,k_-} & = \frac{1}{3} (r_+ - r_-) \abs{\sin\theta_-} (r_-^2 + r_- r_+ + r_+^2 + 3 a^2 \cos^2\!\theta_-), \\
  \Delta L_{i_-,j,k_-} & = \frac{1}{3} \abs{\cos\theta_--\cos\theta_+} \paren[\big]{3r_-^2 + a^2 (\cos^2\!\theta_- + \cos\theta_-\cos\theta_+ + \cos^2\!\theta_+)}, \\
  \Delta L_{i_-,j_-,k} & = (\phi_+ - \phi_-) \abs{\sin\theta_-} (r_-^2 + a^2 \cos^2\!\theta_-).
\end{align} \end{subequations}
Again we bound two of the cell widths:
\begin{subequations} \begin{align}
  \Delta W_r & \geq \int_{r_-}^{r_+} \frac{r+M}{\sqrt{r^2+M^2}} \, \dr = \sqrt{r_+^2+M^2} - \sqrt{r_-^2+M^2} + M \log\paren[\bigg]{\frac{\sqrt{r_+^2+M^2}+r_+}{\sqrt{r_-^2+M^2}+r_-}}, \\
  \Delta W_\theta & \geq \int_{\theta_-}^{\theta_+} r \, \dth = \bar{r} (\theta_+ - \theta_-), \\
  \Delta W_\phi & = \abs{\sin\bar{\theta}} (\phi_+ - \phi_-) \sqrt{\bar{r}^2+a^2+\frac{2Ma^2\bar{r}}{\bar{r}^2+a^2\cos^2\!\bar{\theta}}\sin^2\!\bar{\theta}}.
\end{align} \end{subequations}
The exact midpoints obey the same relations as in Boyer--Lindquist coordinates:
\begin{subequations} \begin{gather}
  \bar{r}^3 + \beta \bar{r} - \frac{1}{2} \paren[\big]{r_-^3+r_+^3+\beta(r_-+r_+)} = 0, \\
  \cos^3(\theta_j) + \frac{4\alpha}{3a^2} \cos\bar{\theta} - \frac{2}{3} (\cos^3\!\theta_- + \cos^3\!\theta_+) - \frac{2\alpha}{3a^2} (\cos\theta_- + \cos\theta_+) = 0, \\
  \bar{\phi} = \frac{1}{2} (\phi_- + \phi_+),
\end{gather} \end{subequations}
with
\begin{align}
  \alpha & = r_-^2 + r_- r_+ + r_+^2, \\
  \beta & = a^2 \paren{\cos^2\!\theta_-+\cos\theta_-\cos\theta_++\cos^2\!\theta_+}.
\end{align}
In practice we therefore use the formulas
\begin{subequations} \begin{align}
  \bar{r} & \approx \frac{1}{2} (r_- + r_+), \\
  \bar{\theta} & \approx \frac{1}{2} (\theta_- + \theta_+), \\
  \bar{\phi} & = \frac{1}{2} (\phi_- + \phi_+).
\end{align} \end{subequations}

The transformations \eqref{eq:to_local} are given by
\begin{subequations} \begin{align}
  \paren[\big]{\tensor{M}{^{\hat{\mu}}_\nu}}_{i_-,j,k} & =
  \begin{pmatrix}
    1/\!\sqrt{g_{11}}        & 0                        & 0             & 0                           \\
    \tensor{M}{^{\hat{x}}_t} & \tensor{M}{^{\hat{x}}_r} & 0             & 0                           \\
    0                        & 0                        & \sqrt{\Sigma} & 0                           \\
    \tensor{M}{^{\hat{z}}_t} & \tensor{M}{^{\hat{z}}_r} & 0             & \tensor{M}{^{\hat{z}}_\phi}
  \end{pmatrix}, \\
  \paren[\big]{\tensor{M}{^{\hat{\mu}}_\nu}}_{i,j_-,k} & =
  \begin{pmatrix}
    1/\!\sqrt{g_{11}}       & 0             & 0             & 0                               \\
    0                       & 0             & \sqrt{\Sigma} & 0                               \\
    0                       & 0             & 0             & \sqrt{\Sigma} \abs{\sin\theta}  \\
    2Mr/\Sigma\sqrt{g_{11}} & \sqrt{g_{11}} & 0             & -a \sin^2\!\theta \sqrt{g_{11}}
  \end{pmatrix}, \\
  \paren[\big]{\tensor{M}{^{\hat{\mu}}_\nu}}_{i,j,k_-} & =
  \begin{pmatrix}
    1/\!\sqrt{g_{11}}       & 0             & 0             & 0                               \\
    0                       & 0             & 0             & \sqrt{\Sigma} \abs{\sin\theta}  \\
    2Mr/\Sigma\sqrt{g_{11}} & \sqrt{g_{11}} & 0             & -a \sin^2\!\theta \sqrt{g_{11}} \\
    0                       & 0             & \sqrt{\Sigma} & 0
  \end{pmatrix},
\end{align} \end{subequations}
where we have
\begin{subequations} \begin{align}
  \tensor{M}{^{\hat{x}}_t} & = 2Mr \sqrt{\Sigma} \paren[\big]{(\Sigma + 2Mr) (r^4 + a^2r^2 + 2Ma^2r + \Delta a^2\cos^2\!\theta)}^{-1/2}, \\
  \tensor{M}{^{\hat{x}}_r} & = \sqrt{\frac{\Sigma+2Mr}{r^2+a^2+(2Ma^2r/\Sigma)\sin^2\!\theta}}, \\
  \tensor{M}{^{\hat{z}}_t} & = -\frac{2Mar\abs{\sin\theta}\sqrt{r^2+a^2+(2Ma^2r/\Sigma)\sin^2\!\theta}}{r^4+a^2r^2+2Ma^2r+\Delta a^2\cos^2\!\theta}, \\
  \tensor{M}{^{\hat{z}}_r} & = -\frac{a(\Sigma+2Mr)\abs{\sin\theta}\sqrt{r^2+a^2+(2Ma^2r/\Sigma)\sin^2\!\theta}}{r^4+a^2r^2+2Ma^2r+\Delta a^2\cos^2\!\theta}, \\
  \tensor{M}{^{\hat{z}}_\phi} & = \abs{\sin\theta} \sqrt{r^2+a^2+(2Ma^2r/\Sigma)\sin^2\!\theta}.
\end{align} \end{subequations}
The inverse transformations \eqref{eq:to_global} are given by
\begin{subequations} \begin{align}
  (\tensor{M}{^\mu_{\hat{\nu}}})_{i_-,j,k} & =
  \begin{pmatrix}
    1/\tensor{M}{^{\hat{t}}_t}                                                 & 0                                                                             & 0                               & 0                             \\
    -\tensor{M}{^{\hat{x}}_t}/\tensor{M}{^{\hat{t}}_t}\tensor{M}{^{\hat{x}}_r} & 1/\tensor{M}{^{\hat{x}}_r}                                                    & 0                               & 0                             \\
    0                                                                          & 0                                                                             & 1/\tensor{M}{^{\hat{y}}_\theta} & 0                             \\
    0                                                                          & -\tensor{M}{^{\hat{z}}_r}/\tensor{M}{^{\hat{x}}_r}\tensor{M}{^{\hat{z}}_\phi} & 0                               & 1/\tensor{M}{^{\hat{z}}_\phi}
  \end{pmatrix}, \\
  (\tensor{M}{^\mu_{\hat{\nu}}})_{i,j_-,k} & =
  \begin{pmatrix}
    1/\tensor{M}{^{\hat{t}}_t}                                                 & 0                               & 0                                                                                & 0                          \\
    -\tensor{M}{^{\hat{z}}_t}/\tensor{M}{^{\hat{t}}_t}\tensor{M}{^{\hat{z}}_r} & 0                               & -\tensor{M}{^{\hat{z}}_\phi}/\tensor{M}{^{\hat{z}}_r}\tensor{M}{^{\hat{y}}_\phi} & 1/\tensor{M}{^{\hat{z}}_r} \\
    0                                                                          & 1/\tensor{M}{^{\hat{x}}_\theta} & 0                                                                                & 0                          \\
    0                                                                          & 0                               & 1/\tensor{M}{^{\hat{y}}_\phi}                                                    & 0
  \end{pmatrix}, \\
  (\tensor{M}{^\mu_{\hat{\nu}}})_{i,j,k_-} & =
  \begin{pmatrix}
    1/\tensor{M}{^{\hat{t}}_t}                                                 & 0                                                                                & 0                          & 0                               \\
    -\tensor{M}{^{\hat{y}}_t}/\tensor{M}{^{\hat{t}}_t}\tensor{M}{^{\hat{y}}_r} & -\tensor{M}{^{\hat{y}}_\phi}/\tensor{M}{^{\hat{y}}_r}\tensor{M}{^{\hat{x}}_\phi} & 1/\tensor{M}{^{\hat{y}}_r} & 0                               \\
    0                                                                          & 0                                                                                & 0                          & 1/\tensor{M}{^{\hat{z}}_\theta} \\
    0                                                                          & 1/\tensor{M}{^{\hat{x}}_\phi}                                                    & 0                          & 0
  \end{pmatrix},
\end{align} \end{subequations}
where the $i,j,k$ indices in the matrix entries are understood to match those on the left-hand side of their respective equation.

\clearpage
\singlespacing
\phantomsection
\addcontentsline{toc}{chapter}{References}
\bibliographystyle{plainnat}
\renewcommand{\bibname}{References}
\bibliography{references}

\end{document}